\documentclass[11pt]{article}

\usepackage{graphicx}  
\usepackage[section]{placeins}

\def\simless{\mathbin{\lower 3pt\hbox
   {$\rlap{\raise 5pt\hbox{$\char'074$}}\mathchar''7218$}}}   
\def\simgreat{\mathbin{\lower 3pt\hbox
   {$\rlap{\raise 5pt\hbox{$\char'076$}}\mathchar''7218$}}}   

\begin{document}

\title{The Equilibrium Structure of CDM Halos}
\author{
Paul R. Shapiro\footnotemark[2]\footnote{Department of Astronomy, University of Texas, 
Austin, 78712, US},
Ilian T. Iliev\footnotemark[1]\footnote{Canadian Institute for Theoretical Astrophysics, 
University of Toronto, 60 St. George Street, Toronto, ON M5S 3H8, Canada},
Hugo Martel\footnotemark[3]\footnote{D\'epartement de physique, de g\'enie 
physique et d'optique, Universit\'e Laval, Sainte-Foy, QC G1K 7P4, Canada}, 
Kyungjin Ahn\footnotemark[2], and 
Marcelo A. Alvarez\footnotemark[2]
}

\maketitle              

\pagestyle{myheadings} 
\thispagestyle{plain}
\markboth{Iliev et al.}{The Equilibrium Structure of CDM Halos}
\setcounter{page}{1}    

\begin{abstract}
Dark-matter (DM) halos are the scaffolding around which galaxies and 
clusters are built. They form when the gravitational instability of
primordial density fluctuations causes regions which are denser than average 
to slow their cosmic expansion, recollapse, and virialize. Understanding
the equilibrium structure of these halos is thus a prerequisite for 
understanding galaxy and cluster formation. Numerical N-body simulations 
of structure formation from Gaussian-random-noise initial conditions in the 
Cold Dark Matter (CDM) universe find a universal internal structure 
for halos.  Objects as different in size and mass as dwarf 
spheroidal galaxies and galaxy clusters are predicted to have 
halos with the same basic structure when properly rescaled, independent of 
halo mass, of the shape of the power spectrum of primordial density 
fluctuations, and of the cosmological background parameters.  This remarkable
universality is a fundamental prediction of the CDM model, but our knowledge
is limited to the ``empirical'' N-body simulation results, with little 
analytical understanding.  We summarize here our attempts to fill this gap,
in an effort to derive and give physical insight to the numerical results
and extend them beyond the range of numerical simulation:  (1) Simulated halos
which form from highly simplified initial conditions involving gravitational
instability in a cosmological pancake show that many of the universal 
properties of CDM halos are generic to cosmological gravitational collapse and
do not require Gaussian-random-noise density fluctuations or hierarchical
clustering.  (2) A fluid approximation derived from the Boltzmann equation 
yields an analytical theory of halo dynamics which can explain many of the 
N-body results if the complex mass assembly history of individual halos is
approximated by continuous spherical infall.  The universal mass growth history
reported for CDM N-body halos corresponds to a time-varying infall rate
which self-consistently determines the shape of the equilibrium halo profile
and its evolution without regard for the complicated details of the merger
process.  (3) The first fully-cosmological, similarity solutions for halo 
formation in the presence of collisionality provide an analytical theory of 
the effect of the self-interacting dark
matter (SIDM) hypothesis on CDM halo density profiles.  Collisions transport
heat inward which flattens the central cusp, but continuous infall pumps 
energy into the halo to stabilize the core against gravothermal catastrophe.
(4) The postcollapse equilibrium halo structure derived
by matching spherical top-hat collapse to a unique, minimum-energy solution of
the isothermal Lane-Emden equation -- the Truncated Isothermal Sphere (TIS) 
model -- reproduces well the average properties of simulated CDM halos.  The
TIS yields a universal, self-similar equilibrium structure whose parameters 
are determined by the halo's total mass and collapse redshift. These latter 
two parameters are statistically correlated, however, since halos of the same 
mass form on average at the same epoch, with small-mass objects forming first 
and then merging hierarchically to form larger-mass objects.  The structural 
properties (e.g. sizes, mass profiles, velocity dispersions and virial 
temperatures) of dark-matter dominated halos of different masses, therefore, 
should reflect this statistical correlation, an imprint of the statistical 
properties of the primordial density fluctuations which formed them. 
When combined with the Press-Schechter (PS) formalism, the TIS model can 
predict these statistical correlations of halo properties analytically, which 
we compare to observational data on these correlations, providing a 
fundamental test of the CDM model, which probes the shape of the power 
spectrum of primordial density fluctuations and the cosmological background 
parameters.  (5) For observational tests which are sensitive to the shape of 
the inner mass profile at the very center, the TIS model provides a convenient 
analytical tool for studying the effect of a small nonsingular core on CDM 
halo predictions.  As an example, strong gravitational lensing properties of 
CDM halos with nonsingular cores are contrasted with those with singular, 
cuspy cores.

\end{abstract}

\section{Introduction}
\label{intro}
Numerical N-body simulations of structure formation in the
Cold Dark Matter (CDM) universe show a cosmic web of lumps, filaments,
and sheets. This web evolves in a self-similar way, with smaller mass
structures forming first and merging to form larger-mass structures
later, in a continuous sequence of mass assembly. At each epoch, the
web produces gravitationally-bound, quasi-spherical {}``halos{}''
in virial and hydrostatic equilibrium. These virialized halos are
the sites of galaxy and cluster formation. Their universal equilibrium
structure is a fundamental prediction of the CDM model, but our knowledge
is limited to the numerical N-body results, with very little analytical
understanding. We shall describe our attempts to fill this gap in what 
follows.

Most of the progress to date on the formation and evolution of virialized,
dark-matter dominated halos in a CDM universe has been via numerical N-body 
simulations of collisionless dark matter involving Gaussian random noise 
initial conditions.  According to those N-body simulations, the 
spherically-averaged mass distribution inside halos is universal, with a 
density profile
which declines with radius, approaching $\rho\propto r^{-3}$ at large
radii, flattening
near the center to $\rho(r)\propto r^{-\alpha}$ with $\alpha<2$.  Two 
``universal'' profiles bracket the results (NFW, \cite{NFW97}; Moore, 
\cite{Metal98}).  The NFW (Moore) profile has an inner density profile 
$\rho\propto r^{-1}(r^{-1.5})$.  The actual value of $\alpha$ is still 
uncertain, even though the N-body results have advanced to the point of 
including millions of simulation particles within the virial radius of a 
given halo (e.g. \cite{FM01,FM03,FKM,Getal00,JS00,JS02,Ketal01,Petal03}).  
Since the true
inner profile may not be a power-law, comparisons of 
the inner slope for different simulation
results are typically referred to a particular radius (e.g. 
$r=0.01r_{\rm vir}$, where $r_{\rm vir}$ is some measure of the outer 
radius of the virialized region).
The N-body results generally support the conclusion that the same halo density
profile applies to objects as different as dwarf galaxies and galaxy
clusters, independent of halo mass, of the shape of the density fluctuation
power spectrum, and of the background cosmology.
This universality is apparent when comparisons of different halo profiles
are made with density expressed in units of the density 
$\rho_{-2}\equiv\rho(r_{-2})$ at the radius $r_{-2}$ at which each profile
has a logarithmic slope of $-2$, while radii for each halo are expressed
in units of $r_{-2}$ \cite{Navarro03}.
An exception to this universality is claimed by \cite{Ricotti}, who reports
that the value of $\alpha$ depends upon halo mass, shallower for
dwarf galaxies than for clusters, reflecting the different slopes of
the power spectrum at the different scales represented by these objects,
but \cite{CKVG} has challenged this claim.

Much attention has been focused on the N-body results for this inner slope,
since the observed rotation curves of dark-matter dominated dwarf and low
surface brightness (LSB) disk galaxies tend to favor mass profiles with
a flat-density core unlike the singular profiles of the CDM N-body 
simulations (e.g. \cite{DEB,DB,FP,MOO,SIM}; but, for a different view, 
see also \cite{SWA1,SWA2}). On the cluster scale, too, there is 
some evidence from observations of strong gravitational lensing of 
background galaxies by foreground clusters which favors a flatter 
inner halo density profile than is found by the CDM N-body 
simulations (e.g. \cite{GAV,SAN,TKD98}; but see also \cite{CMKS02}).
However, the halo mass fraction contained within the 
disputed inner cusp is actually quite small, so there may yet be dynamical
processes not fully accounted for in the pure N-body simulations which
can affect this small central mass without disturbing the overall universality
of the rest of the profile.

Along with their universal mass profiles, CDM N-body halos also exhibit
several universal properties in their phase-space distributions. Over
most of the halo volume inside the virial radius, the DM particles are
approximately isothermal -- i.e. their velocity dispersion is fairly
constant with radius -- with only a relatively small dip in 
``temperature'' toward the center
(e.g. \cite{SantaBarbara, TBW97}). 
Halo particle velocities are also 
approximately isotropic, with only a mild radial bias in the outer
halo, which gives way to increasing isotropy toward the center
(e.g. \cite{carlberg, CKK, ENF, FM01}). 
The spherically-averaged mass motion at each radius is quite small and
``subsonic;'' the halo is not only in a state of global virial 
equilibrium but is close to hydrostatic equilibrium at each radius, too
-- i.e. it satisfies the spherical Jeans equation (e.g. \cite{TBW97}).

Individual halos in CDM N-body simulations evolve over time, on average,
through a continuous sequence of universal-shaped mass profiles of
increasing total mass \cite{TKGK, VDB, WEC}. This Lagrangian mass
evolution can be characterized by a universal mass accretion history:
$M(a)/M(a_{\rm f})$, where $a$ is the cosmic scale factor and $a_{\rm f}$ is some
particular value of $a$, such as that at which $d\ln M /d\ln a=2$ 
\cite{WEC}. As the mass of each halo grows with time due to the average
effect of mergers and smooth infall, so does the concentration
parameter $c$ of its density profile, where $c\equiv r_{\rm vir}/r_{-2}$, 
roughly as $c(a)/c(a_{\rm f})\propto a/a_{\rm f}$ for $a/a_{\rm f}>1$ \cite{WEC},
after hovering at low values $c \le 3-4$ during the initial phase of most
rapid mass assembly prior to $a_{\rm f}$ \cite{TKGK}.

This description of the CDM halos of N-body simulations is a 
spherically-averaged one, which neglects many details. There is
some scatter in the N-body results about this average description,
of course. Individual halos are not truly spherically symmetric, either,
but only approximately so. The neglect of net angular momentum is
probably not a bad first approximation, since the specific angular
momentum is typically found to be far below that required for rotational
support (e.g. \cite{BE, BUL})\footnote{The specific angular momentum 
profile, $j(M)$, of individual CDM halos averaged over spherical shells
encompassing mass M, also has a universal shape, which has been fitted
by $j(M)\propto M^{s}$ with $s=1.3\pm 0.3$ \cite{BUL2}.}. 
However, the spherically-averaged description also averages out
the small-scale density inhomogeneities inside each halo. This small-scale
inhomogeneity may play an important role in the underlying dynamics
which leads to halo formation and evolution in these N-body simulations.
Regardless of its dynamical significance, this halo substructure has also
been the subject of special attention for its own sake, once it was noticed
that the number of subhalos which typically survive their merger into
a larger halo in the N-body results is much larger than the number of
galaxies observed within the Local Group (e.g. \cite{KKVP, MGGLQST}).

While N-body simulations have made the universal equilibrium structure of
halos described above a fundamental prediction of the CDM model, much 
less progress has been made on the analytical side, to derive and
understand the numerical results and extend them beyond the range
of numerical simulation.  Our purpose in what follows is to summarize our
own attacks on this problem by a hierarchy of approximations, each one
simpler than the last, from 3D gas and N-body simulations of halo formation
involving simplified initial conditions (\S \ref{simul_sect}), to
1D, spherically-symmetric, dynamical models involving a fluid approximation
derived from the Boltzmann equation (\S \ref{fluid_approx_leading_sect}),
to a model for the hydrostatic equilibrium of halos which follows from the 
virialization of top-hat
density perturbations (\S \ref{tis_sect}).  As we shall see, the last 
of these, known as the Truncated Isothermal Sphere (TIS) model, not only
provides an excellent match to many of the average properties of the
halos found by the N-body simulations of CDM from realistic initial
conditions, but one which is conveniently coupled to the
Press-Schechter formalism for the mass function of halos in CDM to
predict the observed statistical properties
of galaxies and clusters analytically.  We summarize several 
such applications of this model in \S \ref{virial_plane_subsect}. In 
\S \ref{lensing_sect}, motivated by the prospect that gravitational
lensing will provide a direct observational test of the CDM halo density
profile, we use the analytical theory of the TIS halo, with its small,
flat-density core at the center of a halo which otherwise closely 
resembles the average CDM halos of N-body simulations, to contrast
the lensing properties of nonsingular and singular CDM halos.

\section{N-body Halos from Simplified Initial Conditions:
    3D Halo Formation by Gravitational Instability of Cosmological Pancakes}
\label{simul_sect}

A natural question which emerges from the results of N-body simulations of
structure formation in CDM is whether the universal properties of dark 
matter halos, like their density profiles and mass accretion histories, 
are a consequence of hierarchical clustering from Gaussian-random-noise density
fluctuations or are in fact more general. To address this question,
we have analyzed the results of simulations with initial conditions that 
are much simpler than a CDM power spectrum with Gaussian-random-noise, while 
retaining the realistic features of continuous and anisotropic infall in 
three dimensions.
Each simulation forms a single halo from the gravitational 
instability of a cosmological pancake. Because we focus on one halo at a time, 
we are able to follow its formation and 
evolution, and find that several of the trends reported for halo evolution 
in the CDM simulations are also present in the simplified pancake model.  Thus,
the pancake instability model serves as a convenient test-bed for halo 
formation which shares not only similar boundary conditions with more 
realistic models of halo formation, but also produces halos that have similar 
structural and evolutionary properties. Some of this work was summarized 
previously in \cite{ASM00,ASM01,ASM03}.

\subsection{Halo Formation via Pancake Instability}

\subsubsection{Unperturbed Pancake}
\label{ics}
Consider the growing mode of a single sinusoidal plane-wave density
fluctuation of comoving wavelength $\lambda_{\rm p}$ and dimensionless
wavevector ${\bf k_p=\hat{x}}$ (length unit = $\lambda_{\rm p}$) in an
Einstein-de Sitter universe dominated by cold, collisionless dark matter 
\cite{SSM}.
Let the initial amplitude $\delta_{\rm i}$ at scale factor $a_{\rm i}$ be chosen so
that the first density caustic forms in the collisionless component at scale
factor $a=a_{\rm c}=a_{\rm i}/\delta_{\rm i}$.

\subsubsection{Perturbations} 
\label{ics1}
Pancakes modeled in this way have been shown
to be gravitationally unstable, leading to filamentation and fragmentation
during the collapse \cite{VAL}.  As an example, we shall perturb the 1D
fluctuation described above by adding to the initial primary pancake mode
two transverse, plane-wave density
fluctuations with equal wavelength $\lambda_{\rm s}=\lambda_{\rm p}$, wavevectors
${\bf k_s}$ pointing along the orthogonal unit vectors ${\bf \hat{y}}$ and
${\bf \hat{z}}$, and smaller initial amplitudes, $\epsilon_y\delta_{\rm i}$ and
$\epsilon_z\delta_{\rm i}$, respectively, where $\epsilon_y \ll 1$ and
$\epsilon_z \ll 1$.  A pancake perturbed by two such density modes will be
referred to as $S_{1,\epsilon_y,\epsilon_z}$.  All results presented here
refer to the case $S_{1,0.2,0.2}$.  The initial
position, velocity, density, and gravitational potential are given by

\begin{equation}
r_l=q_l+\frac{\delta_{\rm i} \epsilon_l}{2\pi k_{\rm p}}\sin2\pi k_{\rm p}q_l
\end{equation}
\begin{equation}
v_l=\frac{\epsilon_l}{2\pi k_{\rm p}}\left(\frac{d\delta}{dt}\right)_{\rm i}\sin2\pi k_{\rm p}q_l
\end{equation}
\begin{equation}
\rho=\frac{\overline{\rho}}{1+\delta_{\rm i}(\cos2\pi
k_{\rm p}q_x+\epsilon_y\cos2\pi k_{\rm p}q_y+\epsilon_z\cos2\pi k_{\rm p}q_z)},
\end{equation}
and
\begin{equation}
\phi=\langle\phi\rangle\left( \cos2\pi
k_{\rm p}x+\epsilon_y\cos2\pi k_{\rm p}y+\epsilon_z\cos2\pi k_{\rm p}z\right),
\end{equation}
where $l\equiv x$, $y$, or $z$, $\epsilon_x\equiv 1$, $q_x$, $q_y$, 
and $q_z$ are the unperturbed particle positions, and $\overline{\rho}$ is the 
cosmic mean matter density.

Such a perturbation leads to the formation of a quasi-spherical mass
concentration in the pancake plane at the intersection of two filaments
(Fig.~\ref{pancake1}). 

\begin{figure}[!t]
\centering
\vspace{3.5in}
\caption{Dark matter particles at $a/a_{\rm c}=3$. (see jpg file)
\label{pancake1}}
\end{figure}

\subsection{N-body and Hydrodynamical Simulations} 

The code we use to simulate the formation of the halo couples the Adaptive
SPH (ASPH) algorithm, first described in \cite{OWE,SHA}, to a P$^3$M 
gravity solver.  The ASPH
method improves on standard SPH by introducing nonspherical, ellipsoidal
smoothing kernels to better track the anisotropic flow that generally
arises during cosmological structure formation.  Two simulations were
carried out, one with gas and one without. In both cases, there were
64$^3$ particles of dark matter, while there were also 64$^3$ gas
particles when gas was included.  The P$^3$M grid was 128$^3$ cells in a
periodic cube size $\lambda_{\rm p}$ on a side, with a comoving softening length
of $r_{\rm soft}=0.3\Delta x=0.3\lambda_{\rm p}/128$, where $\Delta x$ is the
cell size. The initial conditions were those described in \S~\ref{ics1}.

The adiabatic pancake problem (i.e. without radiative cooling) is
self-similar and scale-free, once distance is expressed in units of the
pancake wavelength $\lambda_{\rm p}$ and time is expressed in terms of the
cosmic scale factor $a$ in units of the scale factor $a_{\rm c}$ at which
caustics first form in the dark matter and shocks in the gas \cite{SSM}
In the currently-favored, flat,
cosmological-constant-dominated universe, however, this self-similarity is
broken because $\Omega_{\rm M}/\Omega_\Lambda$ decreases with time, where
$\Omega_{\rm M}$ and $\Omega_\Lambda$ are the matter and vacuum energy density
parameters, respectively. For objects which collapse at high redshift in
such a universe (e.g. dwarf galaxies), the Einstein-de Sitter results are
still applicable as long as we take
$(\Omega_{\rm B}/\Omega_{\rm DM})_{\rm EdS}=
(\Omega_{\rm B}/\Omega_{\rm DM})_{\Lambda}$, where
$\Omega_{\rm B}$ and $\Omega_{\rm DM}$ are the baryon and dark matter density
parameters.  If $\Omega_{\rm B}=0.045$, $\Omega_{\rm DM}=0.255$, and
$\Omega_\Lambda=0.7$ at present, then the EdS results are applicable if we
take $\Omega_{\rm B}=0.15$ and $\Omega_{\rm DM}=0.85$, instead.

\begin{figure}[t!]
\centering
\includegraphics[width=5.5in]{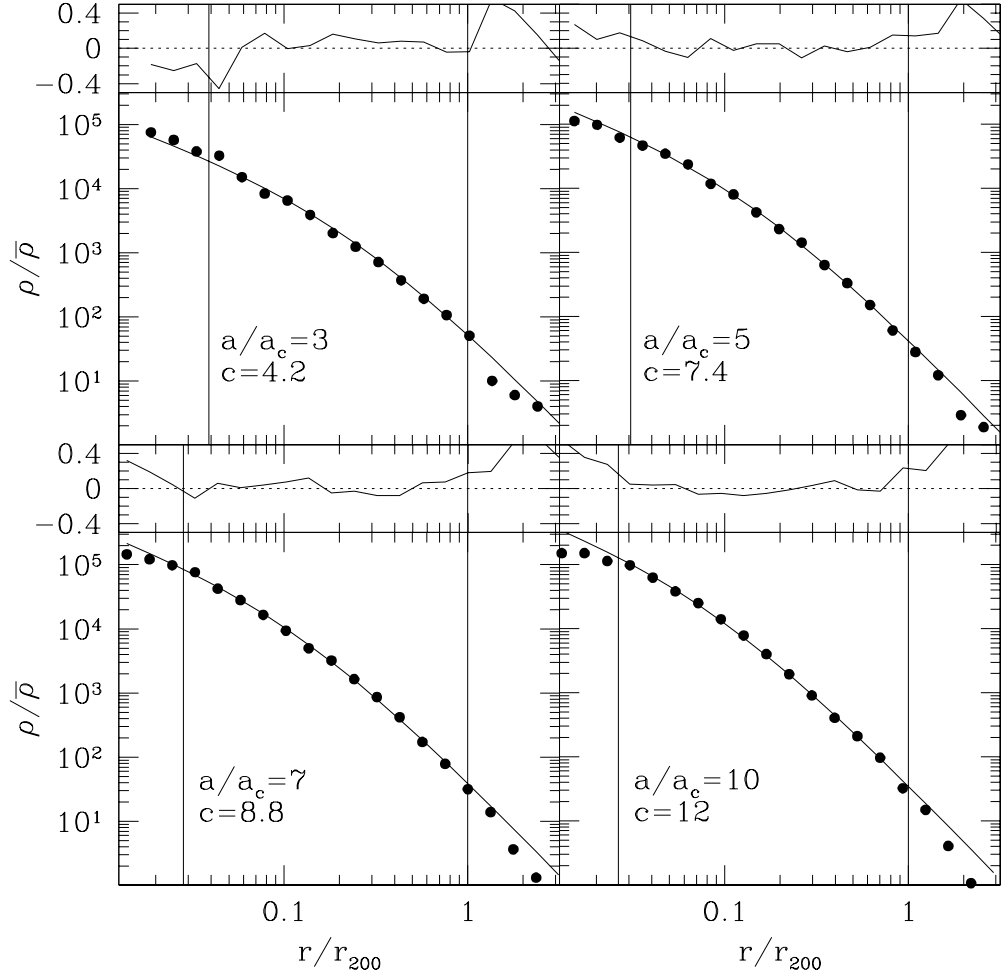}
\caption{Density profile of the dark matter halo as simulated
without gas at four different scale factors, $a/a_{\rm c}=$3, 5, 7, and
10, with spherically-averaged simulation results in radial bins (filled
circles) and the best-fitting NFW profiles (solid curves) for several
epochs, as labeled.  Shown above each panel are fractional deviations
$(\rho_{\rm NFW}-\rho)/\rho_{\rm NFW}$ from the best-fitting NFW profiles for
each epoch.  Vertical lines indicate the location of $r_{\rm soft}$, the
numerical softening-length, and $r_{200}$.
\label{pancake2}}
\end{figure}

\subsection{Profiles}
\label{profs}
\subsubsection{Density}
The density profiles at different epochs for the simulations without gas 
are shown in Fig.~\ref{pancake2}, along with the best-fitting NFW profile 
for each epoch, which has the form 
\begin{equation}
\frac{\rho}{\overline{\rho}}=\frac{\delta_c}{(r/r_{\rm s})(1+r/r_{\rm s})^2},
\end{equation}
with $\delta_c$ given by
\begin{equation}
\delta_c=\frac{200}{3}\frac{c^3}{\ln(1+c)-c/(1+c)},
\end{equation}
where $r_{\rm s}\equiv r_{-2}$, $c=r_{200}/r_{\rm s}$, and $r_{200}$ is the radius 
within which the mean density $\langle \rho\rangle=200\overline{\rho}$. 
The NFW profile has only one free parameter, the
concentration parameter $c$.  Our pancake halo density profiles have
concentrations which range from 3 to 15, increasing with time, and are usually
within 20\% of the best-fit NFW profile at all radii.  
This NFW profile is a fit to N-body results for CDM halos, but it is also 
consistent with the halos which form in simulations using 
Gaussian-random-noise initial fluctuations with small-scale fluctuations 
suppressed \cite{AVI,BOT,CAV,HJS99,KNE}, as is the case for the halos 
that form from pancake instability presented here.  

\begin{figure}[t!]
\centering
\includegraphics[width=2.7in]{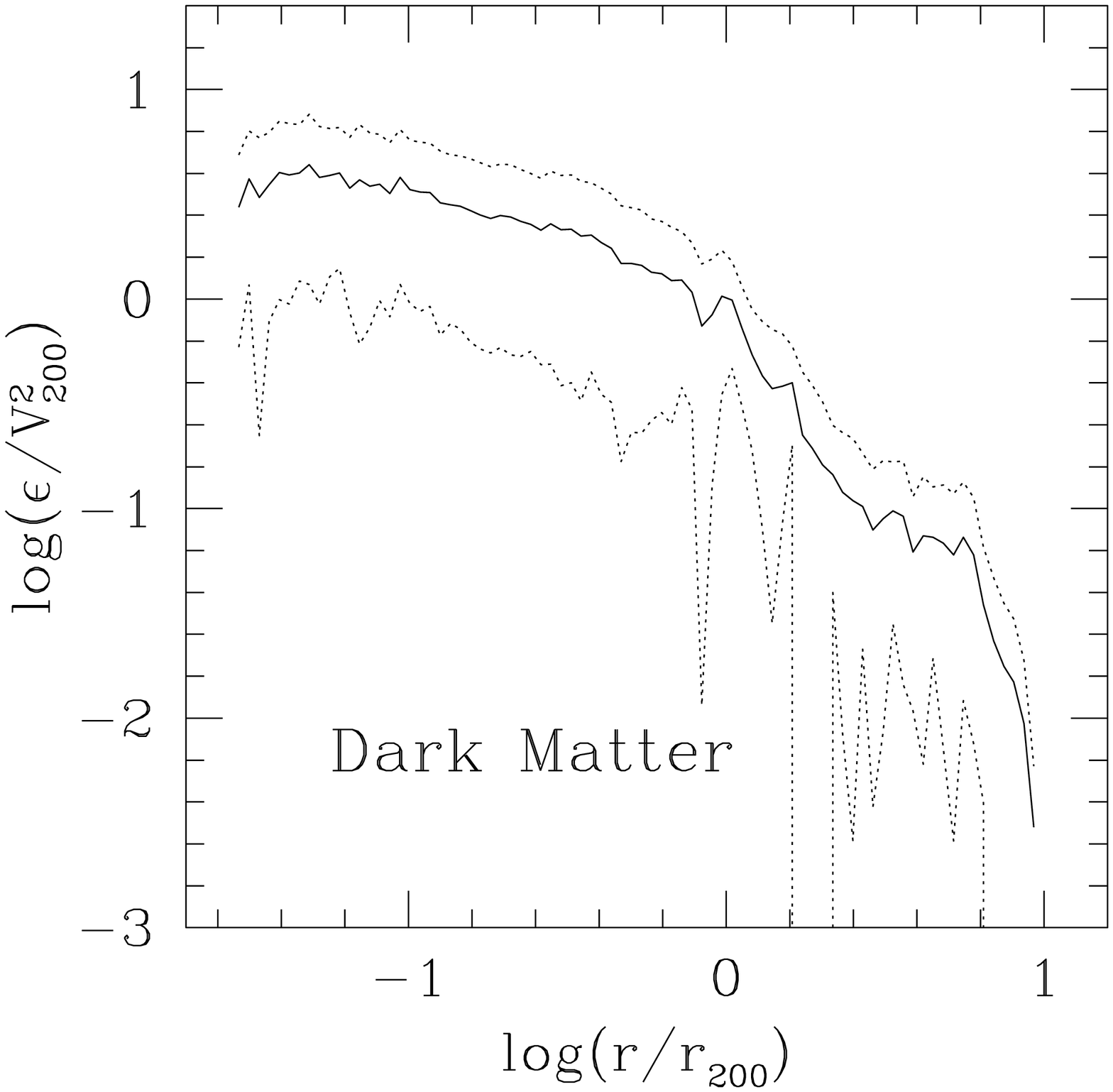}
\includegraphics[width=2.7in]{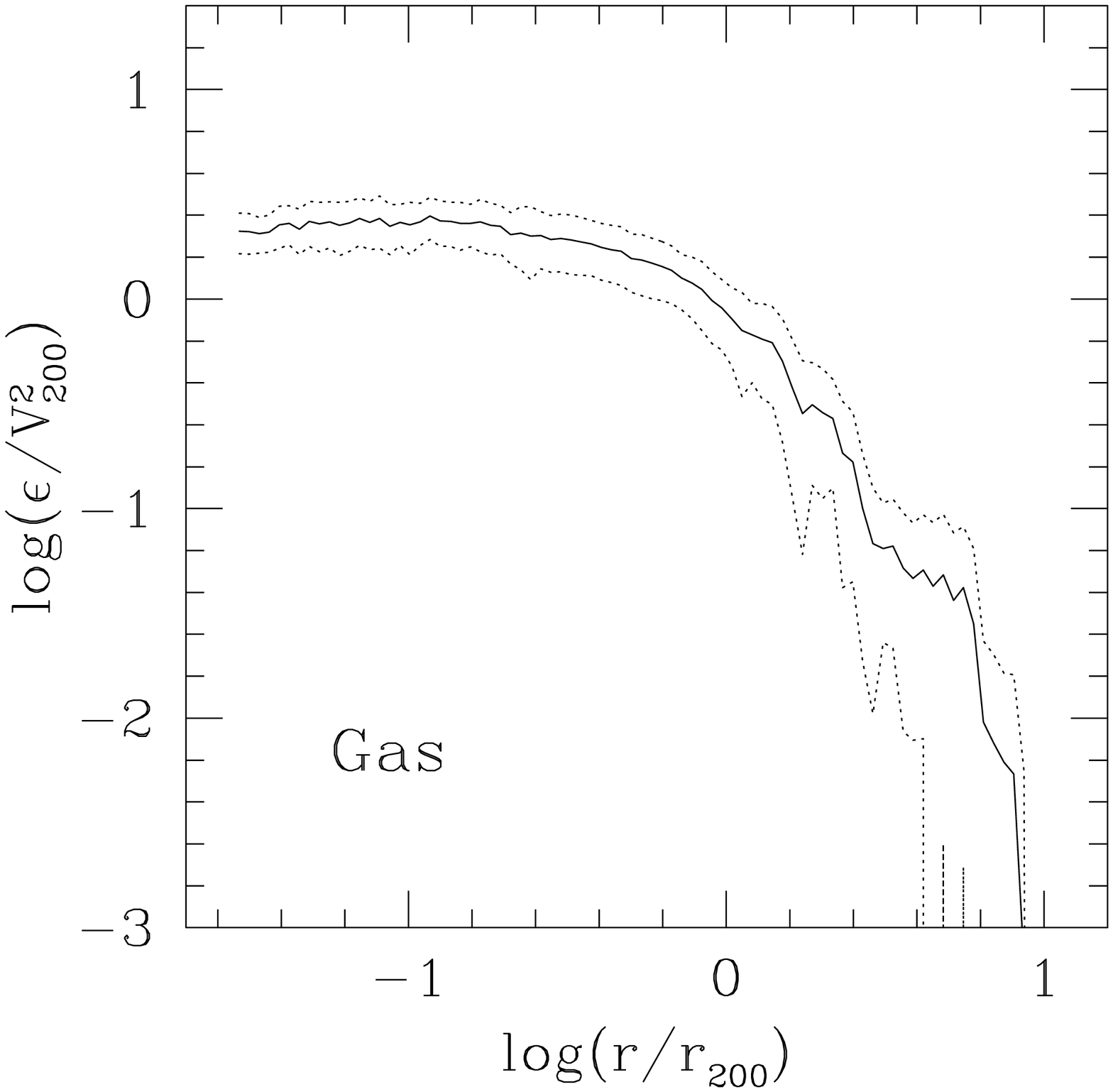}
\caption{
Dimensionless specific thermal energy profiles at $a/a_{\rm c}=7$ in the
(gas+DM)-simulation, for the dark matter (left) and gas (right), with dotted 
lines indicating the rms scatter within each
logarithmic bin. 
\label{pancake3}}
\end{figure}

\subsubsection{Kinetic and Thermal Energy} The profiles of the
dimensionless specific thermal kinetic energy $\epsilon_{\rm DM}=\langle
|{\bf v}-\langle {\bf v}\rangle|^2\rangle/2=3\sigma^2_{\rm DM}/2$ of the
dark matter and the dimensionless specific thermal energy $\epsilon_{\rm
gas}=3k_{\rm B}T/2m$ of the gas are shown in Fig.~\ref{pancake3}, for
the simulation with gas and dark matter at $a/a_{\rm c}=7$.  Although the dark
matter velocity dispersion rises towards the center, the rise is shallow
and the kinetic energy distribution is approximately isothermal inside the
radius $r_{200}$.  As seen from Fig.~\ref{pancake3}, the gas is even more
isothermal than the dark matter, with $\epsilon_{\rm gas}(r_{\rm
soft})/\epsilon_{\rm gas}(r_{200})\simeq 2.5$, while the density varies by
more than three orders of magnitude over the same region. In
Fig.~\ref{pancake4} we show the ratio of the specific thermal energy of
the dark matter to that of the gas.  This ranges from $\epsilon_{\rm
DM}/\epsilon_{\rm gas}\sim 1.6$ in the center to $\epsilon_{\rm
DM}/\epsilon_{\rm gas}\sim 1$ at $r_{200}$.

\subsubsection{Velocity Anisotropy}

\begin{figure}[t!]
\centering
\includegraphics[width=2.7in]{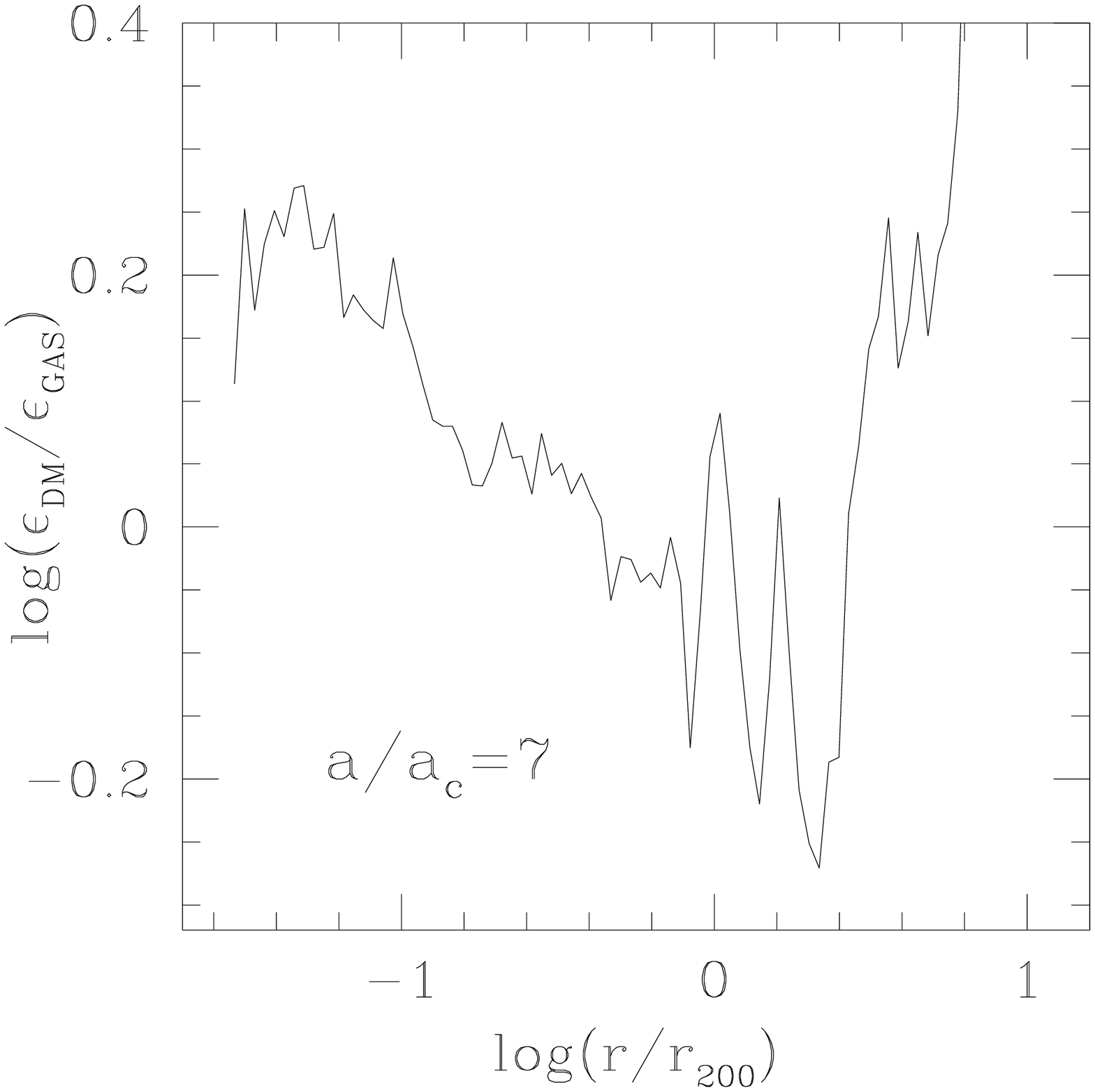}
\includegraphics[width=2.7in]{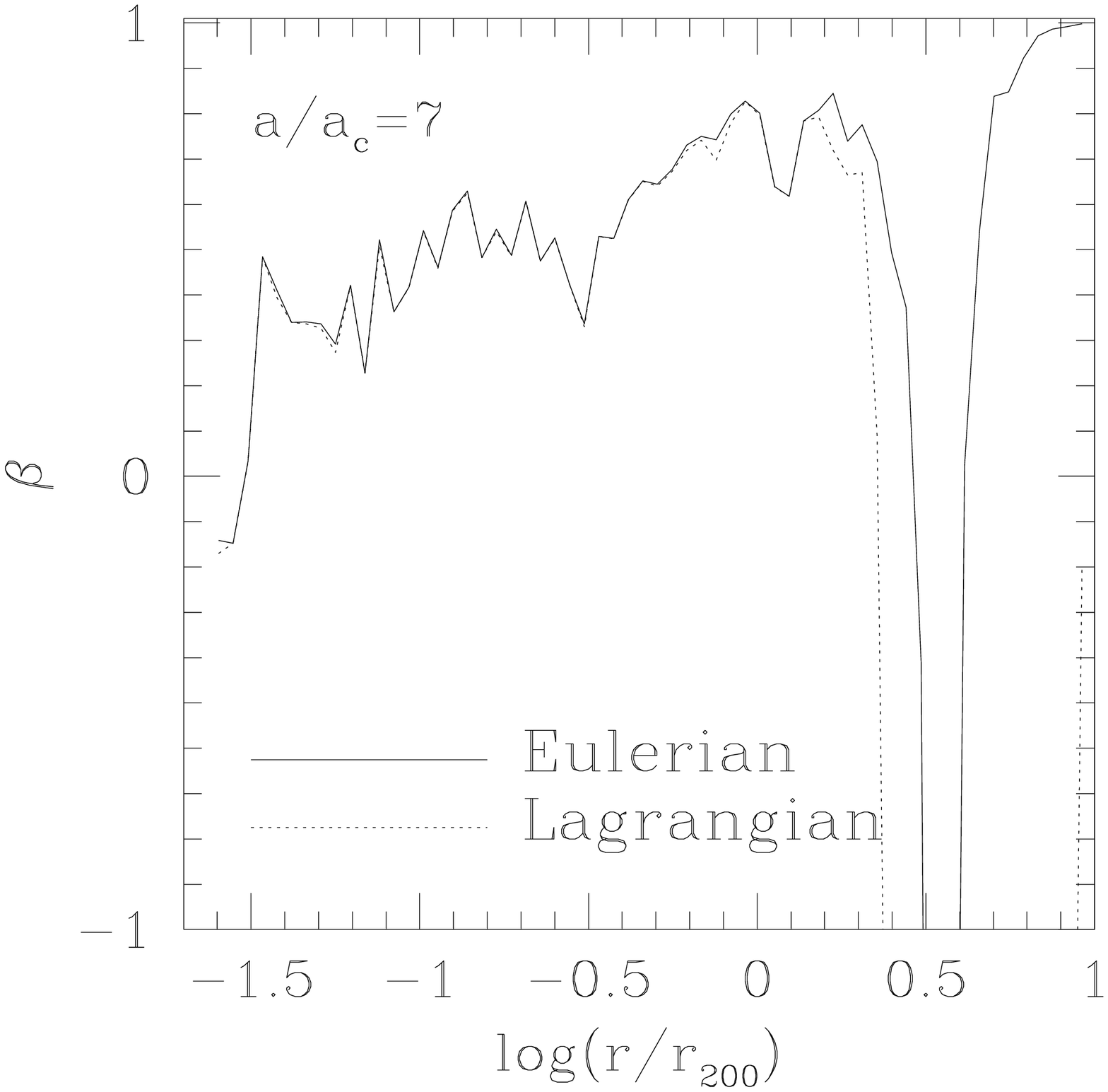}
\caption{(left)
Ratio of dimensionless specific thermal energy profiles at $a/a_{\rm c}=7$ in the
(gas+DM)-simulation. (right) Anisotropy profile, defined two different ways, 
at $a/a_{\rm c}=7$.  The curves are labeled according to the definitions given in 
the text.
\label{pancake4}}
\end{figure}

In Fig.~\ref{pancake4}, we plot the profile of the anisotropy parameter
$\beta$, defined two different ways, according to the frame of reference
in which the velocity dispersion is calculated.  In the Eulerian case,
where the bulk motion of the shell contributes to the anisotropy, it is
defined as $\beta=\beta_{\rm E}\equiv 1-\langle v_t^2\rangle/(2\langle
v_r^2\rangle)$.  In the Lagrangian case, however, the bulk motion of the
shell is subtracted out, $\beta=\beta_{\rm L}\equiv
1-\sigma_t^2/(2\sigma_r^2)$, where $\sigma_i^2=\langle (v_i-\langle
v_i\rangle)^2\rangle$. Because of the symmetry of the pancake initial 
conditions, $\langle v_t^2\rangle=\sigma_t^2$ at all times.  Differences 
between the two definitions therefore arise because of differences in the 
radial component.  The values $\beta=1,0,$ and $-\infty$ correspond to 
motion which is purely radial, fully isotropic, and tangential, respectively.
While the two definitions of anisotropy give nearly indistinguishable
profiles at $r\leq r_{200}$, the two profiles depart significantly at $r >
r_{200}$. This is expected, since the bulk radial motion is
nearly zero inside the halo where equilibrium is a reasonable expectation.  
The two definitions of $\beta$ are identical when the system is in
equilibrium. Outside the halo, however, equilibrium is violated because
the bulk radial motion is not zero, and the difference between the
profiles arises due to the detailed nature of the region outside the halo.

Simulations of CDM typically find values of $\beta$ near 0 at the center,
slowly rising to a value of $\beta\approx 0.5$ at $r_{200}$ 
\cite{carlberg,CKK,ENF,FM01}. As seen from Fig.~\ref{pancake4}, the
halo formed by pancake instability is more anisotropic, with
values of $\beta$ rising from $\sim 0.2$ near the center to $\sim 0.8$ at
$r_{200}$.  This reflects the strongly filamentary substructure of the
pancake within which the halo forms, and perhaps the absence of strong
tidal fields or mergers as well, which might otherwise help convert radial
motions into tangential ones.

\begin{figure}[t!]
\centering
\includegraphics[width=2.7in]{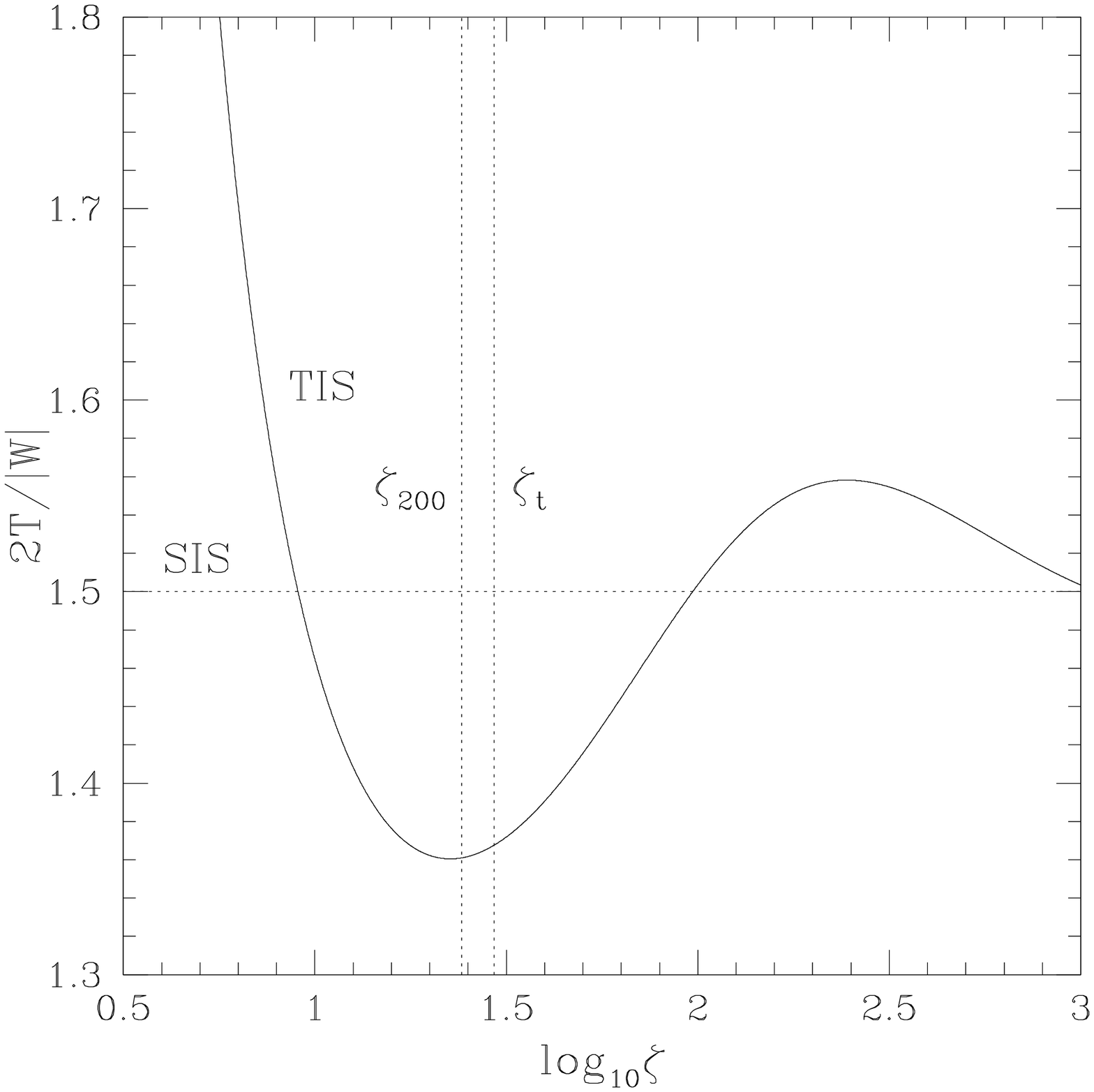}
\includegraphics[width=2.7in]{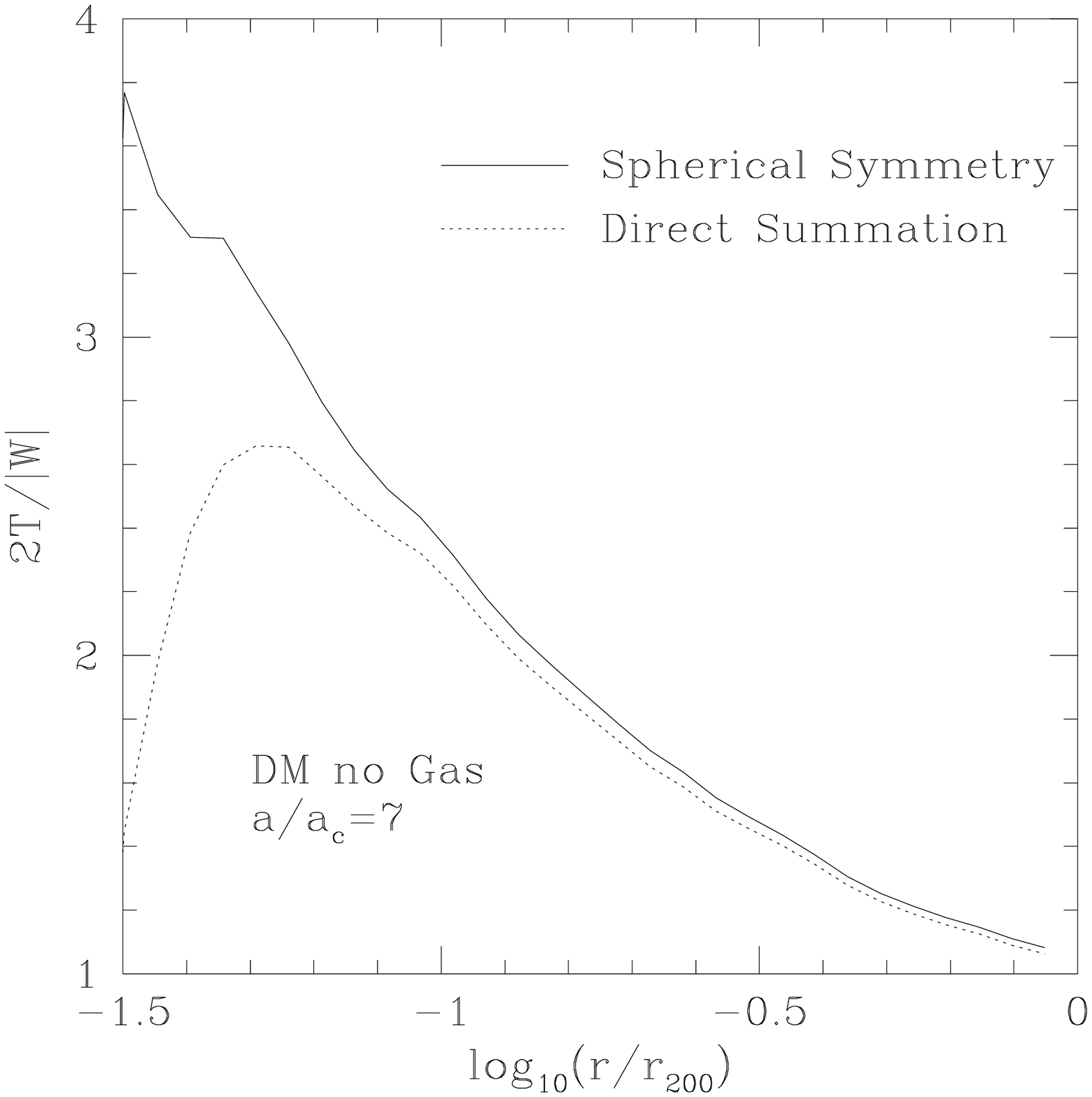}
\caption{(left) Virial ratio versus dimensionless radius $\zeta$ for the TIS
solution. The vertical dotted line on the right corresponds to the
truncation radius, $\zeta_t$, at which the total energy is a minimum
at fixed mass and boundary pressure, while the one on the left
corresponds to the radius within which the mean density is 200 times the
background density. The horizontal dotted line is the virial ratio of the 
singular isothermal sphere. (right) Virial ratio vs. radius calculated two 
different ways: (1) direct summation and (2) assuming spherical symmetry.
\label{pancake5}}
\end{figure}

\subsection{Virial Equilibrium}

A state of equilibrium is commonly assumed in analytical modeling of dark
matter halos \cite{LM}.  
Using $N$-body
simulations, \cite{TBW97} found that halos which formed from CDM initial
conditions roughly obey the Jeans equation for dynamical equilibrium in
spherical symmetry, within a radius of order $r_{200}$, suggesting that
CDM halos are in approximate virial equilibrium.  In what follows, we will
interpret the numerical halo results further by comparison with
equilibrium halo models.

\begin{figure}[!t]
\centering
\includegraphics[width=3.5in]{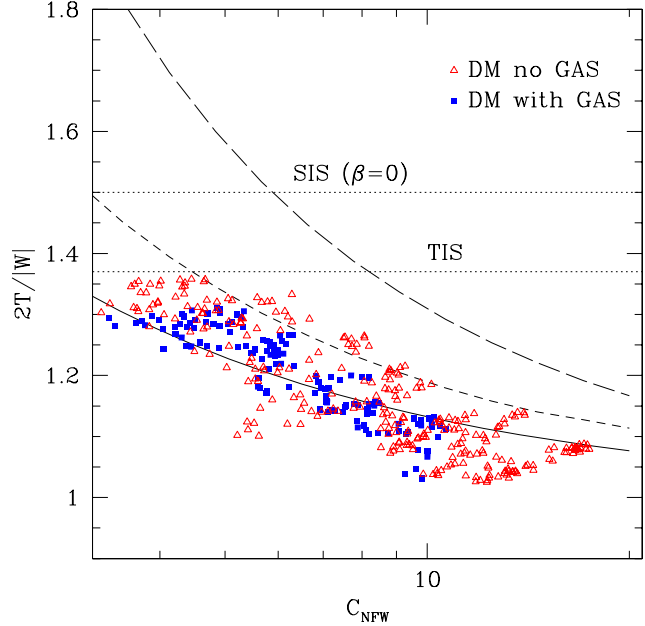}
\caption{Virial ratio versus NFW concentration parameter for the halos as 
simulated with and without gas, as they evolve.  Solid, dashed, and 
long-dashed lines represent expected values for $\beta=0,0.5,1$, respectively.  
Shown also are the values corresponding to the simulated halos, the isotropic
singular isothermal sphere (SIS), and the truncated isothermal 
sphere (TIS).
\label{pancake6}}
\end{figure}

\subsubsection{Virial Ratio}

The scalar virial theorem states that for a self-gravitating system in
static equilibrium ($\langle v \rangle=0$) with no magnetic fields,
$2{\cal T}+W+S_p=0$, where $W$ is the potential energy, ${\cal T}$ is the thermal and
kinetic energy, and $S_p$ is a surface pressure term,
\begin{equation}
S_p=-\int p\bf{r}\cdot\bf{dS}, 
\end{equation} 
where $\bf{dS}$ is the surface area element. If the system is isolated,
there can be no material outside to create a boundary pressure, and we
have $S_p=0$, implying $2{\cal T}/|W|=1$.  Cosmological halos are not isolated
systems, however, so we cannot expect $S_p=0$ and $2{\cal T}/|W|=1$.  In fact,
the presence of infalling matter creates a flux of momentum into the halo
which can act as a surface pressure (even for collisionless matter) in 
the virial theorem.  With infall present, we therefore expect $S_p/|W|<0$, 
implying $2{\cal T}/|W|>1$.

\subsubsection{Singular Isothermal Sphere}
The singular isothermal sphere (SIS) is the singular solution of the
equation of hydrostatic equilibrium with uniform temperature.
The density is given by
\begin{equation}
\rho(r)=\frac{\sigma_0^2}{2\pi Gr^2}=\frac{k_BT}{2\pi Gmr^2},
\label{sisrho}
\end{equation}
where $\sigma_0$ is the 1D velocity
dispersion, and $T$ is the gas temperature.  A more
general class of solutions can be found, however, by allowing the
anisotropy parameter $\beta$ to be nonzero but remain independent of
radius.  The relevant equation is the Jeans equation \cite{BT}, 
\begin{equation}
\label{jeans}
\frac{d}{dr}(\rho\overline{v_r^2})+\frac{2\rho\beta\overline{v_r^2}}{r}=
-\rho\frac{d\Phi}{dr},
\end{equation}
where the anisotropy parameter is now the ``Eulerian" quantity,
$\beta=\beta_{\rm E}$, as defined in \S \ref{profs}. 
We can solve the Jeans equation in this case for 
the velocity dispersion $\sigma_{\beta}$ as a function of $\beta$ 
and the isotropic velocity dispersion $\sigma_0$ for the same mass 
distribution as in equation (\ref{sisrho}), to show
\begin{equation}
\sigma_{\beta}^2=\frac{3-2\beta}{3(1-\beta)}\sigma_0^2.
\label{betasis}
\end{equation}

If the anisotropy $\beta$ does not vary with radius, we can determine the
virial ratio of an SIS for a given value of the anisotropy, as follows.
The potential energy at $r$ is
\begin{equation}
W(r)=\int(\rho\phi)dV=-2M(r)\sigma_0^2.
\label{wsis}
\end{equation}
The kinetic (or thermal) energy ${\cal T}$ is given by 
\begin{equation}
{\cal T}=\frac{1}{2}\int\rho\langle v^2\rangle
dV=\frac{3}{2}\int\rho\sigma^2dV=\frac{3}{2}M(r)\sigma^2.
\end{equation}
Using equation (\ref{betasis}) to relate the actual $\sigma^2$ to the one
in the isotropic case $\sigma_0^2$, one obtains
\begin{equation}
{\cal T}=\frac{3-2\beta}{2(1-\beta)}M\sigma_0^2.
\end{equation}
The virial ratio is therefore
\begin{equation}
\frac{2{\cal T}}{|W|}=\frac{3-2\beta}{2(1-\beta)}.
\end{equation}

\subsubsection{Nonsingular Truncated Isothermal Sphere}

The Truncated Isothermal Sphere (TIS) model for the equilibrium halos which 
form from the collapse and virialization of top-hat density perturbations 
\cite{SIR} (to be discussed here in \S \ref{tis_sect}) yields a virial
ratio which differs significantly from the familiar value of $2{\cal T}/|W|=1$ for
isolated halos (i.e. with zero boundary pressure) with $\beta=0$.  The TIS 
model is the minimum energy solution of the isothermal Lane-Emden equation 
with non-singular, finite boundary condition for the central density. Since 
the TIS is a
unique solution given by the minimum energy at fixed boundary pressure,
the virial ratio is always the same value, namely $2{\cal T}/|W|\simeq1.37$ at
$\zeta_t=29.4$, where $\zeta_t$ is the truncation radius in units of the
core radius.  This value is smaller than that for the isotropic singular
isothermal sphere (with nonzero boundary pressure), $2{\cal T}/|W|=1.5$, and is 
near the global minimum for all
values of $\zeta$, which is $2{\cal T}/|W|\simeq1.36$ and occurs at
$\zeta\simeq22.6$.  At the intermediate radius $\zeta_{200}\simeq 24.2$,
defined to be the radius within which the mean density is 200 times the
background density, the TIS virial ratio has a value $2{\cal T}/|W|\simeq 1.36$.
As seen in Fig.~\ref{pancake5}, the inner core region of the TIS is
dominated by kinetic (or thermal, in the gas case) energy, whereas the
value approaches that of the SIS at large $\zeta$, where the core region
becomes small compared with the size of the TIS and the density profile
asymptotically approaches that for the SIS ($\rho\propto r^{-2}$).

\subsubsection{NFW Halos}

The equilibrium structure of halos with an NFW density profile was
investigated by \cite{LM}.  Using different values of $\beta(r)$, they
found several analytical solutions to the velocity dispersion of the halo
by integrating the Jeans equation for a given $\rho(r)$ and $\beta(r)$. In
order to integrate the Jeans equation to find the velocity dispersion, it
is necessary to set the velocity dispersion at some $r$.  In the absence
of a physical value for $\sigma_r$ at the boundary of the halo, as could
be inferred from some infall solution, the only other reasonable choice is
to have $\sigma_r \rightarrow 0$ as $r \rightarrow \infty$.  The results
obtained by fixing $\sigma_r=0$ at infinity should reflect the general
trends associated with a single halo evolving in a quasi-equilibrium
state.

Shown in Fig.~\ref{pancake6} is the virial ratio versus concentration
parameter for different values of $\beta(r)=\beta_0$, as expected from the
Jeans equation.  The more isotropic the NFW halo, the lower the virial
ratio.  This is consistent with the fact that the surface pressure term is
directly related to the radial velocity dispersion.  The more concentrated 
the NFW halo, the lower the virial ratio.  This is expected, since the 
more isolated a halo, the less relative importance of the boundary 
pressure, implying $2{\cal T}/|W|\rightarrow 1$.

\subsubsection{Simulated Pancake Instability Halos}

Virial ratios $2{\cal T}/|W|$ were calculated for both simulation runs, with and 
without gas. For the case with no gas,
\begin{equation}
{\cal T}(r)=\sum_i \frac{1}{2}m_iv_i^2,
\end{equation}
where the sum is over all particles within $r$.  For the simulation with
gas included,
\begin{equation}
{\cal T}(r)=\sum_i \frac{1}{2}m_iv_i^2 + \sum_i \frac{3}{2}m_ik_{B}T_{i}.
\end{equation}
The potential energy was found using the
assumption of spherical symmetry,
\begin{equation}
W(R)=-\sum_i\frac{GM_im_i}{r_i}f_i,
\end{equation}
where $M_i$ is the mass interior to $r_i$ and $f_i\equiv f(r_i)$ is a function 
which represents the particular form of the softening used in the P$^3$M 
algorithm, and the sum is over all particles within 
$R$.  As shown in Fig.~\ref{pancake5}, the
assumption of spherical symmetry yields 
results which are very close at $R=r_{200}$ to those arrived at from
the more rigorous definition,
\begin{equation}
W(R)\equiv-\frac{1}{2}\sum_i\sum_j\frac{Gm_im_j}{r_{ij}}f_{ij},
\end{equation}
where $r_{ij}=|{\bf r_i}-{\bf r_j}|$ and $f_{ij}\equiv f(r_{ij})$.

Shown in Fig.~\ref{pancake6} are the virial ratios of both halos, with and
without gas, plotted versus their concentration parameter.  Each point 
corresponds to a different time in the halo's evolution, since the 
concentration of each halo changes with time.  While the simulated halos 
lie below the expected curve for NFW halos of comparable anisotropy, the 
trend of smaller virial ratios for more concentrated halos is clearly evident,
though there is significant scatter, particularly in the simulation without 
gas included.

\begin{figure}[!t]
\centering
\includegraphics[width=3.5in]{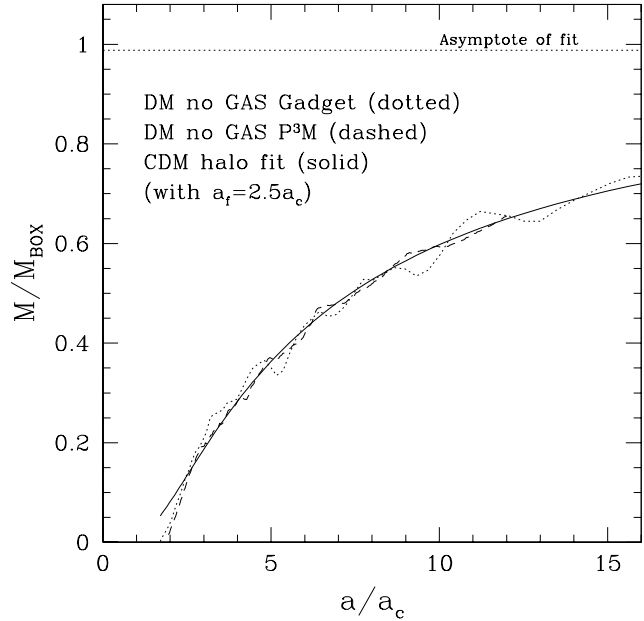}
\caption{
Mass accretion history of simulation with dark matter only (dashed) along
with the best-fit functional form from CDM simulations by \cite{WEC}
(solid) with $S=2$ and $a_{\rm f}=2.5a_{\rm c}$ (see equation~\ref{mass0}). 
Curve labeled ``Gadget'' was for the same initial conditions, run with the 
publicly available code of the same name \cite{SPR}.
\label{pancake7}}
\end{figure}

\subsection{Halo Evolution}

\subsubsection{Mass Accretion History}
\begin{figure}[t!]
\centering
\includegraphics[width=2.7in]{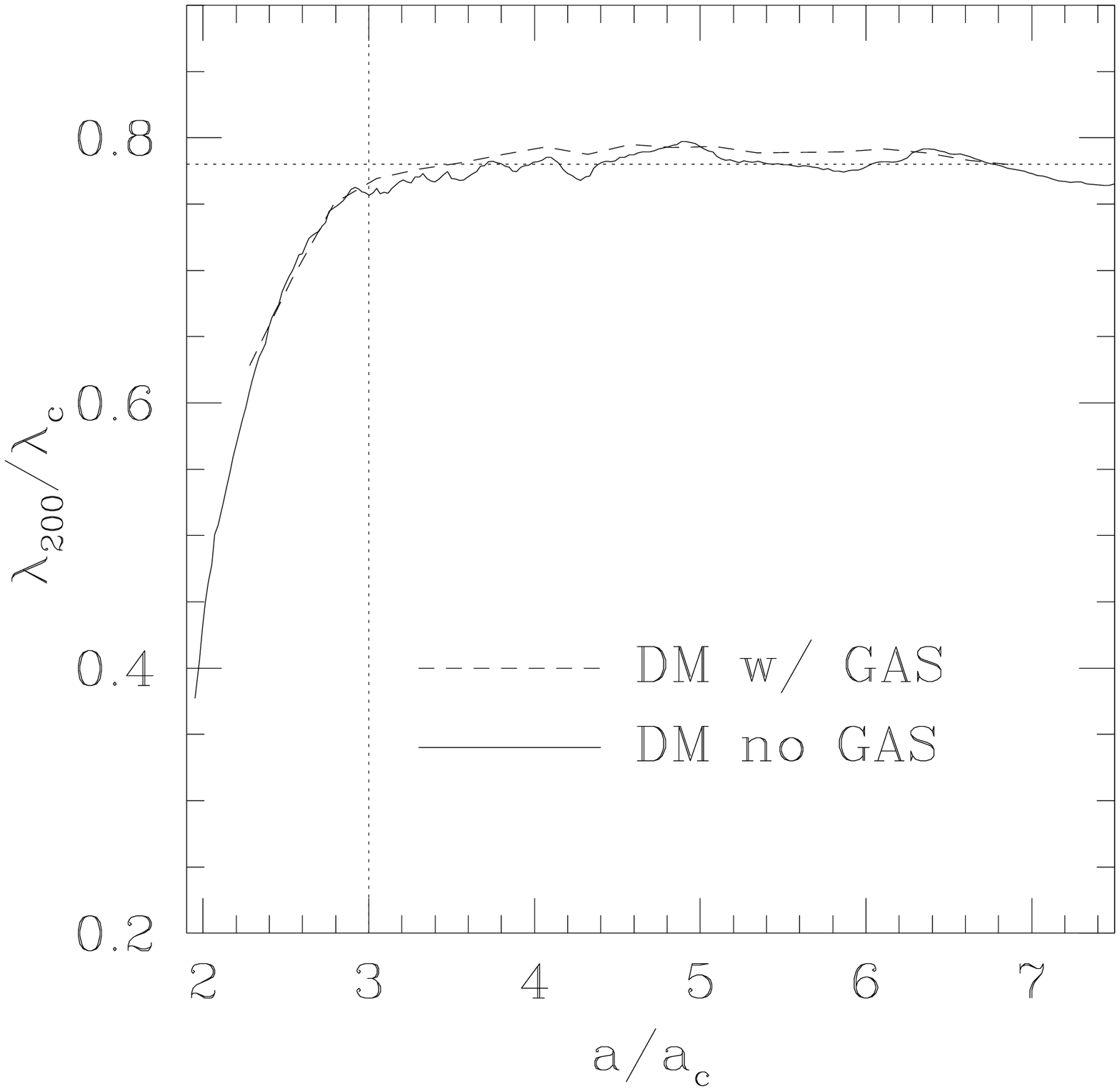}
\includegraphics[width=2.7in]{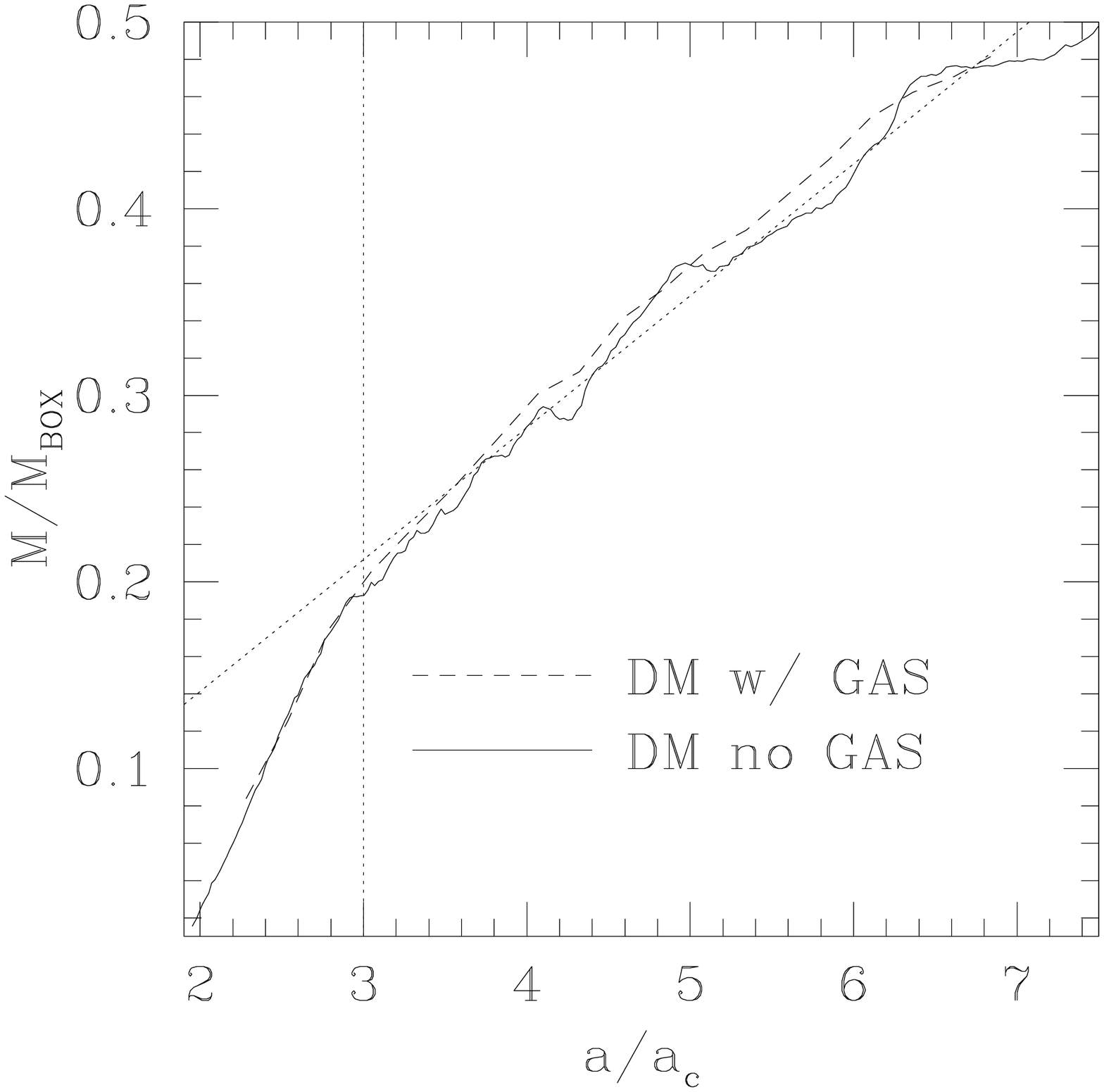}
\caption{Evolution of the dark matter only halo in the intermediate
(self-similar) regime.  (left) Virial radius in units of expected caustic
radius as defined in the text, and the average in the range
$3<a/a_{\rm c}<7.5$ (horizontal dotted line).  (right) Halo mass and the 
linear best-fit (dotted).  The vertical dotted lines indicate the 
virialization epoch at $a=3a_{\rm c}$. 
\label{pancake8}}
\end{figure}

The mass growth of the halo proceeds in three stages (Figs.~\ref{pancake7} 
and \ref{pancake8}). Before $a/a_{\rm c} \sim
3$, the mass within $r_{200}$ grows very quickly, indicating initial
collapse of the central overdensity.  After $a/a_{\rm c} \sim 3$, the infall
rate drops, and can be well-described by $M\propto a$.  Such an infall
rate is reminiscent of self-similar spherical infall onto a point mass 
(e.g. \cite{BERT}).  This infall rate
cannot persist indefinitely, since there is only a finite mass supply to
accrete onto the halo because of the periodic boundary conditions.  As a
consequence, the accretion rate slows after $a/a_{\rm c}=7$.  This is also
expected to occur in halos forming from more realistic initial
conditions, given that neighboring density peaks of a similar mass scale
prevent any one halo from having the infinite mass supply necessary to
sustain this mass accretion rate.  In fact, as shown in Fig.~\ref{pancake7},
the halo can be fit at nearly all times, especially later, by the more general 
fitting function
\begin{equation}
\label{mass0}
M(a)=M_\infty\exp\left[-Sa_{\rm f}/a\right], 
\end{equation}
where $S\equiv [d\ln M/d\ln a]_{a_{\rm f}}$ is the logarithmic 
slope at $a=a_{\rm f}$.  For $S=2$, we find a best-fit value of 
$a_{\rm f}=2.5a_{\rm c}$.  This form was first used 
to fit the evolution of halos formed in CDM simulations (\cite{WEC})
We have identified three distinct phases in the halo evolution:
initial collapse, steady infall, and infall truncation due to finite mass
supply.  In realistic collapse, we see that the halo evolves continuously
from one stage to the next as evidenced by the continuous change in
logarithmic slope of the fitting function, given by

\begin{equation}
\frac{d\ln M}{d\ln a}=S\frac{a_{\rm f}}{a}.
\end{equation}

\begin{figure}[t!]
\centering
\includegraphics[width=5.5in]{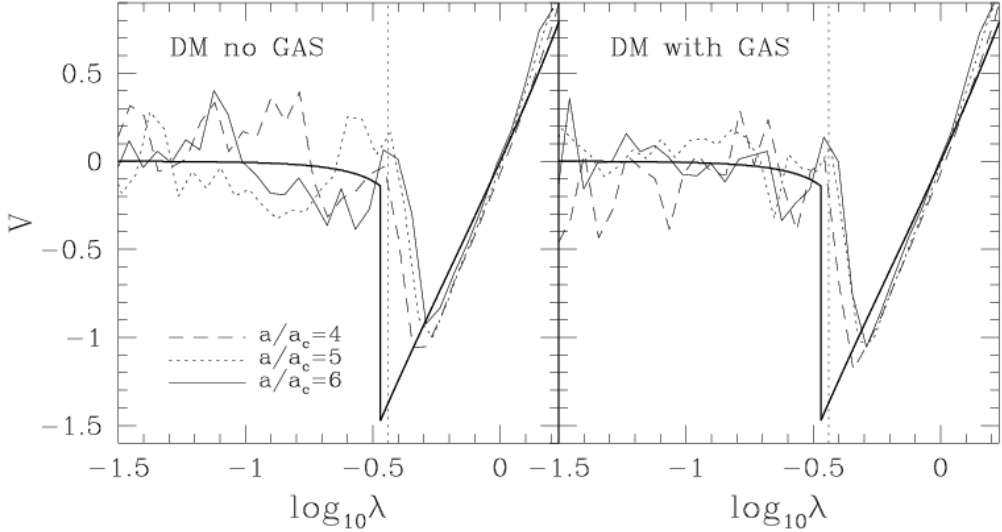}
\caption{
Radial velocity profiles in dimensionless units as in self-similar
spherical collapse.  The thick solid line is the radial velocity profile
for an ideal $\gamma=5/3$ gas as in \cite{BERT}.
\label{pancake9}}
\end{figure}

In the intermediate stage of collapse, the resemblance to self-similar 
spherical infall onto a point-mass perturbation is evident in (Fig.~\ref{pancake8}
and \ref{pancake9}). Plotted in Fig.~\ref{pancake9} is the dark matter radial 
dimensionless velocity profile $V(\lambda)$ for different $a/a_{\rm c}$ 
as simulated with and without gas. The dimensionless velocity is 
\begin{equation}
V\equiv\frac{t}{r_{\rm ta}(t)}v_r=\frac{2}{3H_7}\left(\frac{a}{a_7}\right)^{1/6}
\frac{v_r}{r_{ta,7}},
\end{equation}
where we have used the relations $r_{\rm ta}\propto t^{8/9}$ and $a\propto
t^{2/3}$, $r_{\rm ta}$ is the turnaround radius, $a_7\equiv 7a_{\rm c}$,
$r_{\rm ta,7}\equiv r_{\rm ta}(a_7)$ is set by finding the radius at 
which $v_r=0$ at $a=a_7$, and the Hubble constant $H_7\equiv H(a=a_7)$. 
The dimensionless radius is
\begin{equation}
\lambda=\frac{r}{r_{\rm ta}}=\frac{r}{r_{ta,7}}\left(\frac{a}{7a_{\rm c}}\right)^{-4/3}.
\end{equation}
In the case of self-similar infall, this profile does not change with
time.  As seen in Figure~\ref{pancake9}, $V(\lambda)$ for the simulated halo
follows the self-similar solution closely, with
$\lambda_{200}/\lambda_{\rm c}\simeq0.78$ (see Fig.~\ref{pancake8}), 
where $\lambda_{\rm c}$ is the
radius at the outermost caustic and is approximately where the shock
occurs in the collisional solution, and $\lambda_{200}\equiv
r_{200}/r_{\rm ta}$.

\subsubsection{Density Profile}
\begin{figure}[t!]
\centering
\includegraphics[width=5.3in]{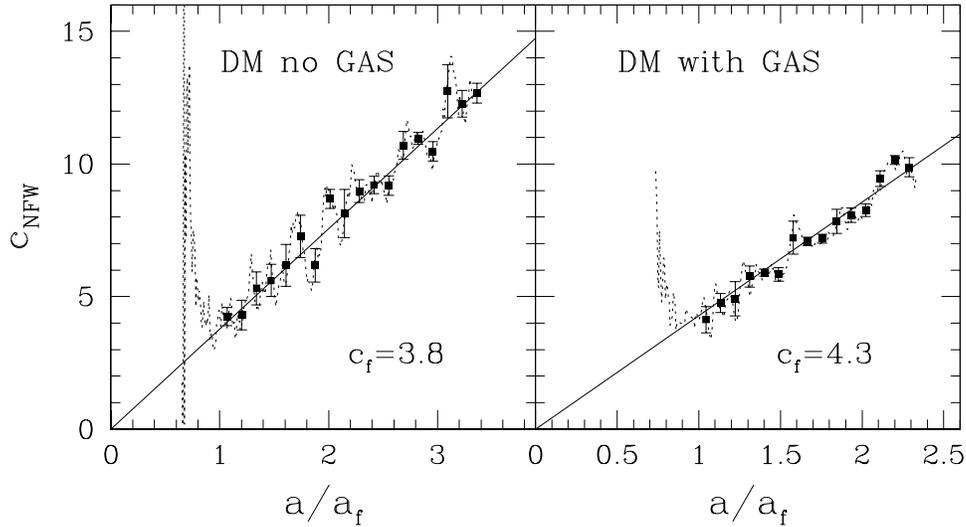}
\caption{Evolution of concentration parameter of the best-fitting NFW profile 
for the dark matter halo in simulations with (right) and without (left) gas.
Dotted line is the actual evolution, with the filled squares showing the mean
concentration binned in scale factor.  Errorbars indicate RMS fluctuations 
within each bin.  The solid lines are the best-fitting linear evolution given
by $c_{\rm NFW}=c_{\rm f}(a/a_{\rm f})$, with $a_{\rm f}=3a_{\rm c}$.
\label{pancake10}}
\end{figure}

Although the halo generally grows by self-similar accretion for
$3<a/a_{\rm c}<7$, the mass density is better fit by an NFW profile with only
one free parameter, $c$.  Shown in Fig.~\ref{pancake10} is the 
concentration parameter versus scale factor.  We find it can be well-fitted 
by
\begin{equation}
c=c_{\rm f}\frac{a}{a_{\rm f}},
\end{equation}
where $a_{\rm f}$ is the scale factor at which the accretion rate becomes
proportional to $a$, marking the end of the collapse phase.  The value
$a_{\rm f}=3a_{\rm c}$ is used here for both
simulation cases and corresponds to the vertical dotted
line in Fig.~\ref{pancake7}.
The solid lines in Fig.~\ref{pancake10} correspond to the best-fit values
$c_{\rm f}=4.3$ and $c_{\rm f}=3.8$ for the cases with and without gas
included, respectively, where each data point was weighted by the goodness
of the corresponding NFW profile fit.

This linear evolution was also found by \cite{BUL} and \cite{WEC}.  In
the latter case, they followed the mass accretion and merger histories of
individual halos in a high-resolution CDM simulation of halos in the mass
range $\approx 10^{11}-10^{12} M_\odot$. The mass accretion histories
allowed them to determine a collapse epoch for each halo, which they
correlated with the halo's concentration.  They found a best-fit slope for
the linear evolution of concentration parameter of $c_{\rm f}=4.1$.  We conclude 
that such an evolution of mass and concentration is generic and not 
limited to halos forming from CDM initial conditions.

\section{The Fluid Approximation: 1D Halo Formation From Cosmological Infall}
\label{fluid_approx_leading_sect}
\subsection{Self-Similar Gravitational Collapse}
\label{ssgc_sect}
\subsubsection{Previous analytical models for halo formation: the spherical
infall model}
\label{sshistory_sect}
Analytical approximations have been developed to model the formation
of halos by the 1D growth of spherical cosmological density
perturbations, in either a collisionless gas or a fluid. We shall need
to refer to some of these solutions to justify our fluid approximation
in \S \ref{fluid_approx_sect}. It is necessary to
begin, therefore, by briefly recounting this earlier work.

\cite{gg} first presented the concept of the so called ``secondary
infall model (SIM)''. 
This SIM refers to the effect of the addition of a point mass to a uniform,
expanding Friedmann-Robertson-Walker universe as a perturbation which
causes the surrounding spherical shells to decelerate relative to the
background universe, until they reach a radius of maximum expansion
and recollapse. Subsequent work generalized this approach to include
spherically-symmetric initial perturbations for which the overdensity
profile depends upon radius or mass as a scale-free power-law.
Along this line, \cite{FG} studied the dynamics
of collisionless CDM halos using a self-similar model, adopting a
scale-free initial overdensity parametrized by its shape:
\(\varepsilon\) in equation (\ref{overmass}). 
\cite{BERT} studied a special case of \cite{FG}
but also extended the analysis to a collisional
fluid. \cite{HS} showed that a power-law power
spectrum would indeed generate a scale-free initial condition,
such as was adopted by \cite{FG}. 
They then argued that the resulting nonlinear structure
would be described by a power-law profile determined by the shape of
the power spectrum.

Previously mentioned works adopted a rather unrealistic condition 
for the collisionless case, that of purely radial motion.
N-body simulations of CDM halo formation find that the virialized
region tends toward isotropic random velocities.
Some attempts to incorporate tangential velocities within the framework
of spherical symmetry have also been made by \cite{afh, hbf, rg}.
Along this line, one may also refer to work by
\cite{popolo, hiotelis, kull, LH} and references therein.

The fluid approximation for collisionless CDM halo formation has emerged
recently. \cite{tca} showed that one could use fluid-like
conservation equations to mimic the radial-only SIM models such as the
\cite{FG} model. \cite{sco} extended this analysis to include
tangential motion. The fluid approximation has been used in the
literature of stellar dynamics, but its application to CDM halo
dynamics is rather new, and as will be described in \S 
\ref{fluid_approx_sect}, it simplifies the description of dark matter dynamics
substantially. 

Self-similar spherical infall models have also been used to study the effect of
 incorporating additional baryonic physics.
\cite{abn}, \cite{bert-cool} and \cite{owv} have studied the effect of gas
cooling on galaxy formation using self-similar models. Because the
system was tuned to maintain self-similarity in the presence of
cooling, the cooling function used is not perfectly physical. However,
as shown by \cite{abn}, one can use these models to test one's
hydrodynamic code in the presence of cooling. And these models
capture the generic behavior of galactic dynamics in the presence of
realistic cooling.

Our model \cite{AS, as04}, which will be described in \(\S\)
\ref{selfinteract_sect},  is the first
self-similar model to include the effective heat conduction resulting from
SIDM collisionality. It utilizes the fluid approximation to describe SIDM
halo dynamics, justifying its validity rigorously. The resulting
formalism is similar to that of self-similar cooling models such as
\cite{abn}. However, the focus is on dark matter physics rather than
on baryonic physics. Thus, this is the first self-similar model in the 
presence of dark-matter-based heat
conduction. It may also be used as a testbed for hydrodynamic simulations
incorporating heat conduction. 

\subsubsection{Halo formation from scale-free linear perturbations}
\label{fg_sect}
In the Einstein-de Sitter (EdS) background universe, an initial linear
perturbation whose mass profile is spherically symmetric and has a 
scale-free, power-law form 
\begin{equation}
\label{overmass}
\frac{\delta M}{M}\propto M^{-\varepsilon }
\end{equation}
results in structure formation which is self-similar (\cite{FG}). Each spherical
mass shell around the center expands until it reaches a maximum radius
(turnaround radius \(r_{ta}\)), and recollapses. 
For a given \( \varepsilon  \), we have 
\begin{equation}
\label{rta}
r_{ta}\propto t^{\xi },
\end{equation}
where
\begin{equation}
\label{xi}
\xi =\frac{2}{3}\left( \frac{3\varepsilon +1}{3\varepsilon }\right)
\end{equation}
(\cite{FG}).
Since there are no characteristic length or time scales for this
problem other than the turn-around radius \(r_{ta}\) and the Hubble
time \(t\), the gravitational collapse which ensues from this
scale-free initial condition must be self-similar as long as the
background universe is Einstein-de Sitter, in the absence of physical processes
which introduce additional scales (e.g. SIDM collisionality). 

In general, 
if the unperturbed matter is a cold fluid, the infall which results from this
perturbation is highly supersonic and is terminated by a strong
accretion shock which thermalizes the kinetic energy of collapse. The
accretion shock radius is guaranteed by self-similarity to be a fixed
fraction of \(r_{ta}(t)\) at all times. The mean density of the
postshock region is, therefore, always a fixed multiple of the cosmic
mean matter density. For most cases of interest, this postshock region
is close to hydrostatic. For a collisionless gas, a similar
description applies as long as the infalling matter initially had
small (or no) random motions. In that case, each mass shell collapses
supersonically as a single stream until it encounters a region of
shell-crossing and density caustics, which encompasses all previously
collapsed (i.e. interior) mass shells. All collapsed mass shells
inside this region oscillate about the center. The radius of this
region of shell-crossing, given by the outermost density caustic, is
analogous to the shock radius in the fluid case.

\begin{figure}[!t]
\centering
\includegraphics[width=4in]{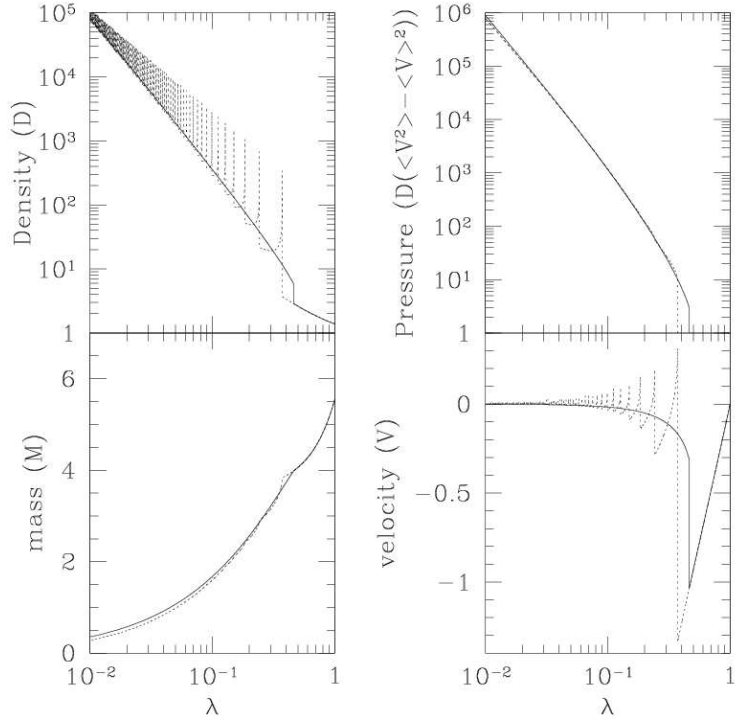}
\caption{Self-similar collisionless halo formation for
\(\varepsilon=1\): Comparison of the skewless-fluid approximation to the exact
collisionless solution by \cite{BERT}. Solid lines represent the
solution obtained from fluid approximation in the radial
direction, while dotted lines represent the collisionless Bertschinger
solution. Spikes in the density plot
simply represent infinite values, corresponding to caustics,
and therefore there is no physical
significance in the height of these spikes. However, spikes in the
velocity plot are finite and real. Note that solid lines do not
represent the \(\gamma=5/3\) fluid Bertschinger solution.}
\label{fig-bert}
\end{figure}

\begin{figure}[!t]
\centering
\includegraphics[width=4in]{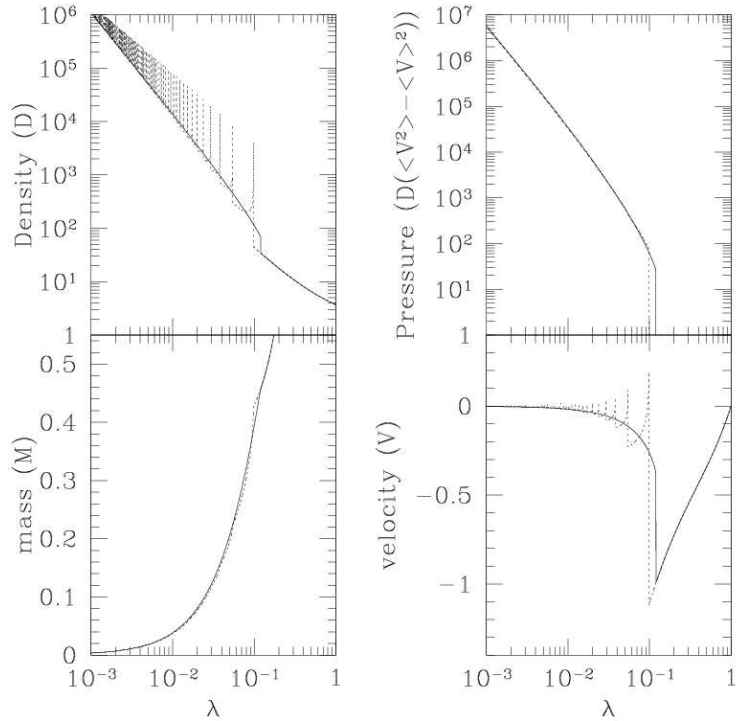}
\caption{Same as Fig. \ref{fig-bert}, but \(\varepsilon=1/6\).
Note again that solid lines were not generated from the 
\(\gamma=5/3\) fluid approximation, but rather from the radial-only fluid
approximation).
}
\label{fig-e16}
\end{figure}

Results for the purely collisionless case were presented for several values
of \(\varepsilon\) by \cite{FG} and for \(\varepsilon=1\) by \cite{BERT}
(where the latter included a fluid component, as well).
Figures \ref{fig-bert} and \ref{fig-e16} show the exact similarity solutions
we have derived (\cite{as04})
for the purely collisionless cases with \(\varepsilon=1\) and
\(\varepsilon=1/6\), respectively. As we describe below in \(\S\)
\ref{hs_sect}, 
these values roughly bracket the range relevant to cosmological halos
in a CDM universe. We will refer to these solutions again for
comparison in deriving our fluid approximation in \(\S\)
\ref{fluid_approx_sect}. 

\subsubsection{Halo formation from peaks of the Gaussian random noise
primordial density fluctuations}
\label{hs_sect}

\begin{figure}[!t]
\centering
\includegraphics[width=2.5in]{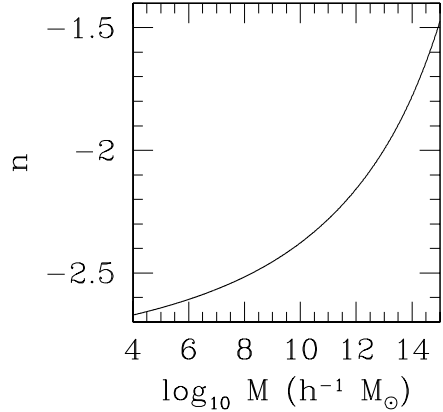}
\caption{Effective index of the power spectrum (\(P(k)\propto k^{n}\))
vs. halo mass for the \( \Lambda \)CDM universe.}
\label{fig-neff}
\end{figure}

The theory of halo
formation from peaks in the density field which results from
Gaussian-random-noise 
initial density fluctuations draws an interesting connection between the
average density profile around these peaks and the shape of the
fluctuation power spectrum.
According to \cite{HS}, local maxima of Gaussian
random fluctuations in the density can serve as the progenitors of
cosmological structures. They show that high density peaks (\( \nu
\geq 3 \), where \(\nu\) corresponds to \(\nu \sigma_{M}\) peak) have a
simple power-law 
profile\footnote{\cite{bbks} also get a similar result: local density
maxima have a triaxial density profile,
but as \( \nu  \) increases it becomes more and more spherical with
a profile converging to equation (\ref{overden}).}
\begin{equation}
\label{overden}
\Delta_{0}(r)\propto r^{-(n+3) },
\end{equation}
where \(\Delta_{0}(r)\) is the accumulated overdensity
inside radius $r$, and \(n\) is the effective index of the power
spectrum \(P(k)\) approximated as a power-law \(P(k)\propto k^{n}\) at
wavenumber $k$ which corresponds to the halo mass through
\begin{equation}
\label{tophatmass}
M=\frac{4\pi}{3}\bar{\rho}_{0}R^{3},
\end{equation}
where \(\bar{\rho}_{0}\) is the present cosmic mean matter density and \(R\) is 
the comoving, spatial top-hat filter radius\footnote{
As long as one is
concerned about the average profile, equation (\ref{overden}) holds
for any value of \(\nu\). 
For small \(\nu\), however, random dispersion around this average profile 
becomes substantial, limiting the generality of equation (\ref{overden}).}.
The overdensity \(\Delta_{0}(r)\) is
equivalent to the fractional mass perturbation \(\delta M/M\) inside
radius \(r\), 
\begin{equation}
\label{overden-m}
\Delta_{0}(r)=\delta M/M \propto M^{-\frac{n+3}{3}}.
\end{equation}
From equations (\ref{overmass}) and (\ref{overden-m}), we deduce that
the power-law power spectrum naturally generates a scale-free initial
condition with 
\begin{equation}
\label{epsilon-n}
\varepsilon=(n+3)/3.
\end{equation}

According to this model, halos of a given
mass \(M\) originate from density perturbations given by equation
(\ref{overden-m}) with \(n\) determined by the primordial power
spectrum after it is transferred according to the parameters of the
background universe and the nature of the dark matter. We plot this
effective \(n\) as a function of halo mass in Fig. \ref{fig-neff} for
the currently-favored \(\Lambda\)CDM universe. The value of \(n\simeq-2.5\) 
is a reasonable approximation for galactic halos
[i.e. \(n\simeq-2.5\pm0.1\) for \(M\simeq10^{8\pm2}M_{\odot}\), while
  \(n\simeq-2.5\pm0.2\) for \(M\) in the range from \(10^{3}M_{\odot}\)
  to \(10^{11}M_{\odot}\)]. For halos in the cluster mass range,
\(M\sim10^{15}M_{\odot}\), \(n\simeq-1.5\).

\subsection{Fluid Approximation of Collisionless CDM Halo Dynamics}
\label{fluid_approx_sect}
This approach is used to simplify our understanding of the formation of CDM
halos. We show that the fluid conservation equations for a gas with adiabatic
index \( \gamma =5/3 \) are a good approximation to the
dynamics of CDM halos and self-interacting dark matter halos, which
will be treated in detail in \S \ref{selfinteract_sect}. 
This may
bother some readers since, strictly speaking, the collisionless nature of CDM 
prohibits the use of such an approximation. However, a couple
of simple assumptions enable us to treat CDM halo dynamics with the usual
fluid conservation equations. For a more detailed description, see \cite{as04}.

We first assume spherical symmetry, and define the average of physical
quantities as follows:
\begin{equation}
\label{rho}
\rho =\int fd^{3}v,
\end{equation}
\begin{equation}
\label{aver}
\langle A\rangle \equiv \frac{\int Afd^{3}v}{\int fd^{3}v}=\frac{1}{\rho }\int Afd^{3}v,
\end{equation}
\begin{equation}
\label{velo}
u\equiv \langle v_{r}\rangle ,
\end{equation}
\begin{equation}
\label{pr}
p_{r}\equiv \rho \langle (v_{r}-\langle v_{r}\rangle )^{2}\rangle ,
\end{equation}
\begin{equation}
\label{ptheta}
p_{\theta }\equiv \rho \langle (v_{\theta }-\langle v_{\theta }\rangle )^{2}\rangle =\rho \langle v_{\theta }^{2}\rangle ,
\end{equation}
\begin{equation}
\label{pphi}
p_{\phi }\equiv \rho \langle (v_{\phi }-\langle v_{\phi }\rangle )^{2}\rangle =\rho \langle v_{\phi }^{2}\rangle ,
\end{equation}
 where \( f \) is the distribution function defined such that \(
f(\textbf{r},\,\textbf{v})d^{3}r d^{3}v = \) mass within an infinitesimal
volume \(d^{3}r d^{3}v\) at \((\textbf{r},\,\textbf{v})\),  \( \rho  \) is the
density, \( \langle A\rangle  \) is the average value of a certain
quantity \( A \), \( u \) is the radial bulk velocity, \( p_{r} \)
is the ``effective radial pressure'', and \( p_{\theta } \) is the
``effective tangential pressure''. Note that 
\(\langle v_{\theta }\rangle = \langle v_{\phi }\rangle = 0\) and
\(p_{\theta}=p_{\phi}\)
because of spherical symmetry. Anisotropy 
in the velocity dispersion occurs in general for collisionless systems --
i.e. \( p_{r}\neq p_{\theta } \), 
or anisotropy parameter \( \beta \neq 0 \), where 
\( \beta \equiv 1-\frac{p_{\theta}}{p_{r }} \)
-- implying that \( p_{r} \) and \( p_{\theta } \) should be treated 
separately.
In a highly collisional system, which is well described by fluid
conservation equations, \( p_{r}=p_{\theta } \) and the usual
pressure \( p=p_{r}=p_{\theta } \). 

A self-gravitating system of collisionless particles can be described by 
the collisionless Boltzmann equation 
\begin{equation}
\label{boltzmann}
\frac{df}{dt}\equiv\frac{\partial f}{\partial t}+{\bf v}\cdot{\bf\nabla}f
         -{\bf\nabla}\Phi\cdot\frac{\partial f}{\partial {\bf v}}=0.
\end{equation}
In spherical symmetry, \(f=f(|\textbf{r}|,\,\textbf{v})\), equation 
(\ref{boltzmann}) becomes
\begin{eqnarray}
0 & = & \frac{\partial
f}{\partial t}+v_{r}\frac{\partial f}{\partial r}+\left(
\frac{v_{\theta }^{2}+v_{\phi }^{2}}{r}-\frac{\partial \Phi }{\partial
r}\right) \frac{\partial f}{\partial v_{r}} \nonumber \\ 
&  &+\frac{1}{r}\left( v_{\phi }^{2}\cot \theta -v_{r}v_{\theta
}\right) \frac{\partial f}{\partial v_{\theta }} \nonumber \\ 
&  & -\frac{v_{\phi }}{r}\left( v_{r}+v_{\theta }\cot \theta \right)
\frac{\partial f}{\partial v_{\phi }}, \label{sph-boltz} 
\end{eqnarray}
where \( \Phi \) satisfies the Poisson equation,
\( {\nabla}^{2} \Phi = 4 \pi G \rho  \) \cite{BT}.
By multiplying equation (\ref{sph-boltz}) by \(v_{r}^{m}v_{\theta}^{n}\),  where
\(m,\,n\) are integers, and integrating over velocity $d^3v$, we can form 
a set of moment equations. Moment equations from the lowest order are
\begin{equation}
\label{mass}
\frac{\partial \rho }{\partial t}+\frac{\partial }{r^{2}\partial r}(r^{2}(\rho u))=0,
\end{equation}
\begin{equation}
\label{momentum}
\frac{\partial }{\partial t}(\rho u)+\frac{\partial }{\partial r}(p_{r}+\rho u^{2})
+\frac{2}{r}(p_{r}-p_{\theta }+\rho u^{2})=-\rho \frac{Gm}{r^{2}},
\end{equation}
\begin{equation}
\label{energyr}
\rho \frac{D}{Dt}\left( \frac{p_{r}}{2\rho }\right) +p_{r}\frac{\partial u}{\partial r}=\Gamma _{1},
\end{equation}
\begin{equation}
\label{angular}
\rho \frac{D}{Dt}\left( \frac{p_{\theta }}{2\rho }\right) +\frac{p_{\theta}u}{r}=\Gamma _{2},
\end{equation}
 \[
\vdots \]
 where \( m \) is the mass enclosed by a shell at radius \( r \), 
\( \frac{D}{Dt}\equiv \frac{\partial }{\partial t}
+u\frac{\partial }{\partial r}, \)
and 
\begin{equation}
\label{gamma1}
\Gamma_{1} = 
\frac{\rho
}{r}\left\langle 2\left( v_{r}-\left\langle v_{r}\right\rangle \right)
v_{\theta }^{2}\right\rangle
-\frac{1}{2r^{2}}\frac{\partial }{\partial r}
\left( r^{2}\rho \left\langle \left( v_{r}
-\left\langle v_{r}\right\rangle \right) ^{3}\right\rangle \right) ,
\end{equation}
\begin{equation}
\label{gamma2}
\Gamma _{2}=
-\frac{1}{4r^{4}}\frac{\partial }{\partial r}\left( r^{4}\rho
\left\langle \left( v_{r}-\left\langle v_{r}\right\rangle \right)
v_{\theta }^{2}\right\rangle \right) . 
\end{equation}
Equations (\ref{mass}-\ref{angular}) are 
conservation equations of mass, momentum,
``radial'' energy, and angular momentum, respectively.
Note that equations (\ref{mass})-(\ref{gamma2}) are all in exact
form, and the hierarchy of equations is not closed in principle. 

Now we make a further simplification that the distribution of \(v_{r}\) is
skewless -- \(v_{\theta}\) and \(v_{\phi}\) are 
naturally skewless because of
spherical symmetry. In other words, we assume that \(v_{r}\) has a
symmetric distribution around \(\langle v_{r} \rangle\). It is not
straightforward to show that \(\Gamma_{1}\) and \( \Gamma_{2}\) are
negligible in equations (\ref{energyr}) and (\ref{angular}). However,
we demonstrate that the assumption of ``skewlessness'' in the fluid 
approximation yields results which are in good agreement 
with the purely collisionless CDM structure for specific examples.
Equations (\ref{mass}) - (\ref{energyr})
with the condition $p_{\theta}=0$ and $\Gamma_{1}=0$ can be used to solve
purely radial problems, such as the self-similar spherical infall
problems with similarity
solutions by \cite{BERT} ($\varepsilon=1$) and \cite{FG} ($\varepsilon=1/6$).
Comparisons of the fluid approximation in these cases, as shown in 
Figures \ref{fig-bert} and \ref{fig-e16}, reveal an excellent agreement
with the true collisionless solutions.
The difference observed at caustics -- places
where the density becomes infinite -- is negligible, because caustics do not 
affect the overall dynamics of the halo.  Since the skew-free assumption 
naturally neglects dynamically 
unimportant structure (e.g. caustics) while accurately reproducing the profile
of the exact solution in these radial cases, it may also be applied to describe
 CDM halos, in which particles also have a tangential motion.

The final assumption is that the velocity distribution is isotropic, or
\(p_{r} = p_{\theta }\). This is an empirical assumption: CDM
halos in cosmological N-body simulations show mild anisotropy. For
instance, \cite{carlberg} show that CDM halos in their numerical
simulation can be well-fitted by a fitting formula
\begin{equation}
\label{anisotropy}
\beta(r)=\beta_{m}\frac{4r}{r^{2}+4},
\end{equation}
where \(r\) is in units of \(r_{200}\), and \(\beta_{m}\approx 0.5\) in 
N-body simulations \cite{carlberg,CKK}.

With these assumptions -- spherical symmetry, skew-free velocity
distribution and isotropic velocity dispersion -- and with the
operation (equation \ref{energyr}) + 2\(\times\)(equation
\ref{angular}), the usual energy conservation equations are
obtained. They are
\begin{equation}
\label{mass-cdm}
\frac{\partial \rho }{\partial t}+\frac{\partial }{r^{2}\partial r}(r^{2}(\rho u))=0,
\end{equation}
\begin{equation}
\label{momentum-cdm}
\frac{\partial }{\partial t}(\rho u)+\frac{\partial }{\partial r}(p+\rho u^{2})+\frac{2}{r}\rho u^{2}=-\rho \frac{Gm}{r^{2}},
\end{equation}
\begin{equation}
\label{energy-cdm}
\frac{D}{Dt}(\frac{3p}{2\rho })=-\frac{p}{\rho }\frac{\partial }{r^{2}\partial r}(r^{2}u),
\end{equation}
which are identical to the fluid conservation equations for a \(\gamma=5/3\)
gas in spherical symmetry. 

In this section, we showed how one could use the usual fluid
conservation equations to approximate collisionless CDM halo
dynamics. \S \ref{UMH} and \ref{selfinteract_sect}
describe its practical application. 

\subsection{Halo Formation by Non-Self-Similar Infall: Mass Assembly 
History and the Origin of CDM N-body Halo Profiles}
\label{UMH}
N-body simulations of CDM have not only found a universal halo density profile,
but have also found that the masses and concentrations of individual N-body
CDM halos grow over time according to simple universal formulae \cite{WEC}.  
Very similar results were also found for N-body simulations of halos formed by
the instability of cosmological pancakes (see \S \ref{simul_sect}). 
In what follows we use the fluid approximation to show that this universal 
time-dependent halo density profile can be understood as the dynamical 
outcome of continuous infall according to the universal mass accretion 
history \cite{AAS}.

\begin{figure}[!t]
  \begin{center}
  \includegraphics[width=5.5in]{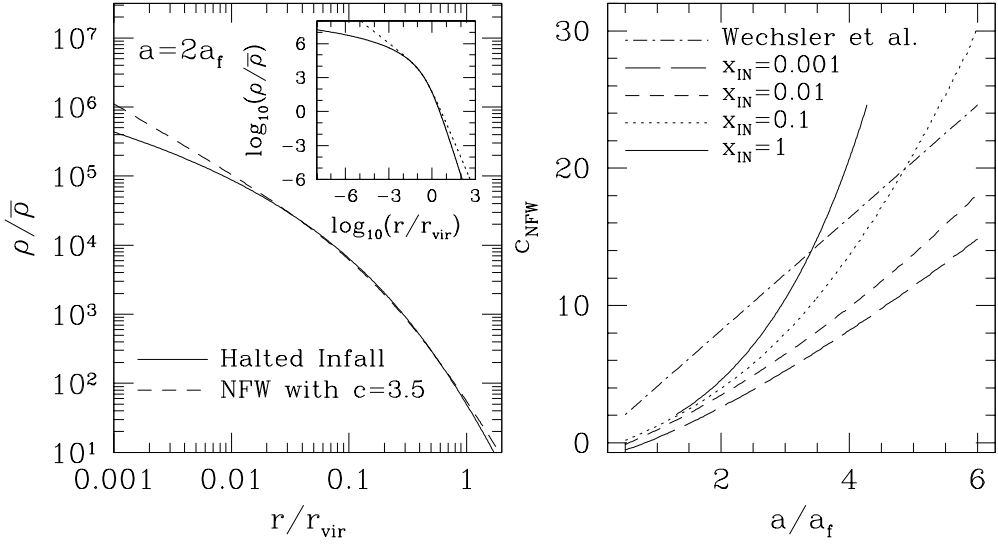}
  \end{center}
  \caption{(left) Density profile from halted infall model along with
best-fitting NFW profile for this profile at present.  Inset in
upper-right shows same over much larger range. (right) Evolution of NFW
concentration parameter in the halted infall model, compared with empirical 
relation of \cite{WEC} for CDM N-body halos. Different line types
indicate different ranges $x_{\rm in}<x<1$, within which halo was fit to
an NFW profile, where $x\equiv r/r_{\rm vir}$, $r_{\rm vir}\equiv r_{200}$.
\label{mah1}}
\end{figure}

\begin{figure}[!t]
  \begin{center}
  \includegraphics[width=3.7in]{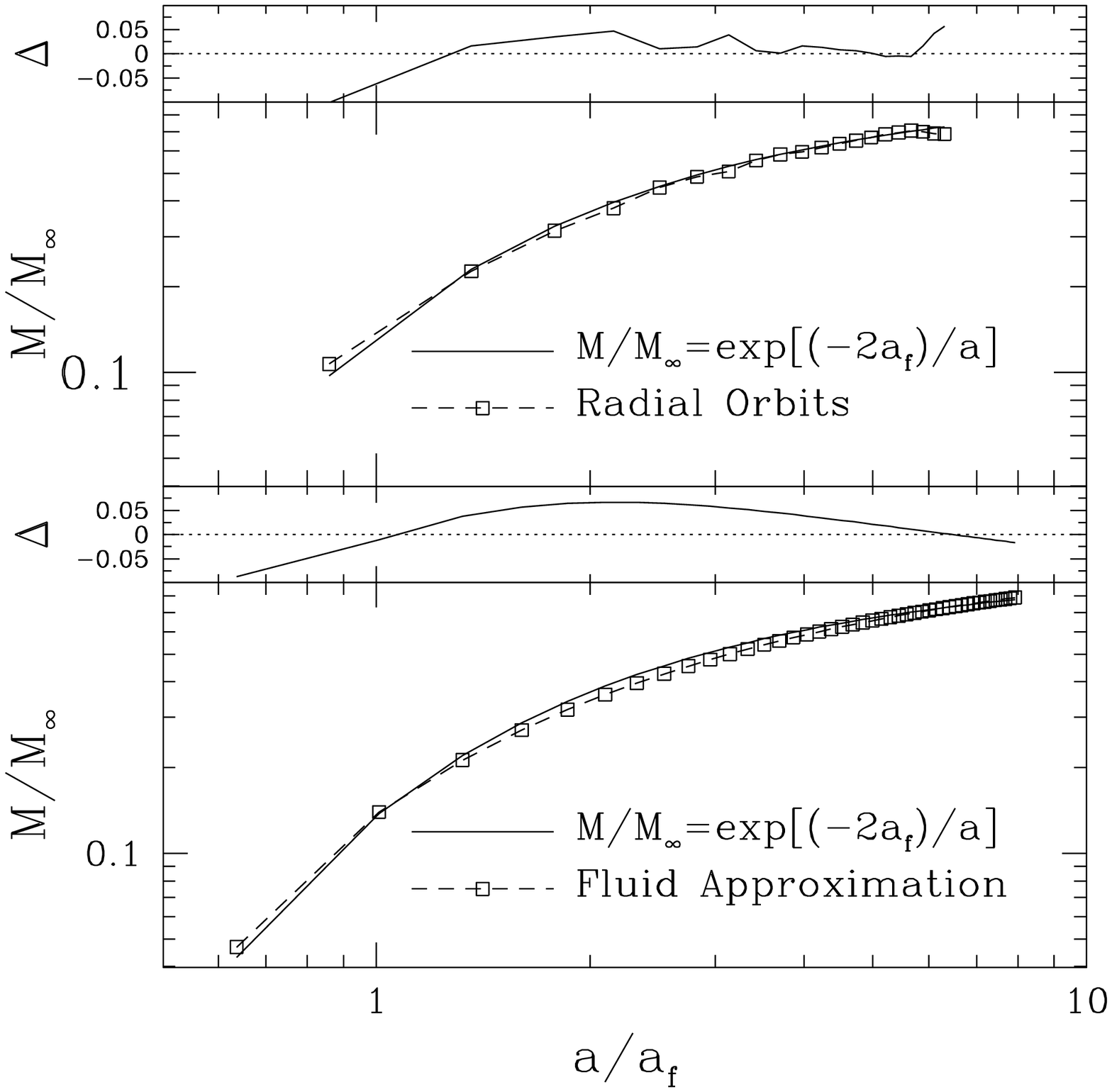}
  \end{center}
  \caption{Evolution of mass for the radial orbits (top) and fluid 
approximation (bottom) simulations, compared with the empirical relation of
\cite{WEC} for CDM N-body results.  Shown above each are the fractional 
deviations $\Delta \equiv (M_{\rm exact}-M)/M$.
\label{mah2}}
\end{figure}

\subsubsection{Models and Initial Conditions}

We attempt to understand the form and evolution of dark matter halos with
three spherically-symmetric models; halted infall, radial orbits, and a
fluid approximation.  Each model assumes that the mass $M_{\rm vir}$
within an overdensity $\Delta_{\rm vir}$ follows the relation given by
\cite{WEC}
\begin{equation} 
\label{wec}
M_{\rm vir}(a)=M_\infty\exp\left[-Sa_{\rm f}/a\right], 
\end{equation}
where $S$ is the logarithmic
mass accretion rate $d{\rm ln}M_{\rm vir}/d{\rm ln}a$ when $a=a_{\rm f}$.
Here and in \cite{WEC}, $S=2$. Such a relation is claimed to be a good 
fit
to the evolution of halos of different masses and formation epochs.  We
use $\Delta_{\rm vir}=200$, so that the halo has a mass $M_{\rm 200}$ and
radius $r_{200}$.  We have found an initial perturbation profile
consistent with equation (\ref{mah1}) (for EdS),
\begin{equation}
\frac{\delta M}{M}\equiv\frac{M-\overline{M}}{\overline{M}}=\delta_{\rm i}
{\rm ln}\frac{M}{M_\infty},
\label{init}
\end{equation}
where $\delta_{\rm i}$ depends on the initial scale factor $a_{\rm i}$,
$a_{\rm f}$, and $\delta_{\rm vir}$, and $\overline{M}$ is the unperturbed
mass.  The parameter $b=1$ if pressure or shell crossing are not present
outside of $r_{\rm vir}$.  If they are present outside the halo however,
the initial perturbation is not guaranteed to lead to the correct mass
accretion rate.  In our radial orbit and fluid approximation calculations below,
where shell crossing and pressure are indeed present outside of $r_{\rm vir}$, 
we have found that the resulting mass is close to that of equation
(\ref{wec}) if $b$ is allowed to vary as a fitting parameter 
($b=1(0.7)$ in the fluid approximation (radial orbits) calculations).

\begin{figure}[!t]
  \begin{center}
  \includegraphics[width=3.7in]{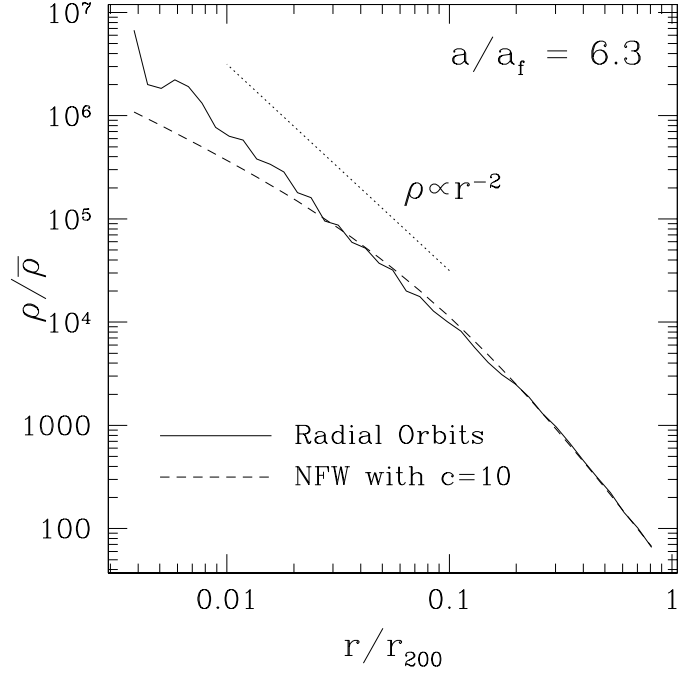}
  \end{center}
  \caption{Density profile at the end of the radial orbit simulation.
\label{mah3}}
\end{figure}

\subsubsection{Halted Infall Model} 
\label{eqm}

In the simplest model, we have assumed that infalling shells come to 
an abrupt halt upon crossing into the halo, so that the velocity is zero 
for $r<r_{\rm vir}$.  The mass of the halo is
\begin{equation} 
M_{\rm vir}(a)=\frac{4\pi}{3}\delta_{\rm vir}\overline{\rho}r_{\rm vir}^3, 
\label{vmass}
\end{equation} 
where $\overline{\rho}$ is the cosmic mean mass density at that epoch.  Mass
continuity implies the density $\rho_{\rm vir}$ just inside the virial
radius is related to the rate of halo mass and radius increase according to
\begin{equation}
\label{jump}
\frac{dM_{\rm vir}}{da}=4\pi\rho_{\rm vir}r_{\rm vir}^2\frac{dr_{\rm vir}}{da}.
\end{equation}
Differentiating equation (\ref{vmass}) and combining with equations 
(\ref{wec}) and (\ref{jump}), one obtains 
\begin{equation}
\label{rhovir}
\frac{\rho_{\rm vir}}{\bar{\rho}_0}=\delta_{\rm
vir}a^{-3}\left[1+\frac{3a}{Sa_{\rm f}}\right]^{-1},
\end{equation} 
where $\bar{\rho}_0$ is the mean background density at $a=1$.  The virial
radius is given by
\begin{equation}
\label{rvir}
\frac{r_{\rm vir}}{r_{\rm vir,0}}=a {\rm
exp}\left[\frac{-Sa_{\rm f}}{3}\left(\frac{1}{a}-1\right)\right].
\end{equation}
Equations (\ref{rhovir}) and (\ref{rvir}) are parametric in $a$, implying
a radial density profile $\rho(r)=\rho_{\rm vir}(r_{\rm vir})$ that is
frozen in place as matter crosses $r_{\rm vir}$.
Taking the limit in which $a\rightarrow\infty$, the outer density profile 
approaches $\rho\propto r^{-4}$ at late times, consistent with finite 
mass, while the inner slope becomes asymptotically flat.  
\begin{figure}[!t]
  \begin{center}
  \includegraphics[width=3.7in]{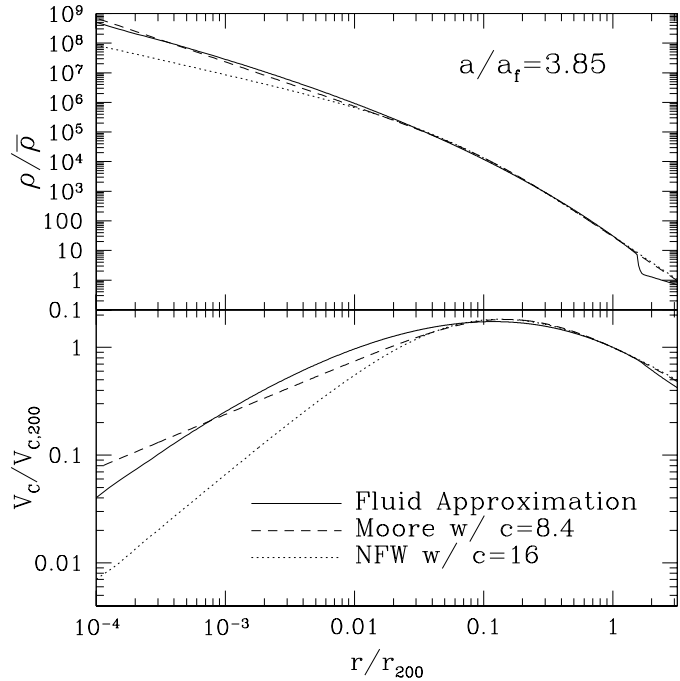}
  \end{center}
  \caption{(top) Density profile at the end of the isotropic fluid 
        calculation. (bottom) Circular velocity profile.\label{mah4}}
\end{figure}
The NFW profile is 
\begin{equation}
\label{nfw}
\frac{\rho(x)}{\overline{\rho}}=\frac{\delta_{\rm vir}g(c)}{3x(1+cx)^2},
\end{equation}
where
\begin{equation}
g(c)=\frac{c^2}{{\rm ln}(1+c)-c/(1+c)},
\end{equation}
and $x\equiv r/r_{\rm vir}$.  Combining equations (\ref{rhovir}) and
(\ref{nfw}) with $x=1$, yields an equation for the
evolution of concentration with scale factor (see Fig. \ref{mah1}),
\begin{equation}
\label{cofa}
\frac{a}{a_{\rm f}}=S\left[\frac{(1+c)^2}{g(c)}-\frac{1}{3}\right].
\end{equation}

\subsubsection{Radial Orbits Model}
\label{rom}

We use a finite-difference spherical mass shell code to follow the
evolution of a small amplitude initial perturbation given by 
equation~(\ref{init}), which is chosen so that the
resulting virial mass will evolve according to equation (\ref{wec}).  
The shell code has an
inner reflecting core and the results presented here used 20,000 shells.
The resulting evolution of halo mass and the comparison of the halo 
density profile with the NFW profile are shown in Figures~\ref{mah2}
and \ref{mah3}.

\subsubsection{Fluid Approximation Model}
\label{fam}
 
\begin{figure}[!t]
  \begin{center}
  \includegraphics[width=3.7in]{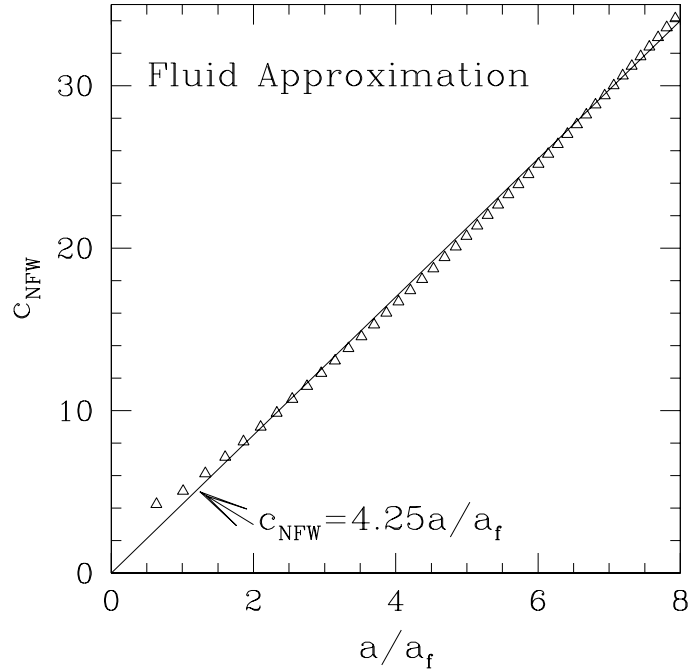}
  \end{center}
  \caption{Evolution of concentration parameter with scale factor in the 
isotropic fluid calculation.\label{mah5}}
\end{figure}

As mentioned earlier in \S \ref{fluid_approx_sect}, the
collisionless Boltzmann equation in spherical symmetry yields fluid
conservation equations ($\gamma=5/3$) when random motions are isotropic.  
Halos in N-body simulations have somewhat radially-biased random motion, but the
bias is small, especially in the center. Outside the virialized halo, in the 
infall region, the radial bias is irrelevant, since the motion is highly 
supersonic and random motions do not affect the dynamics there. This isotropic
fluid model is therefore a better
approximation to halo formation in N-body simulations than one with purely
radial motion. We use a 1-D, spherical, Lagrangian hydrodynamics code as
in \cite{TW}, using 1,000 zones logarithmically spaced in mass.  The
initial conditions were chosen in the same way as those for the radial
orbit model (Eq. (\ref{init})), with zero initial temperature. Results are 
plotted in Figures~\ref{mah2}-\ref{mah5}.  

\subsubsection{Results} 
Our results can be summarized as follows:
\begin{itemize}
\item The halted infall model of \S \ref{eqm} does not reproduce the 
linear evolution of concentration parameter with scale factor reported by 
\cite{WEC}, but can be fit by an NFW profile over a limited range of radii 
and scale factors (Fig. \ref{mah1}).  
\item We have derived an initially linear perturbation profile that is a 
good model for the spherically-averaged initial condition that leads to 
the Lagrangian mass evolution of CDM halos found in N-body simulations by 
\cite{WEC}. 
\item Starting from this perturbation, the radial orbit model of section 
\ref{rom} fails to reproduce the inner slope of the NFW profile, approaching 
$\rho\propto r^{-2}$ instead, consistent with the argument of \cite{RT} 
(Fig. \ref{mah3}).
\item The fluid approximation of \S \ref{fam}, however, leads to 
a halo that is well-fitted by the NFW and Moore 
profiles for all radii resolved by N-body simulations ($r/r_{200}\geq 
0.01$) (Fig. \ref{mah4}).
\item In addition, the evolution of the NFW concentration parameter in the 
fluid approximation is a close match to that of \cite{WEC}, with $c_{\rm
NFW}=4.25a/a_{\rm f}$ a good fit (Fig. \ref{mah5}); $c_{\rm NFW}=4.1a/a_{\rm f}$ was
the relation reported by \cite{WEC}.
\end{itemize}

The fluid approximation model reproduces the N-body results remarkably
well, once the mass accretion history is given.  We are thus led to
conclude that complicated merging processes are not necessary in order to
understand the overall structure and evolution of the halo mass
distribution, and that it is largely determined by the mass accretion
history.

\subsection{Structure of Self-Interacting Dark Matter Halos}
\label{selfinteract_sect}

As we will demonstrate below, halo models with a small flat core, such as
the Truncated Isothermal Sphere (TIS) halo equilibrium model described in \S~\ref{tis_sect}
agree well with the statistical properties of observed halos and with the averaged 
properties of simulated halos. However, the dynamical origin of such an equilibrium 
structure and the particular question of how the flat density core can form are 
not addressed by an equilibrium analysis. Halos in high-resolution N-body 
simulations in the $\Lambda$CDM model, on the other hand, show cuspy density profiles.

Among several suggestions made to resolve this problem, self-interacting
dark matter (SIDM) has drawn substantial attention. It was suggested by 
\cite{SS} that the purely collisionless nature of CDM be replaced
by SIDM which interacts by non-gravitational, microscopic interaction
(e.g. elastic scattering). Various attempts have been made to incorporate
SIDM in semi-analytical studies (e.g. \cite{BSI})
as well as in N-body simulations (e.g. \cite{B00,Dave,KW,Yoshida}),
and they all show that flat density cores can arise. However,
new theoretical problems have arisen for these SIDM halos: 1) study of 
``isolated'' (i.e. no cosmological infall) halos revealed unstable cores
(\cite{B00,KW}) and 2) cosmological N-body simulations
(\cite{Dave,Yoshida}) showed a more prominent
flattening effect for high-mass halos 
(\( M_{\rm halo}\simeq 10^{12}M_{\odot } \))
than for low-mass halos (\( M_{\rm halo}\simeq 10^{9}-10^{10}M_{\odot } \)),
which appears to contradict observations which show that the flattening effect
is most prominent for dwarfs and low surface brightness (LSB)
galaxies with \( M_{\rm halo}\simeq 10^{9}-10^{10}M_{\odot } \).

The semi-analytical study by \cite{BSI} applied the fluid approximation
to an isolated halo, properly accounting for the effective ``conduction''
arising from the SIDM collisionality. They showed that SIDM halos
could have a lifetime long enough to survive in a Hubble time, if initially
the scattering mean free path is much greater than the size of the halos.
However, as was found in \cite{B00, KW}, the scattering mean free path
can be smaller than the halo size if the scattering cross-section is
large enough. In this case, $isolated$ SIDM halos would still be unstable. 

We show that these
problems can be naturally resolved \cite{AS,as04}. We partially adopt
the formalism by \cite{BSI} and improve upon it by properly including
cosmological infall. 
We have thereby derived an analytical,
cosmological similarity solution.
Our approach is as follows. As shown in \(\S\)
\ref{hs_sect}, the power spectrum
in the mass range of dwarfs and LSBs is well described by a power
law \( P(k)\propto k^{n} \) with \( n\simeq -2.5 \). From equation
(\ref{overden-m}), we then have
\begin{equation}
\label{overden-sidm}
\delta M/M \propto M^{-1/6}.
\end{equation}
This also results in 
\begin{equation}
\label{rta_sidm}
r_{\rm ta}\propto t^{2},
\end{equation}
where \(r_{\rm ta}\) refers to a ``turn-around'' radius at which
radial velocity vanishes. For any value of \(n\), 
\(r_{\rm ta}\) serves as a natural length scale for self-similar accretion.

\begin{figure}[!t]
\centering
\includegraphics[width=3.5in]{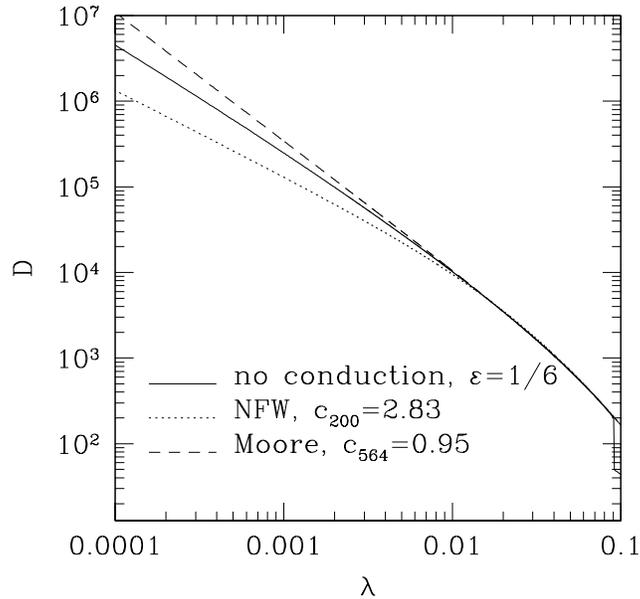}
\caption{Halo density profile produced by fluid
approximation with \( \varepsilon=1/6 \) for standard CDM halos,
compared to the best-fitting NFW and Moore profiles. 
The adiabatic solution, whose inner slope is about -1.27, is a good
fit to density profiles of CDM halos from N-body simulations.}
\label{fig-adia}
\end{figure}

This is not true for SIDM halos because the SIDM interaction
brings in a new length scale which does not in general preserve this
self-similarity. However, \( n=-2.5 \) is a magic number which preserves
the self-similarity because in this case the new length scale is always
a fixed fraction of the halo size. Specifically, SIDM 
introduces a conductive heat flux (\cite{BSI})
\begin{equation}
\label{heatflux}
f=-\frac{3ab\sigma }{2}\sqrt{\frac{p}{\rho }}
\left( a\sigma^{2}+\frac{4\pi G}{p}\right)^{-1}
\frac{\partial }{\partial r}\left(\frac{p}{\rho }\right)
\end{equation}
 where \( \sigma  \) is the interaction cross-section, \( \rho  \)
is the density, and \( p=\rho \left< v-\left< v\right>\right>^{2} \) is 
the effective ``pressure''
(\cite{AS, as04} considered an elastic scattering case, 
in which \( a=2.26 \)
and \( b=1.002 \)). The thermal energy changes according to 
\( \rho \frac{\partial e}{\partial t}\propto r^{2}_{\rm ta}t^{-5} \),
while the conductive heating rate is 
\( \nabla \cdot \textbf{f}\propto r_{\rm ta}^{3}t^{-7}. \)
Self-similarity is preserved only when these terms have the same time
dependence, or \( r_{\rm ta}\propto t^{2} \). From eq (\ref{rta}), this condition
is equivalent to \( n=-2.5 \).

\begin{figure}[!t]
\centering
\includegraphics[width=4.5in]{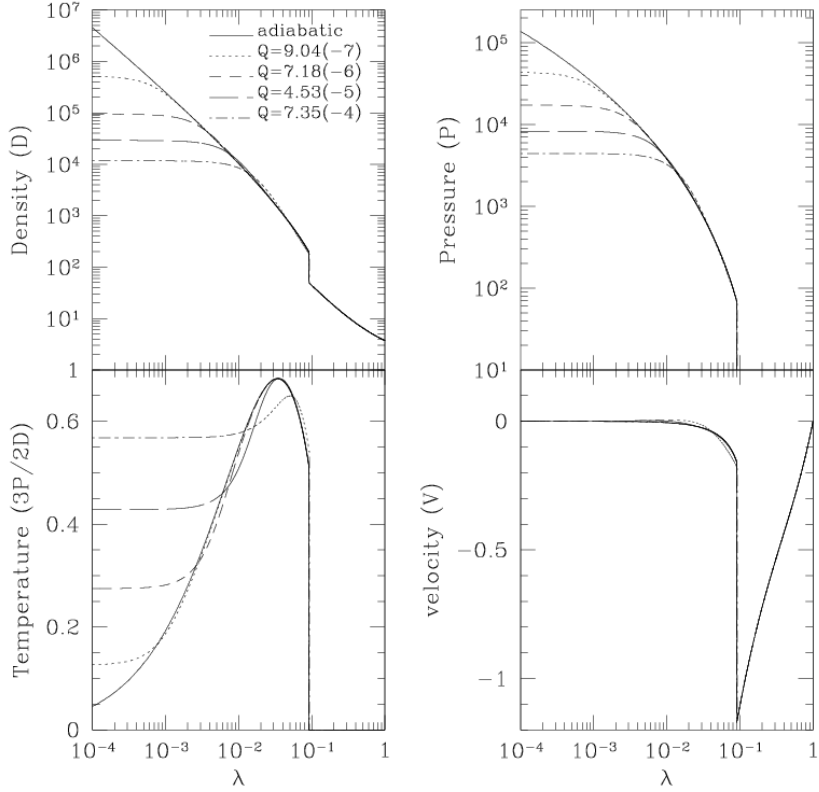}
\caption{Profiles of the low-\(Q\) solutions.}
\label{fig-low}
\end{figure}

\begin{figure}[!t]
\centering
\includegraphics[width=4.5in]{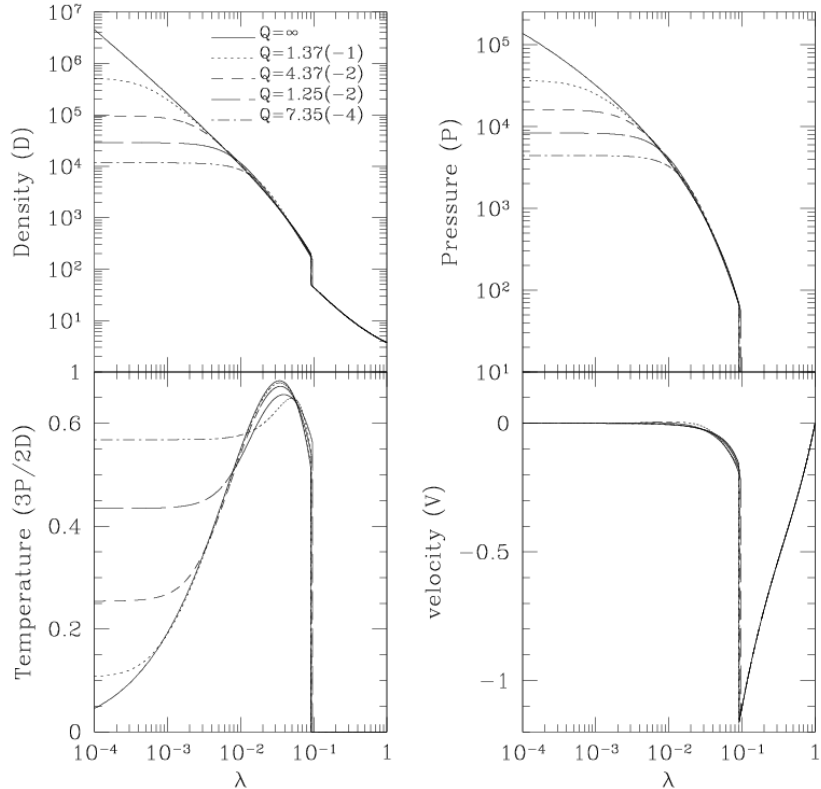}
\caption{Profiles of the high-\(Q\) solutions.}
\label{fig-high}
\end{figure}

\begin{figure}[!t]
\includegraphics[width=5.5in]{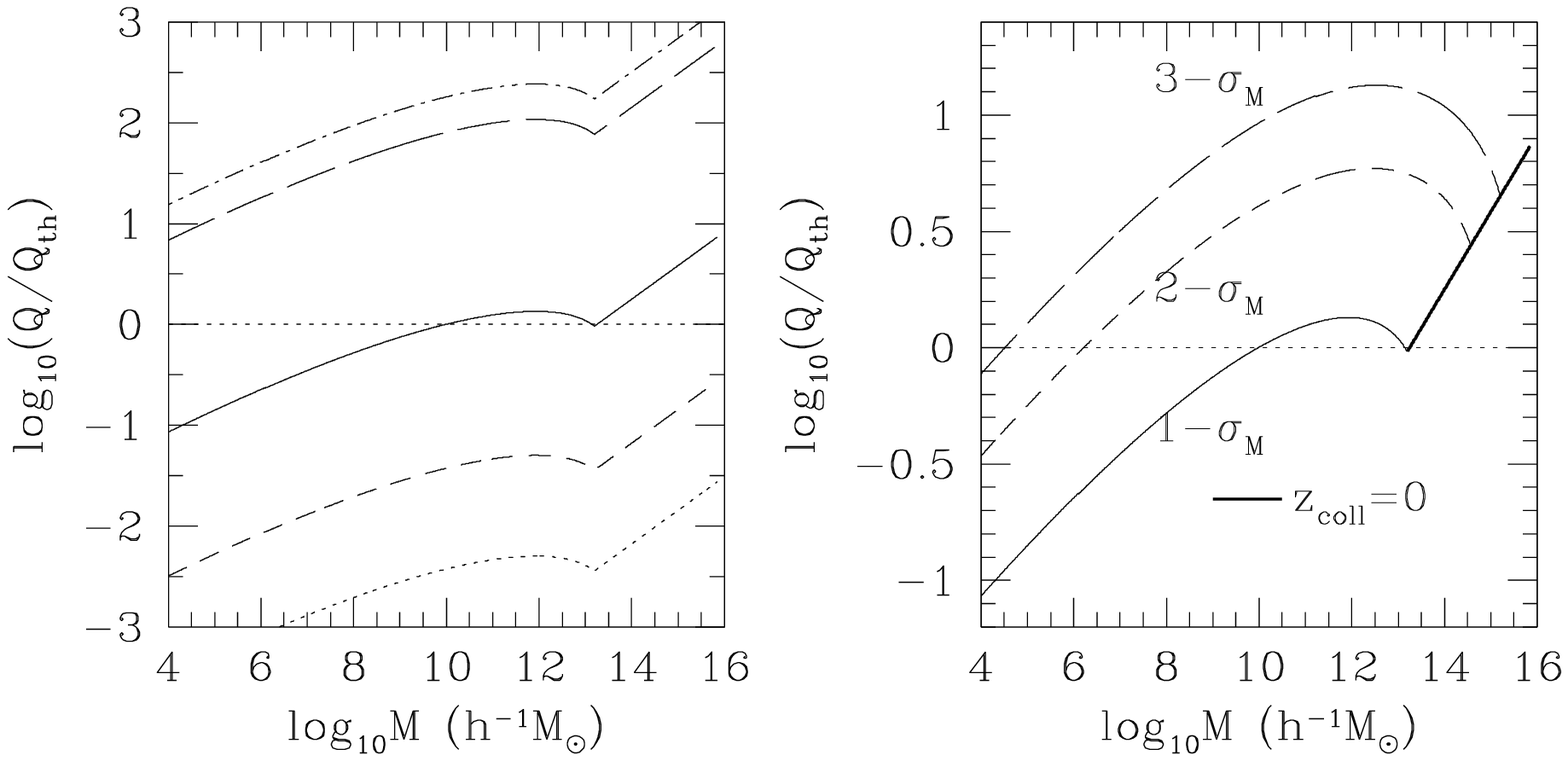}
\caption{Collisionality parameter \(Q\) vs. halo mass $M$. (left) 
``Typical'' halos, well represented statistically by a 
\(1-\sigma_{M}\) fluctuation. Different curves correspond to different
scattering cross section \(\sigma\)
(\(\sigma=0.56,\,5.6,\,200,\,1.2\times10^{4},\,2.7\times10^{4}\) from
bottom to top). (right) Different \(\nu-\sigma_{M}\) fluctuations for
\(\sigma=200 cm^{2} g^{-1}\). }
\label{fig-q}
\end{figure}

As shown in \S \ref{hs_sect}, dwarfs and LSBs are well described by 
self-similar
solutions with this requirement. This implies that the core of an SIDM halo 
in that mass range 
would be stable, since
the core's size would grow in proportion to the virial radius
\( r_{\rm v} \), or the turnaround radius \( r_{\rm ta} \). In other words,
the collapse of halo cores predicted for isolated SIDM halos is entirely
prevented by cosmological infall, which resolves the issue of stability
of SIDM halo cores.

Before treating soft-core solutions, we describe properties of this
self-similar solution without SIDM collisionality, i.e. the case for
$f=0$.
The solution obtained by applying the fluid approximation -- equations 
(\ref{mass-cdm} - \ref{energy-cdm}) -- to the $\varepsilon=1/6 (n=-2.5)$ case
yields a profile which resembles CDM halos in many aspects. First, as 
shown in Figure \ref{fig-adia}, it has
a density cusp with a logarithmic slope $\approx -1.27$ at 
$ 4 \times 10^{-3} < r/r_{200} < 1.4 \times 10^{-2}$, which is between
$-1$ (NFW profile) and $-1.5$ (Moore profile)\footnote{
This solution is 
well-fitted by NFW profiles with very low concentration parameter 
$c\approx 3$. $\varepsilon=1/6$ corresonds to a fast mass accretion rate,
$d \ln M / d \ln a = 6$, which is typical for the earliest epoch of mass 
accretion for standard CDM halos at $a/a_{c}<1$. Therefore, this solution 
can be said to capture the very early formation epoch of CDM halos.
}. Second, the
temperature profile is very similar to that of CDM halos. The temperature
is zero at the center, rises to a maximum as radius increases until a point
where the temperature starts to drop again (see the solid line in the
temperature plot of Figure \ref{fig-low}). This validates the use
of our model for the study of SIDM halos and even makes it possible to
correctly calculate the physical effect of SIDM collisionality.

In the presence of SIDM collisionality, similarity solutions 
reveal soft-cores. Different solutions arise for different values of
the dimensionless collisionality parameter
\( Q\equiv \sigma \rho _{b}r_{s} \), where \( \rho _{b} \) is the
cosmic mean matter density and $r_{s}$ is the effective shock radius.
\( Q \) is constant during the matter-dominated
epoch, \( z\ge 1, \) because \( \rho _{b}\propto t^{-2} \) and 
\( r_{v}\propto t^{2} \).

Fig.~\ref{fig-low} and Fig.~\ref{fig-high} show profiles for different
\( Q \)'s. One would 
naively expect that as \( Q \) (\( \sigma  \)) increases, the central
density decreases because of the increasing flattening effect. However,
if \( Q \) is too large, the mean free path between collisions is smaller 
than the size of the halo, which reduces the flattening
effect. Therefore, there are two opposite regimes, low-\( Q \) (long
mean free path; Fig~\ref{fig-low}) and high-\( Q \) (short mean free path; 
Fig~\ref{fig-high}), which are separated by some threshold \( Q_{\rm th} \). We
found that \( Q_{\rm th}=7.35\times 10^{-4} \).

The behavior of high-\(Q\) solutions also gives us a clue as to how a constant
cross section \(\sigma\) may still resolve the problem 2). By using the
Press-Schechter formalism, \cite{AS, as04} 
relate \(Q\) to the halo mass \(M\). When one matches 
\(Q (M\simeq 10^{9} - 10^{10} M_{\odot}) \)
to \(Q_{\rm th} \), which is well-fitted by the empirical rotation fitting
formula for dwarfs and LSBs (described in the following paragraph),
one finds that \(\sigma\simeq 200\, \rm cm^{2} g^{-1}\) and collisionality
increases as mass increases, which indicates that halos more massive
than dwarfs and LSBs will have \(less \) density-flattening. This 
argument is consistent with previous N-body simulations
by \cite{Yoshida} and \cite{Dave}: they used
\(\sigma \simeq 0.1-10\, \rm  cm^{2} g^{-1} \), which is in the low-\(Q\)
regime. In the low \(Q\) regime, increasing \(Q\) means more 
density-flattening. They simply did not push \(\sigma\) to higher values,
which would have covered the high-\(Q\) regime (See Fig.~\ref{fig-q}).

\begin{figure}[!t]
\centering
\includegraphics[width=3.5in]{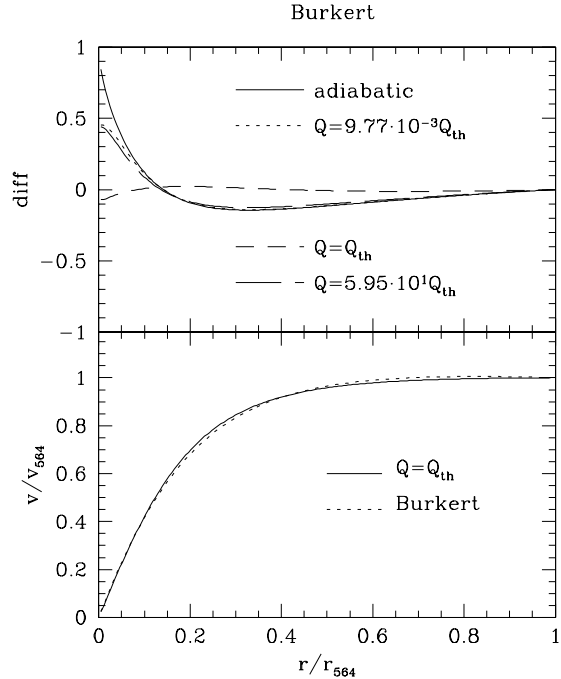}
\caption{Rotation curve fitting: The \(Q_{\rm th}\) solution is the one that 
is best-fitted by the Burkert profile.}
\label{fig-rotation}
\end{figure}

Since the \( Q_{\rm th} \) solution shows the most prominent density-flattening
effect, it can be said to be the most relaxed state possible for SIDM
halos. We found that in fact 
the solution with \( Q=Q_{\rm th} \)
is well described by the TIS profile \cite{as04}. We fitted rotation curves of
various \( Q \) solutions with the empirical rotation curve fitting
formula found by Burkert (\cite{B}; Burkert profile) and found that the
\( Q=Q_{\rm th} \) profile is best fit by the Burkert profile 
(See Fig.~\ref{fig-rotation}). 
As shown in \cite{ISb}, the TIS profile is also well-fitted by the Burkert
profile. The fact that \( Q_{\rm th} \) is best fit by the Burkert profile
(and therefore the TIS profile) suggests that in fact \( Q_{\rm th} \)
is the most relaxed system among the various \( Q \) solutions, since
the requirement for TIS, the minimum energy, is usually met by the
most relaxed system.

\section{The Truncated Isothermal Sphere (TIS) Model}
\label{tis_sect}
We have developed an analytical model for the postcollapse equilibrium
structure of virialized objects which condense out of a 
cosmological background universe, either matter-dominated or flat with
a cosmological constant \cite{SIR,ISb}. The model is based upon the assumption
that cosmological halos form from the collapse and virialization of
``top-hat'' density perturbations and are spherical, isotropic, and 
isothermal. This leads to a unique, nonsingular
TIS, the minimum-energy solution of the Lane-Emden equation
(suitably modified for non-zero cosmological constant $\Lambda\neq0$). 
The size $r_t$ and velocity dispersion $\sigma_V$
are unique functions of the mass and redshift of formation of the object
for a given background universe. Our TIS density profile flattens to a 
constant central value, $\rho_0$, which is 
roughly proportional to the critical density of the universe at the 
epoch of collapse, with a small core radius $r_0\approx r_t/30$ (where 
$\sigma_V^2=4\pi G\rho_0r_0^2$ and 
$r_0\equiv r_{\rm King}/3$, for the ``King
radius'' $r_{\rm King}$, defined by \cite{BT}, p. 228).
The density profiles for gas and dark matter are assumed to be the same 
(no bias), with gas temperature $T=\mu m_p\sigma_V^2/k_B$.
While this TIS density profile is obtained numerically by solving a
differential equation, it is well-fitted by the following approximation:
\begin{equation}
\label{tisfit}
\rho(r)=\rho_0\left({A\over a^2+r^2/r_0^2}-{B\over b^2+r^2/r_0^2}\right),
\end{equation}
where $A=21.38$, $B=19.81$, $a=3.01$, $b=3.82$ \cite{SIR,ISb}.
\begin{figure}[t!]
\centering
\includegraphics[width=5in]{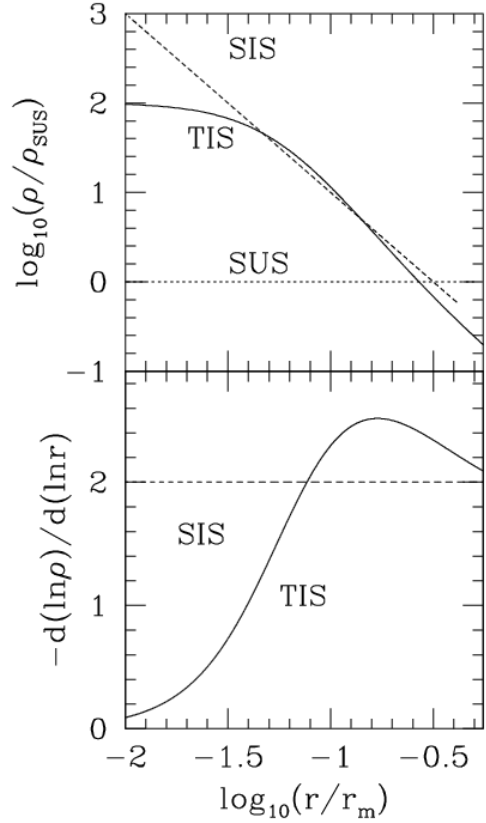}
\caption{(top) Density profile of TIS which forms from the virialization
of a top-hat density perturbation in a matter-dominated universe. Radius $r$ is in
units of $r_m$ - the top-hat radius at maximum expansion. Density $\rho$ is
in terms of the density $\rho_{SUS}$ of the SUS approximation for 
the virialized, post-collapse top-hat. (bottom) Logarithmic slope of density 
profile.    \label{profile}} 
\end{figure}

\begin{table}
\begin{minipage}[c]{\textwidth}
\caption{A comparison of three approximations for the postcollapse equilibrium
structure of top-hat density perturbations in a low density universe.}
\begin{tabular}{@{}lcccc}
\\[1mm]
&SUS\footnote{A top-hat perturbation of a given mass collapses at a given redshift
in a background universe with given values of $\Omega_0$ and $\lambda_0$; all of these 
values are held fixed in this comparison of the three approximations.}
&SIS\footnote{These SIS numbers are an approximation which ignores the small
modification of the Lane-Emden equation solution to take account of $\Lambda\neq0$, but
accounts for the more important effects of $\Lambda\neq0$ on top-hat evolution,
energy conservation and the virial theorem.}&TIS (SCDM;$\Lambda$CDM, $z=0$)
\\
\hline
$\eta/\eta_{\rm SUS}$&1&0.833&1.11;1.07\\[1mm]
$
{T}/{T_{\rm SUS}}$&1&3&2.16;2.19\\[1mm]
$\displaystyle{{\rho_0}/{\rho_t}}$& 1&$\infty$&514;530\\[1mm]
$\displaystyle{{\langle\rho\rangle}/{\rho_t}}$&1&3&3.73;3.68\\[1mm]
$\displaystyle{{r_t}/{r_0}}$& -- NA --&$\infty$&29.4;30.04\\[1mm]
${\Delta_c}/{\Delta_{\rm c,SUS}}$&1&
1.728&0.735;0.774\\[1mm]
$K/|W|$&0.5&0.75&0.683;0.690
\\\hline
\end{tabular}
\end{minipage}
\end{table}

These TIS results differ from those of the more familiar approximations in
which the virialized sphere resulting from a top-hat perturbation  
is assumed to be either the standard uniform sphere (SUS)
or else a singular isothermal sphere (SIS).
We summarize their comparison in Fig.~\ref{profile} and Table 1, where $\eta$ is the 
final radius of the virialized sphere in units of the top-hat radius $r_m$
at maximum expansion (i.e. $\eta_{\rm SUS}=0.5$), $\rho_t\equiv\rho(r_t)$, 
$\langle\rho\rangle$ is the average density of the virialized spheres, 
$\Delta_c=\langle\rho\rangle/\rho_{\rm crit}(t_{\rm coll})$,
and $K/|W|$ is the ratio of total kinetic (i.e. thermal) to gravitational potential
energy of the spheres.   

For all cases of current astronomical interest, an excellent approximation to the
exact results in \cite{ISb} (Paper II) for the dependence of the TIS halo parameters on 
the halo mass, collapse redshift and the background universe is given by \cite{ISb} 
according to
\begin{eqnarray}
\label{r_m_approx}
r_m &=&    337.7 \left(\frac {M_0}{10^{12}M_{\odot}}\right)^{1/3}[F(z_{\rm coll})]^{-1}
		h^{-2/3}\,\, {\rm kpc},\\ 
r_t &=&  187.2 \left(\frac {M_0}{10^{12}M_{\odot}}\right)^{1/3}
	[F(z_{\rm coll})]^{-1} h^{-2/3}\,\, {\rm kpc},\\
r_0&=&	 6.367\left(\frac {M_0}{10^{12}M_{\odot}}\right)^{1/3}[F(z_{\rm coll})]^{-1}
		h^{-2/3}\,\, {\rm kpc},\\
T &=&	 7.843\times 10^5\displaystyle{\left(\frac{\mu}{0.59}\right)\left(\frac {M_0}{10^{12}M_{\odot}}\right)^{2/3}
	F(z_{\rm coll})h^{2/3}\,\, {\rm K},}\\
\sigma_V^2 &=&
	1.098\times10^4\left(\frac {M_0}{10^{12}M_{\odot}}\right)^{2/3}
	F(z_{\rm coll})h^{2/3}\,\, {\rm km^2\,s^{-2}},\\
\label{rho0_approx}
\rho_0&=& 	1.799\times10^4[F(z_{\rm coll})]^3\rho_{\rm crit}(z=0)\nonumber\\
	&=&3.382\times 10^{-25}[F(z_{\rm coll})]^3h^2 \,\,{\rm g\,cm^{-3}}.
\end{eqnarray}
where 
\begin{equation}
F(z_{\rm coll})\equiv\displaystyle{\left[\frac{h(z_{\rm coll})}{h}\right]^2
\frac{\Delta_{\rm c,TIS}(z_{\rm coll},\lambda_0)}{\Delta_{\rm c,TIS}(\lambda_0=0)}}
=\displaystyle{\left[\frac{\Omega_0}{\Omega(z_{\rm coll})}\frac{\Delta_{\rm c,SUS}}{18\pi^2}\right]^{1/3}}
(1+z_{\rm coll}). 
\end{equation}
For the EdS case, $F=(1+z_{\rm coll})$, while for an open, matter-dominated universe
and a flat universe with a cosmological constant, $F\rightarrow\Omega_0^{1/3}(1+z_{\rm coll})$ 
at early times [i.e. $x\rightarrow0$].
Here $\mu$ is the mean molecular weight, where
$\mu= 0.59\, (1.22)$ for an ionized (neutral) gas of H and He with 
$[He]/[H]=0.08$ by number. $\Delta_{\rm c,SUS}$ is well-approximated by $\Delta_{\rm c,SUS}=18\pi^2+c_1x-c_2x^2$,
where $x\equiv\Omega(z_{\rm coll})-1$, $\Omega(z_{\rm coll})=\Omega_0(1+z)^3\left[h/h(z)\right]^2$, where
$[h(z)/h]^2=\Omega_0(1+z)^3+\lambda_0$ (or $\Omega_0(1+z)^3+(1-\Omega_0)(1+z)^2$) and 
$c_1=82\, (60)$ and $c_2=39\,(32)$ for the
flat (open) cases, $\Omega_0+\lambda_0=1$ ($\Omega_0<1,\lambda_0=0$), respectively 
\cite{BN}.

\subsection{TIS Model vs. Numerical CDM Simulations}
\label{simul_tis_subsect}
\begin{figure}[t!]
\centering
\includegraphics[height=2.7in]{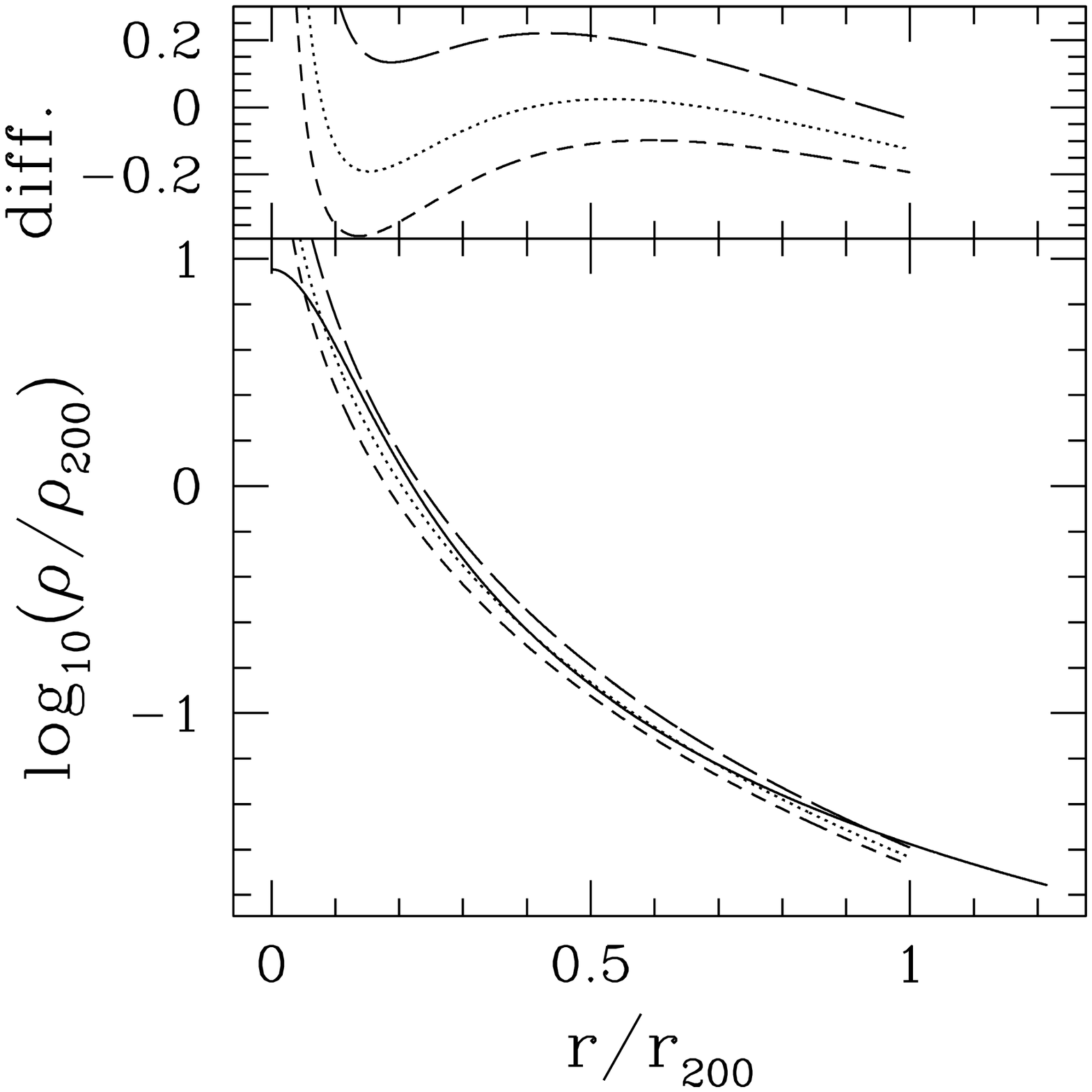}
\includegraphics[height=2.7in]{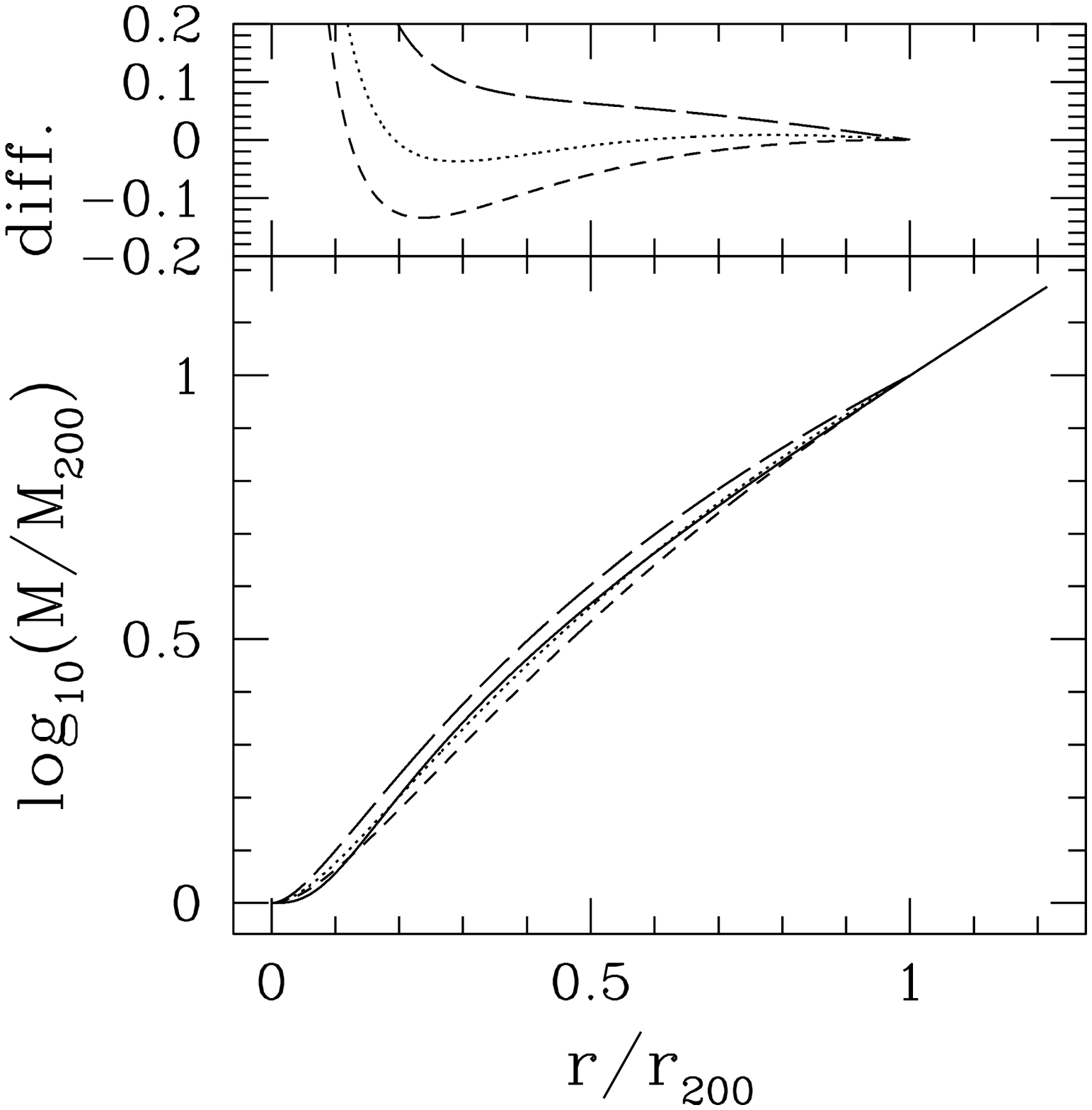}
\caption{Profiles of density (left) and integrated mass (right), for 
TIS (solid) and NFW with $c_{\rm NFW}=4$ (short-dashed), $5$ (dotted)
and $7$ (long-dashed) with same $(r_{200},M_{200})$.    \label{profileNFWTIS}} 
\end{figure}

The TIS model reproduces many of the average properties of the halos 
in numerical CDM simulations quite well, suggesting that it is a useful 
approximation for the halos which result from more realistic initial 
conditions: 

(1) The TIS mass profile agrees well with the fit to
N-body simulations by \cite{NFW96} (``NFW'') (i.e. fractional
deviation of $\sim20\%$ or less) at all radii outside of a few TIS core radii
(i.e. outside King radius or so), for NFW concentration parameters
$4\leq c_{\rm NFW}\leq7$ (Fig.~\ref{profileNFWTIS}). The flat density 
core of the TIS halo differs from
the singular cusp of the NFW profile at small radii, but this involves 
only a small fraction of the halo mass, thus not affecting their good agreement
outside the core.  
As a result, the TIS central density $\rho_0$ can be used to characterize the  
core density of cosmological halos, even if the latter have singular profiles
like that of NFW, as long as we interpret $\rho_0$, in that case,
as an average over the innermost region.

\begin{figure}[t!]
	\centering
    \includegraphics[width=2.7in]{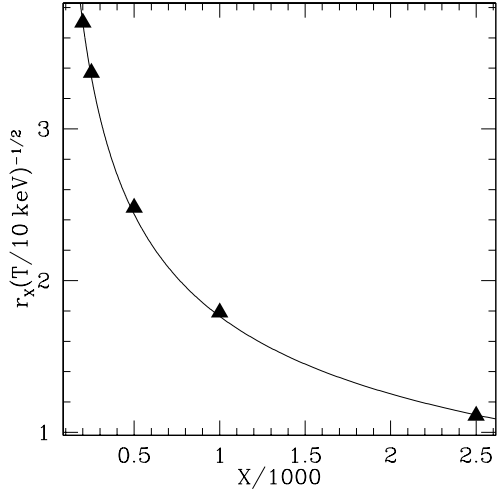}
    \caption{ Cluster radius-temperature virial relation
for CDM simulation results (at $z=0$) as fit by EMN (triangles) and as predicted by TIS 
(solid curve).
    \label{EMN}}
\end{figure}

\indent (2) The TIS halo model predicts the internal structure of X-ray
clusters found by gas-dynamical/N-body simulations of cluster 
formation in the CDM model.  Our TIS model
predictions, for example, agree astonishingly well with the
mass-temperature and  radius-temperature virial relations and integrated 
mass profiles derived empirically from the simulations of cluster formation
by \cite{EMN,ME} (EMN).
Apparently, these simulation results are not sensitive
to the discrepancy between our
prediction of a small, finite density core and
the N-body predictions of a density cusp for 
clusters in CDM. Let $X$ be the average overdensity inside radius $r$
(in units of the cosmic mean density) $X\equiv{\langle\rho(r)\rangle}/{\overline{\rho}}$. The
radius-temperature virial relation is defined as 
$r_X\equiv r_{10}(X)\displaystyle{\left( T/{10\, {\rm keV}}\right)^{1/2}}\,{\rm Mpc}$,
and the mass-temperature virial relation by
$M_X\equiv M_{10}(X)\displaystyle{\left( T/{10\, {\rm keV}}\right)^{1/2}}
	\,h^{-1} 10^{15}\,M_\odot$.
A comparison between our predictions of 
$r_{10}(X)$ and the results of EMN is given in Fig.~\ref{EMN}. 
EMN obtain $M_{10}(500)=1.11\pm 0.16$ and $M_{10}(200)=1.45$, while our TIS solution
yields $M_{10}(500)=1.11$ and $M_{10}(200)=1.55$.

(3) The TIS halo model also successfully reproduces  the mass - velocity 
dispersion relation for clusters in CDM N-body simulations and 
its dependence on redshift for different background cosmologies. 
N-body simulation of the Hubble volume 
[(1000 Mpc)$^3$] by the Virgo Consortium \cite{Eetal02} 
yields the following empirical relation:
\begin{equation}
\sigma_V=(1080\pm65)\left[
{h(z)M_{200}}/{10^{15}M_\odot}\right]^{0.33}{\rm km/s},
\end{equation}
where $M_{200}$ is the mass within a sphere with average
density 200 times the cosmic mean density, and 
$h(z)=h_0\sqrt{\Omega_0(1+z)^3+\lambda_0}$
is the redshift-dependent Hubble constant (assuming a flat background universe). 
Our TIS model predicts:
\begin{equation}
\sigma_V=1103\left[
{h(z)M_{200}}/{10^{15}M_\odot}\right]^{1/3}{\rm km/s},
\end{equation}
in excellent agreement with simulations.

\begin{figure}[t!]
	\centering
    \includegraphics[width=2.7in]{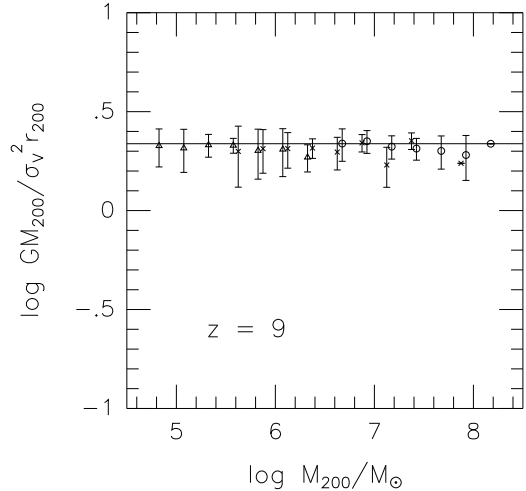}
    \caption{Virial relation $GM_{200}/(\sigma_V^2r_{200})$
predicted by the TIS model (horizontal line) vs. mass for halos from N-body 
simulations \cite{S01} [with $1\sigma$ error bars]. 
    \label{vir_rel_fig}}
\end{figure}

(4) The TIS model successfully predicts the average virial ratio,
$K/|W|$, of halos in CDM simulations.  An equivalent TIS
quantity, $GM_{200}/(r_{200}\sigma^2_V) = 2.176$, is 
plotted for dwarf galaxy minihalos at $z=9$ in Fig.~\ref{vir_rel_fig}, for
simulations described in \cite{S01} and \cite{ISMS03}, 
showing good agreement between TIS and N-body halo results. A similar plot,
but of  $K/|W|$ for such halos, was shown by \cite{JKH01} based 
upon N-body simulations, in which the average $K/|W|$ is close to 0.7, as 
predicted by the TIS model (Table 1). Those authors were apparently unaware 
of this TIS prediction since they compared their results with the SUS value
of $K/|W|$, $0.5$, and interpreted the discrepancy incorrectly as an indication that
their halos were not in equilibrium.  

\subsection{TIS Model vs. Observed Halos}
\label{observ_tis_subsect}
\begin{figure}[t!]
	\centering
    \includegraphics[width=3.7in]{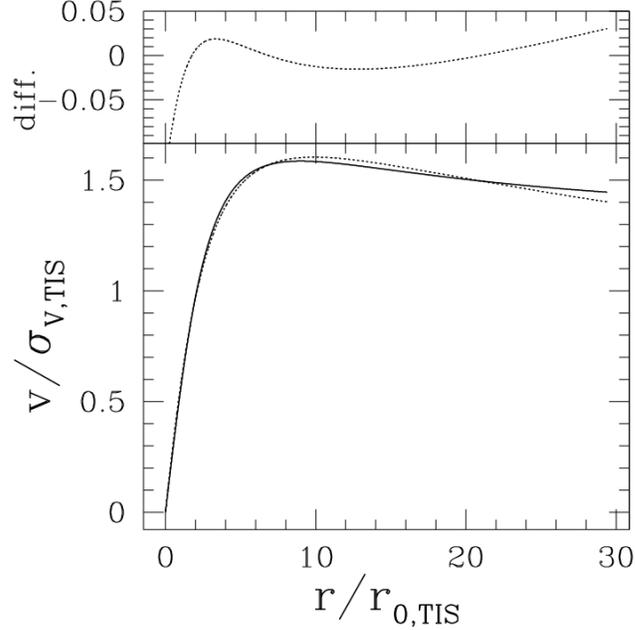}
    \caption{Rotation Curve Fit. 
Solid line = Best fit TIS; Dashed line = Burkert profile.
    \label{burk_fit}}
\end{figure}

(1) The TIS profile matches the mass profiles of dark-matter-dominated dwarf 
galaxies deduced from their observed rotation curves \cite{ISa}. 
Equation~(\ref{tisfit}) can be integrated to yield an analytical
fitting formula for the TIS rotation curve, as well, given by
\begin{equation}
\label{fit_v}
\displaystyle{\frac{v(r)}{\sigma_V}
=\left\{A-B
+\frac{1}{\zeta}\left[bB\tan^{-1}\left(\frac{\zeta}{b}\right)\right.\right.}\nonumber\\
\displaystyle{\left.\left.-aA\tan^{-1}\left(\frac{\zeta}{a}\right)\right]\right\}^{1/2}},
\end{equation}
where $\zeta=r/r_0$ and $v(r)=(GM[\leq r]/r)^{1/2}$. The observed rotation curves of
dwarf galaxies are well fit by a density profile
with a finite density core given by \cite{B}:
\begin{equation}
\rho(r)=\frac
{\rho_{0,B}}{(r/r_c+1)(r^2/r_c^2+1)}
\end{equation}
for which the rotation curve is given by
\begin{equation}
\label{v_B}
\displaystyle{\frac{v_{\rm B}(r)}{v_{\rm *,B}}}=
	\displaystyle{\left\{\frac{\ln\left[(\zeta_B+1)^2
	(\zeta_B^2+1)\right]
	-2\tan^{-1}(\zeta_B)}{\zeta_B}\right\}^{1/2}},
\end{equation} 
where $\zeta_B\equiv r/r_{\rm 0,B}$ and 
$v_{\rm *,B}\equiv(\pi G\rho_{\rm 0,B}r_{\rm 0,B}^2)^{1/2}$. 
The TIS model gives a nearly perfect fit to this profile 
(Fig.~\ref{burk_fit}), with best fit parameters
$\displaystyle{{\rho_{0,B}}/{\rho_{0,{\rm TIS}}}=1.216,\,
{r_{c}}/{r_{0,{\rm TIS}}}=3.134}$.
This best-fit TIS profile correctly predicts $v_{\rm max}$, the maximum rotation
velocity, and the radius, $r_{\rm max}$, at which it occurs in the Burkert profile:
 $\displaystyle{{r_{\rm max,B}}/{r_{\rm max,TIS}}
	=1.13,\,}$ and  $\displaystyle{{v_{\rm max,B}}/{v_{\rm max,TIS}}=1.01}$.
(i.e. excellent agreement).

\begin{figure}[t!]
	\centering
    \includegraphics[width=3.5in]{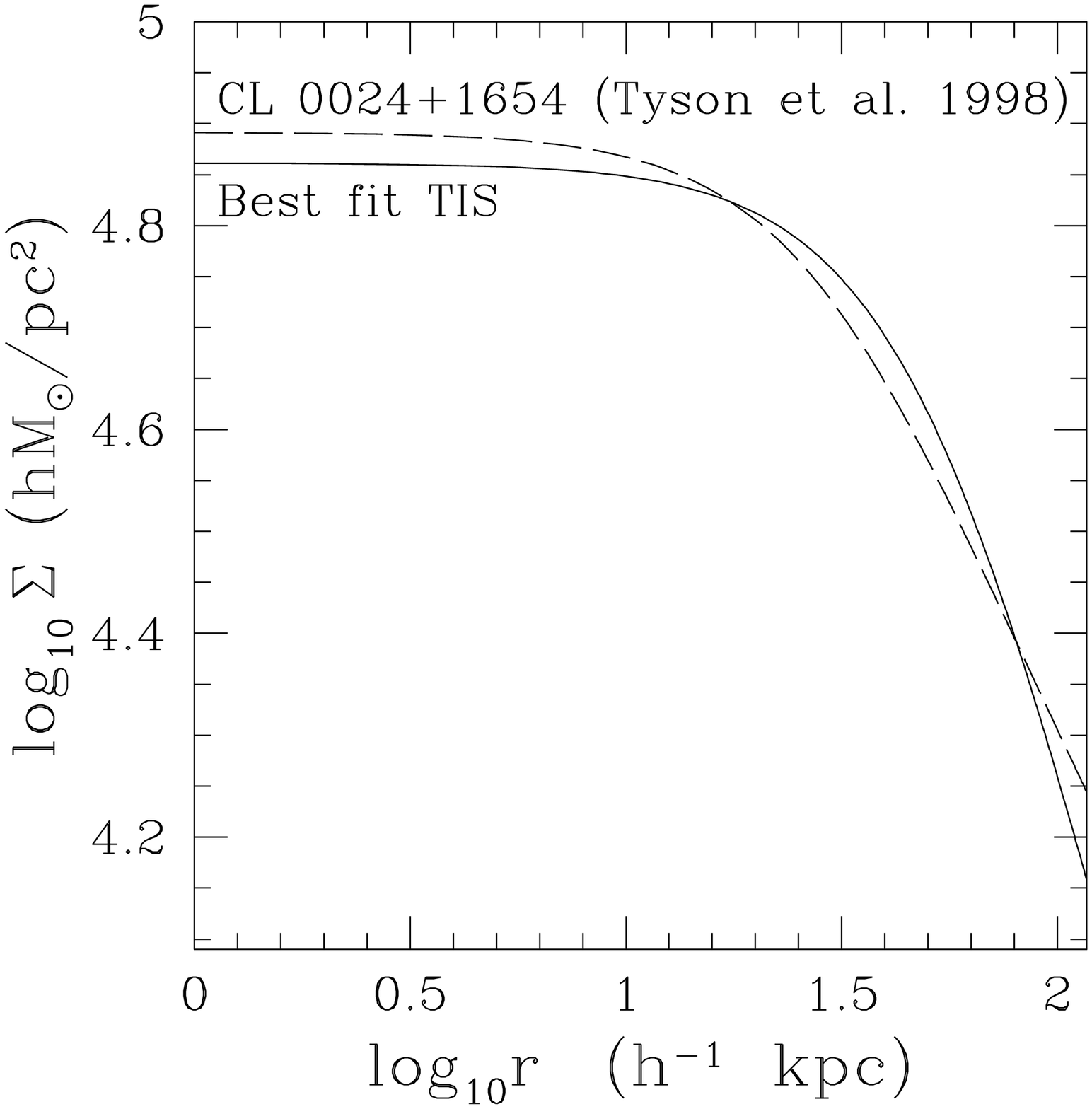}
    \caption{Projected surface density of cluster CL 0024+1654 inferred from lensing 
measurements, together with the best-fit TIS model.}
    \label{CL00_fig}
\end{figure}

{(2) The TIS halo model can explain the mass profile with a flat density core 
measured by \cite{TKD98} for cluster CL 0024+1654
at $z=0.39$, using the strong gravitational lensing of background galaxies by the 
cluster to infer the cluster mass distribution \cite{SI00} 
(see, however \cite{CMKS02} for an alternative view on the structure of this cluster).}
The TIS model not only provides a good fit to the projected 
surface mass density distribution of this cluster within the arcs 
(Fig.~\ref{CL00_fig}), but also predicts the overall 
mass, and a cluster velocity dispersion in close agreement with the value 
$\sigma_v=1150$ km/s measured by \cite{DSPBCEO}.

\subsection{Making Tracks on the Cosmic Virial Plane}
\label{virial_plane_subsect}
The TIS model yields ($\rho_0,\sigma_{\rm V},r_t,r_0$) uniquely as
functions of ($M,z_{\rm coll}$). This defines a ``cosmic virial plane" in 
($\rho_0,r_0,\sigma_{\rm V}$)-space and
determines halo size, mass, and collapse redshift for each point on the plane. 
In hierarchical clustering models like CDM, $M$ is statistically
correlated with $z_{\rm coll}$. This determines the distribution of points on 
the cosmic virial plane.
 We can combine the TIS model with the Press-Schechter (PS) approximation for
$z_{\rm coll}(M)$ -- typical collapse epoch for halo of mass $M$ -- to
predict correlations of observed halo properties.

\begin{figure}[t!]
\centering
\includegraphics[height=3.65in]{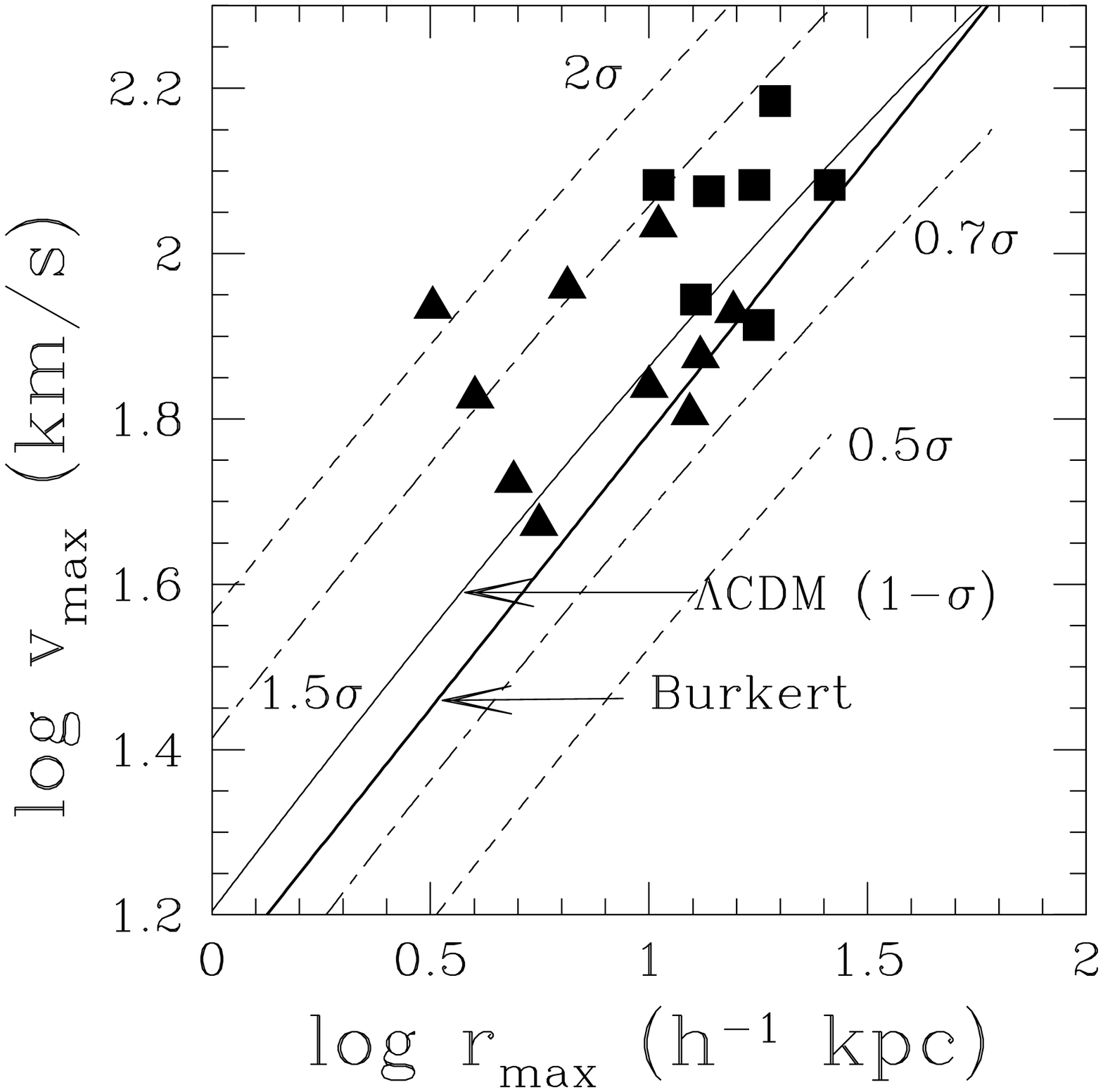}
\caption{$v_{\rm max}$--$r_{\rm max}$ correlation predicted by (TIS+PS) model 
for $\Lambda$CDM [COBE normalized, $\Omega_0=1-\lambda_0=0.3, h=0.65$; no tilt,
i.e. primordial power-spectrum index $n_p=1$),
for halos formed from $\nu-\sigma$ fluctuations, as labelled with $\nu$-values.
All curves are (TIS+PS) results, except curve labelled
``Burkert'' is a fit to data \cite{B}.
Observed dwarf galaxies (triangles) and LSB galaxies (squares) from 
\cite{KKBP}.
}
\label{vmax_rmax}
\end{figure}

\begin{figure}[t!]
\centering
\includegraphics[height=3.65in]{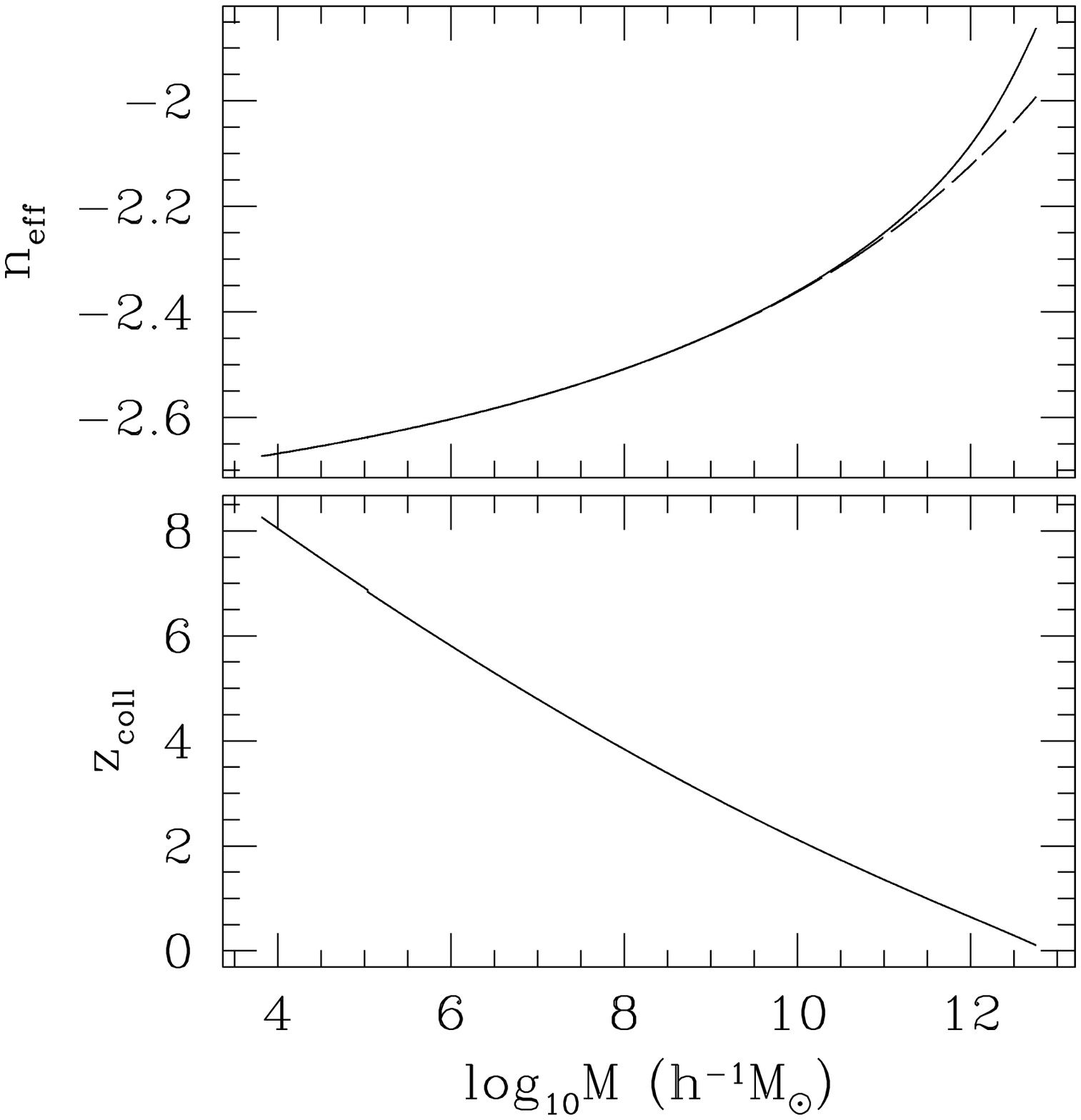}
\caption{ (a)(upper)
	Effective logarithmic slope, $n_{\rm eff}$, of the density 
	fluctuation power spectrum versus halo mass $M=M_\star$ (i.e. for
	$1$-$\sigma$ fluctuations) for $\Lambda$CDM, based upon $y_F$ (solid curve)
	and the approximation which uses $y_\sigma$, instead (dashed).    
	(b) (lower) Typical collapse redshift $z_{\rm coll}$ 
        for halos of mass $M_\star$ in $\Lambda$CDM.
\label{n_eff_lam}}
\end{figure}

According to the PS approximation, the fraction 
of matter in the universe which is condensed into objects of mass 
$\geq M$ at a given epoch is
$f_{\rm coll}(\geq M)=
 	{\rm erfc}\left(\nu/2^{1/2}\right)$,
where $\nu\equiv{\delta_{\rm crit}}/\sigma(M)$, 
$\sigma(M)$ is the standard deviation of the density fluctuations  
at that epoch, according to linear theory, when filtered on mass
scale $M$, and  
$\delta_{\rm crit}$ is the fractional overdensity of 
a top-hat perturbation when this linear theory is extrapolated to the 
time of infinite collapse in the exact nonlinear solution. The ``typical''
collapse epoch for a given mass is that for which $\sigma(M)=\delta_{\rm crit}$
(i.e. $\nu=1$). For a given $z_{\rm coll}$, this defines a typical mass scale:
$M_\star\equiv M(\nu=1)$.

If we approximate the power-spectrum of density fluctuations at high
redshift (e.g. just after recombination) as a power-law in wavenumber $k$,
$P(k)\propto k^{n}$, 
and define a mass $M\propto k^{-3}$, then
 $F\propto M^{-(3+n)/6}$ if 
$n=n_{\rm eff}\equiv-3(2y_F+1)$, where 
$y_F\equiv(d\ln F/d\ln M)_{\rm exact}$ at the relevant
mass scale. For all masses in the EdS case and for masses which collapse
early in the low-density,   matter-dominated and flat, universes, $y_F$ reduces 
to $y_\sigma\equiv(d\ln \sigma/d\ln M)_{\rm exact}$, and  
$(1+z_{\rm coll})\propto\sigma(M)\propto M^{-(3+n)/6}$, where $\sigma(M)$ is
evaluated at the same cosmic time for all masses.
The dependence of $n_{\rm eff}$ and $z_{\rm coll}$ on $M$
for $1$-$\sigma$ fluctuations is shown in Figure~\ref{n_eff_lam} for $\Lambda$CDM,
along with 
the approximate $n_{\rm eff}$ which results if $y_F$ is replaced by $y_\sigma$, 
which shows that the latter is a very good approximation for all masses 
$M<10^{12}M_\odot h^{-1}$.

{(1) The combined (TIS+PS) model explains the observed correlation of 
the maximum circular velocity $v_{\rm max}$ and its location $r_{\rm max}$
in the rotation curves of dwarf spiral and LSB galaxies}, with preference for the 
currently favored
$\Lambda$CDM model with no tilt of the power spectrum of the primordial density 
fluctuations \cite{ISa} (Fig.~\ref{vmax_rmax}). According to \cite{MB00},
the observed $v_{\rm max}-r_{\rm max}$ correlation can be expressed as follows:
\begin{equation}
v_{\rm max}=9.81(r_{\rm max}/1\,{\rm kpc})^{2/3}\rm km s^{-1}.
\label{vm-rm_fit}
\end{equation}
Our results for the $\Lambda$CDM case indicate that the galaxies which
make up the $v_{\rm max}-r_{\rm max}$ data points in Figure~\ref{vmax_rmax} 
collapsed at redshifts $1< z_{\rm coll}<6$
with masses in the range 
$8\times10^9< M_0/(M_\odot h^{-1})< 3\times10^{11}$.
Hence, the precollapse fluctuation growth rate is approximately 
EdS, and we can let $\Omega(z_{\rm coll})=1$. In that case, 
$(1+ z_{\rm coll})\propto\sigma(M)\propto M^{-(3+n)/6}$,
$r_{\rm max}\propto M^{(5+n)/6}\Omega_0^{-1/3}$ and 
$v_{\rm max}\propto M^{(1-n)/12}\Omega_0^{1/6}$, which combine to yield  
\begin{equation}
\label{vm_rm_scaling_analyt}
\displaystyle{v_{\rm max}=
	v_{\rm max,*}\left({r_{\rm max}}/{r_{\rm max,*}}\right)^{(1-n)/[2(5+n)]}},
\end{equation}
where $v_{\rm max,*}$ and $r_{\rm max,*}$ are for a 1-$\sigma$ 
fluctuation of fiducial mass $M_*$, 
with $(1+z_{\rm coll})=(1+z_{\rm rec})\sigma(M_*,z_{\rm rec})/\delta_{\rm crit}$,
where $\delta_{\rm crit}=1.6865$ and $\sigma(M_*,z_{\rm rec})$ is the value of
$\sigma(M_*)$ evaluated at the epoch of recombination $z_{\rm rec}$ [i.e. 
early enough that $(1+z)\sigma$ is independent of $z$].
Over the relevant mass range $M=10^{10\pm1}{ h^{-1}M_\odot}$, 
$n_{\rm eff}\approx-2.4\pm0.1$ 
for our $\Lambda$CDM case. For $M_*=10^{10}h^{-1}M_\odot$, our 
COBE-normalized, flat $\Lambda$CDM case ($\Omega_0=0.3,h=0.65$)
yields $(1+z_{\rm rec})\sigma(M_*,z_{\rm rec})=5.563$,
so $(1+z_{\rm coll})=3.30$, $v_{\rm max,*}=53.2\,{\rm km\,s^{-1}}$ and 
$r_{\rm max,*}=5.59\,h^{-1}{\rm kpc}$. With these values and 
$n_{\rm eff}=-2.4$, equation~(\ref{vm_rm_scaling_analyt}) yields the
TIS model analytical prediction 
\begin{equation}
v_{\rm max}=(13.0\, {\rm km\,s^{-1}})(r_{\rm max}/1\,{\rm kpc})^{0.65},
\end{equation}
remarkably close to the observed relation in equation~(\ref{vm-rm_fit}).

\begin{figure}[t!]
\centering
    \includegraphics[width=2.7in]{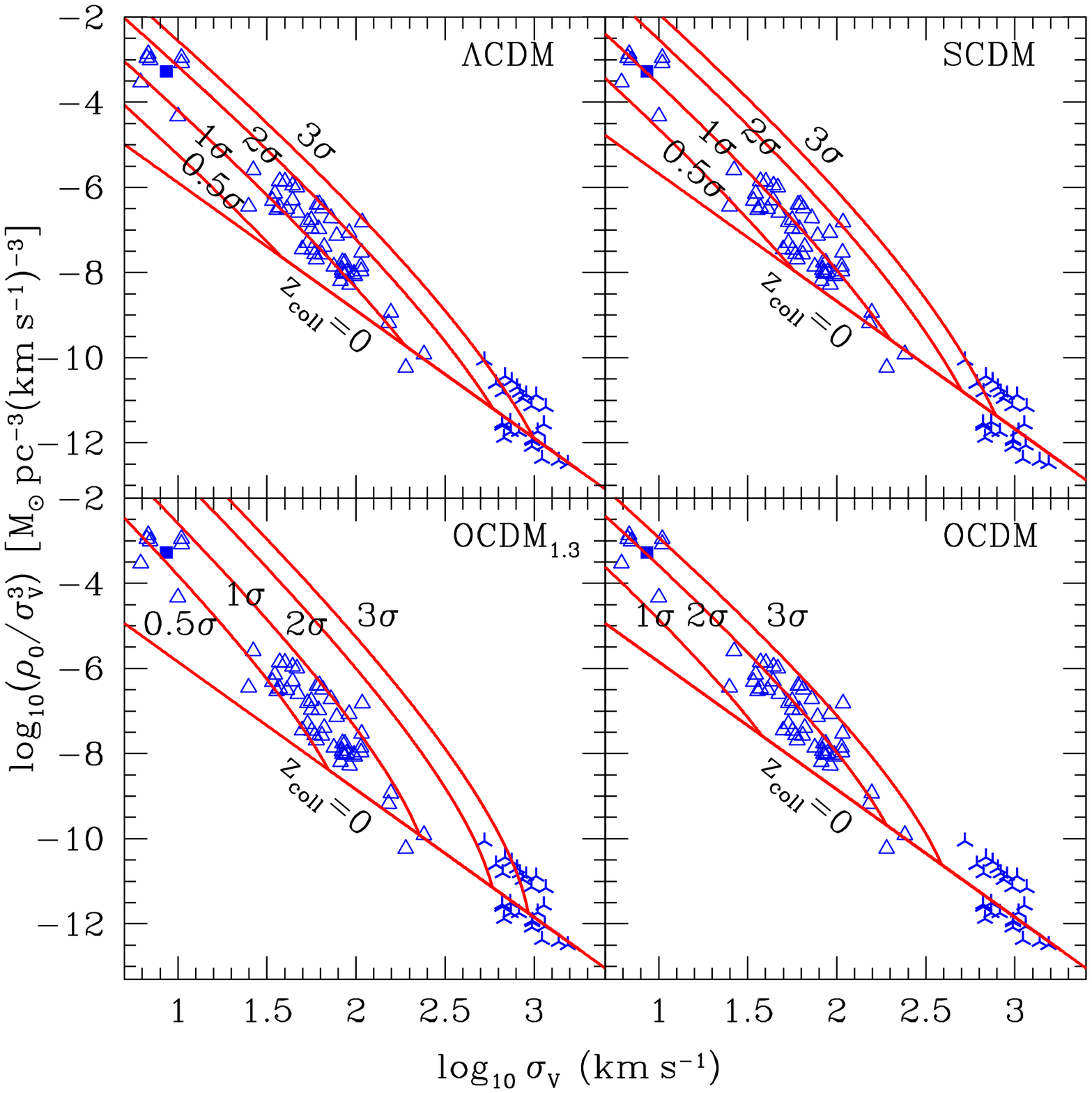}
    \includegraphics[width=2.7in]{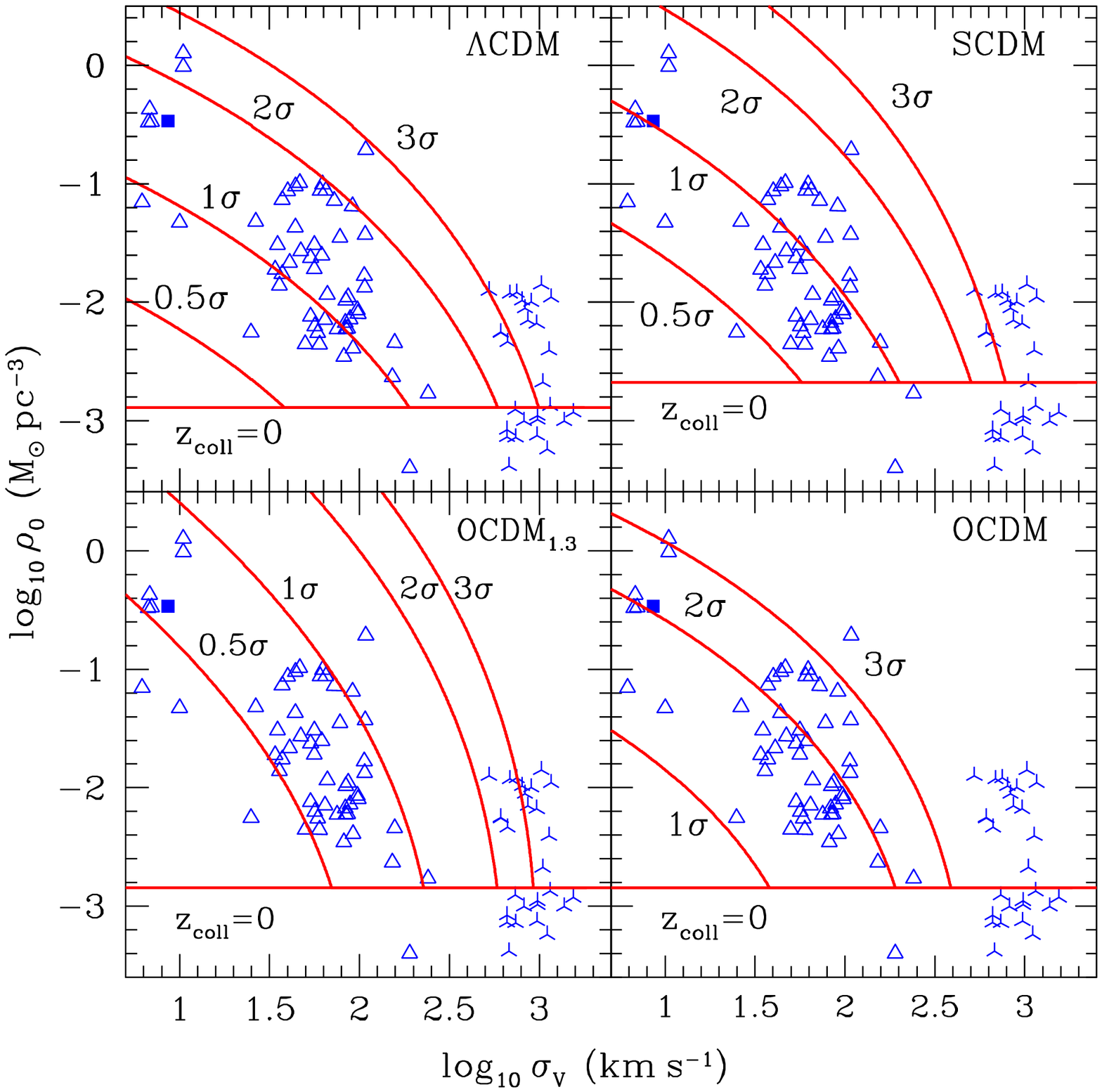}
\caption{
(a) (left) Maximum phase-space density $Q=\rho_0/\sigma_V^3$
versus velocity dispersion $\sigma_V$ for halos observed today,
as predicted for various CDM universes by the
TIS model combined with the Press-Schechter
approximation (TIS + PS) (solid curves) for halos formed from
$\nu$-$\sigma$ fluctuations, as labelled with the values of $\nu$, for 
$\nu =$
0.5, 1, 2, and 3.  In each panel, lines representing halos of different
mass which collapse at the same redshift are shown for the case 
$z_{\rm coll} = 0$, 
as labelled.  Each panel represents different assumptions for the 
background universe and primordial density fluctuations, 
as labelled: COBE-normalized
$\Lambda$CDM ($\lambda_0=0.7$, $\Omega_0=0.3$) (upper left), 
cluster-normalized SCDM
($\Omega_0=1$) (upper right), and COBE-normalized OCDM ($\Omega_0=0.3$)
(lower panels), all assuming
$h=0.7$ and primordial power spectrum index $n_p =$1 (i.e. untilted),
except for OCDM$_{1.3}$, for which $n_p =$1.3.
Data points represent observed
galaxies and clusters, taken from the following sources: 
(1) 49 late-type spirals of type Sc-Im and 7 dSph galaxies from 
\cite{KF96,KF01} (open triangles); 
(2) Local Group dSph Leo I from \cite{MOVK} (filled square);
(3) 28 nearby clusters, $\sigma_V$ from \cite{G} and \cite{JF}, 
and $\rho_0$ from \cite{MME} (crosses).
(b) (right) Same as (a), except for 
$\rho_0$ vs. $\sigma_V$.  \label{phase_rho}}
\end{figure}

(2) The TIS+PS model also predicts the correlations of 
central mass and phase-space densities, $\rho_0$ and 
$Q\equiv\rho_0/\sigma_V^3$, 
of dark matter halos with their velocity dispersions $\sigma_V$,
with data for low-redshift dwarf spheroidals 
to X-ray clusters again most consistent with $\Lambda$CDM with no tilt
\cite{SI02} (Fig.~\ref{phase_rho}). 
\footnote{There have also been claims that $\rho_0=$const for all cosmological halos,
independent of their mass, as expected for certain types of SIDM
\cite{FDCHA,KKT}.
This claim, however, is not supported by most current
data (Fig.~\ref{phase_rho}).}
A fully analytical approximation for these correlations was also derived in 
\cite{SI02} by combining the TIS+PS model with the power-law model for $P(k)$ 
described above, which yields $\rho_0\propto M^{-(n+3)/2}$, $\sigma_V\propto M^{(1-n)/12}$,
and $Q\propto M^{-(n+7)/4}$. These combine to give
\begin{equation}
\label{scaling_Q}
Q=Q_{\star}(\sigma_V/\sigma_{V,\star})^{\alpha-3},
\end{equation} 
and
\begin{equation}
\label{scaling_rho0}
\rho_0=\rho_{0,\star}(\sigma_V/\sigma_{V,\star})^{\alpha},
\end{equation}
where $\alpha\equiv 6(n+3)/(n-1)$,  
except that $\alpha=0$ for $M>M_\star(z=0)$ (e.g.for untilted $\Lambda$CDM
$M_\star(z=0)\approx 10^{13}M_{\odot}$), 
for which $z_{\rm coll}=0$ is assumed.

\begin{figure}[t!]
	\centering
    \includegraphics[width=2.7in]{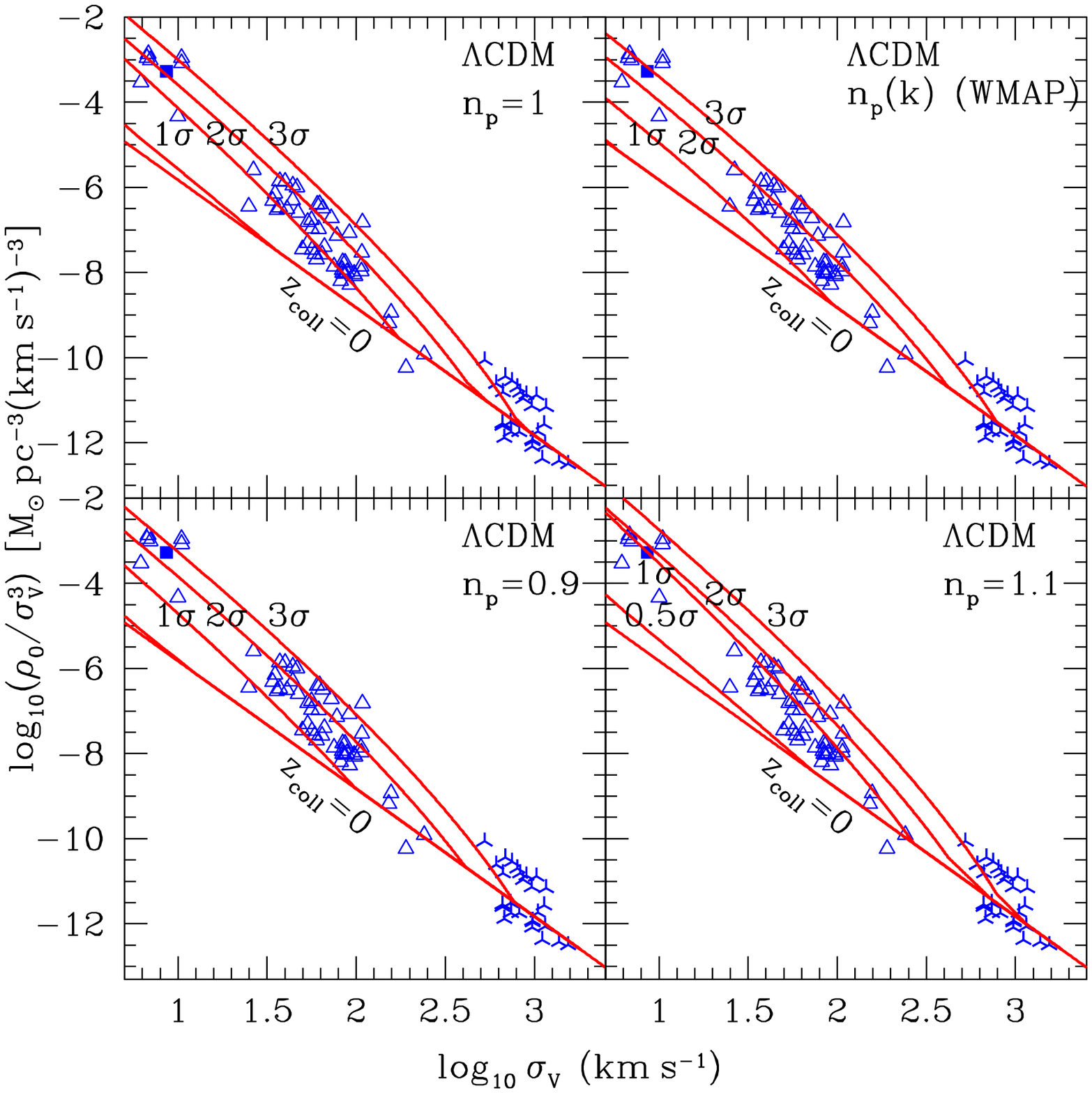}
    \includegraphics[width=2.7in]{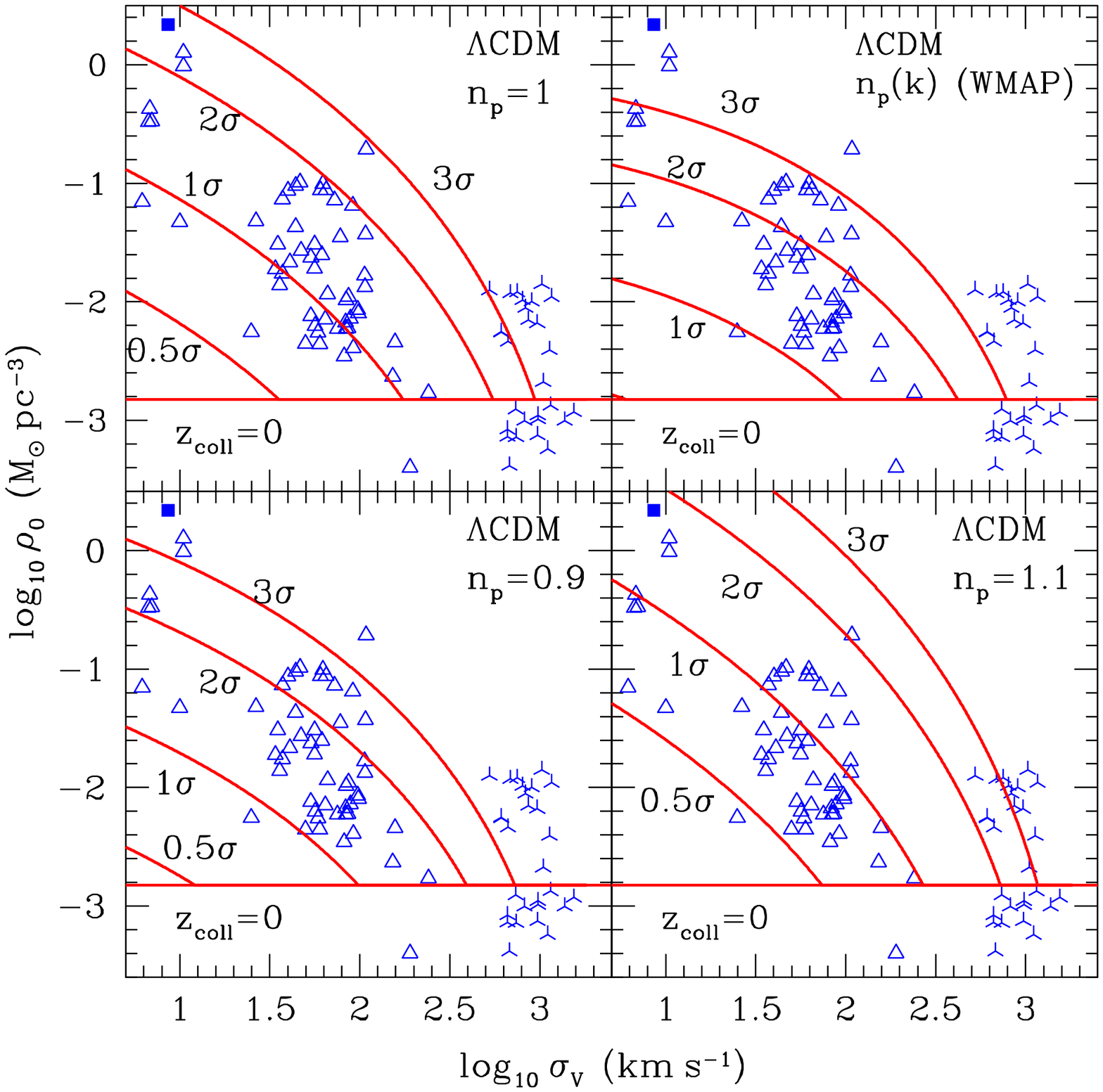}
    \label{lcdm_4panel}
\caption{Same as Fig.~\ref{phase_rho}, but for $\Lambda$CDM variants: (top left) 
untilted $\Lambda$CDM ($n_p=1$), (top right) $\Lambda$CDM with running power-spectrum 
index $n_p=n_p(k)$ according to the best WMAP fit, (bottom left) $\Lambda$CDM with $n_p=0.9$,
and (bottom right) $\Lambda$CDM with $n_p=1.1$.\label{phase_rho_lcdm}}
\end{figure}
Similarly, we can compare different flavors of the $\Lambda$CDM model, with no tilt
($n_p=1$), slight tilts ($n_p=0.9$ or 1.1) and with the ``running'' power-spectrum index
suggested by the first year Wilkinson Microwave Anisotropy Probe results 
\cite{Setal03} $\left[n_p(k)=0.93-0.015\ln\left(\frac{k}{0.05\,\rm Mpc^{-1}}\right)\right]$  
(Fig.~\ref{phase_rho_lcdm}). In this case the untilted $\Lambda$CDM model is
again moderately preferred.

\section{Gravitational lensing by CDM halos: Singular vs. nonsingular profiles}
\label{lensing_sect}

\subsection{INTRODUCTION}

The gravitational lensing of distant sources has in recent years become
one of the most powerful tools in observational cosmology\cite{BS01,So01}. 
Since the
effects of gravitational lensing depend upon the redshift of the source,
the cosmological background, and the distribution of matter in the universe,
they can be used to constrain the cosmological parameters and the 
primordial power spectrum of density fluctuations from which structure
originates. In addition, many of the effects produced by gravitational lenses,
such as image multiplicity, separations, and time delay,
depend strongly upon the matter distribution inside the lenses.
Hence, measurements of these effects can
provide a unique tool for probing the matter
distribution inside collapsed objects like galaxies and clusters,
providing the only direct measurement of their dark matter content, 
and constraining the theory of their 
formation and evolution.

Until recently, the internal structure of halos adopted in lensing
studies was generally some gravitational equilibrium 
distribution, either singular or
nonsingular (e.g., King model, singular isothermal sphere, 
pseudo-isothermal sphere), 
not necessarily motivated directly by the
theory of cosmological halo formation 
\cite
{Betal91,HK87,J91,J92,K95,NW88,PMM98,Petal01a,Petal01b,RM01,TOG84,Yetal80}. 
As the theory
of halo formation in the CDM model has advanced in recent years, however,
the halo mass profiles adopted for lensing models have been refined to
reflect this theory.
Numerical simulations of large-scale
structure formation in Cold Dark Matter (CDM) universes predict that
galaxies and clusters have a singular density profile
which approaches a power law $\rho\propto r^{-n}$ at the center,
with the exponent $n$ ranging from 1 to 1.5 
\cite{CL96,FM97,FM01,FM03,Getal00,HJS99,Metal98,Metal99,NFW96,NFW97,JS00,Ketal01,Petal03,TBW97}.
These results are in apparent conflict with observations
of rotation curves of dark-matter-dominated dwarf 
galaxies and low surface brightness galaxies,
which favor a flat-density core \cite{BS99,M01,Metal98,Petal99}.
On the scale of clusters of galaxies, observations of 
strong gravitational lensing
of background galaxies by foreground clusters also favor the presence of a
finite-density core in the centers of clusters \cite{TKD98,GAV,SAN}. 

Lensing by the two kinds of halo mass profiles, singular versus flat-core,
will be different. This has led to attempts to predict the
differences expected if the halos have the
singular cusp of the Navarro-Frenk-White (NFW) or Moore profiles or else
a profile with a flat core \cite{KM01,K95,LO02,RM01,TC01,WTS01}. 
Several authors have studied the effect of lensing by halos with
a flat-density core \cite{J91,J92,K95,PMM98,Petal01a,Petal01b} 
or by NFW or Moore profiles
that have been generalized, so
that the inner slope of the density profile is arbitrary
\cite{KM01,LO02,RM01,WTS01}.
These particular density profiles are essentially mathematical
conveniences without physical motivation. There is no underlying
theoretical model in these cases that was used to predict the value of the
core radius or the departure of the inner slope of the density profile
from the value found by N-body simulations of CDM.
By contrast, the TIS model is based on a set of physical
assumptions concerning the origin, evolution, and
equilibrium structure of halos in CDM universes. Observations of gravitational
lenses have the potential to distinguish between the TIS profile
and singular ones like the NFW profile, as several observable properties of 
gravitational lenses will be strongly affected by the presence, or
absence of a central cusp in the density profile. 
One example of an
important observable that can distinguish between various density profiles
is the parity of the number of images. Lenses with nonsingular density
profiles, such as the TIS, obey the {\it odd number theorem}. The number of
images of a given source is always odd, unless the source is extended and
saddles a caustic (see \cite{SEF92},
p.~172). Lenses with singular profiles, like the singular isothermal sphere,
the NFW profile, or the Moore profile, need not obey this theorem, even for
point sources. Most observed multiple-image gravitational lenses
have either 2 or 4 images, and this may argue against 
profiles with a central core \cite{RM01}.
There are, however, other possible explanations for the 
absence of a third or fifth
image. That image tends to be very close to the optical axis, 
and might be hidden behind the lens itself. Also, it 
is usually highly demagnified, and might be too faint to be seen. 

In this section, we derive all the lensing properties of the TIS. We also
compare the TIS with three other density profiles: The Navarro-Frenk-White
(NFW) density profile, the Singular Isothermal Sphere (SIS), 
and the Schwarzschild Lens\footnote{A point mass}.
To compare the lensing properties of these various lens models, we
focus on one particular cosmological model, the currently favored
COBE-normalized $\Lambda$CDM model with $\Omega_0=0.3$, $\lambda_0=0.7$,
and $H_0=70\,\rm km\,s^{-1}Mpc^{-1}$ (this model is also cluster-normalized).

\subsection{THE DENSITY PROFILES}

\subsubsection{The Radial Density Profiles}

We will compute the lensing properties of halos with the TIS density profile,
as well-approximated by equation~(\ref{tisfit}),
and compare them with the properties derived for three comparison profiles.
The first one is the Navarro, Frenk, and White (NFW) density profile,
\begin{equation}
\label{nfwfit}
\rho(r)={\rho_{\rm NFW}^{\phantom2}\over(r/r_{\rm NFW}^{\phantom2})
(r/r_{\rm NFW}^{\phantom2}+1)^2}\,,
\end{equation}

\noindent
where $\rho_{\rm NFW}^{\phantom2}$
is a characteristic density and $r_{\rm NFW}^{\phantom2}$ is a 
characteristic radius. We will also consider the Singular Isothermal Sphere
(SIS) density profile,
\begin{equation}
\label{sisfit}
\rho(r)={\sigma_V^2\over2\pi Gr^2}\,,
\end{equation}

\noindent where $\sigma_V^{\phantom2}$ is the velocity dispersion and $G$ is
the gravitational constant. This model might not represent actual halos
very well, but it is a well-studied profile that has important
theoretical value. Finally, for completeness, we will also consider
the Schwarzschild lens,
\begin{equation}
\label{schfit}
\rho(r)=M_{\rm Sch}\delta^3(r)\,
\end{equation}

\noindent where $M_{\rm Sch}$ is the lens mass and $\delta^3$ is the
three-dimensional delta function.

In order to compare the predictions for halo lensing for different
halo density profiles, we must relate the parameters which define one profile
to those which define another. For this purpose, we shall use
the same approach as \cite{WB00}. 
The virial radius
$r_{200}$ of a halo located at redshift $z$ is defined, as usual, 
as being the radius inside which the mean density is equal to 200 times
the critical density $\rho_c(z)\equiv3H^2(z)/8\pi G$ at that redshift
[where $H(z)$ is the Hubble parameter]. The mass $M_{200}$ inside
that radius is given by
\begin{equation}
\label{m200}
M_{200}={800\pi\rho_c(z)r_{200}^3\over3}\,.
\end{equation}

\noindent
When comparing the lensing properties of different density profiles, we 
will consider halos that are located at the same redshift 
$z=z_L$ (the lens redshift) and have the same
value of $r_{200}$. By definition, these halos will also have the same
value of $M_{200}$. By stretching the terminology,
we will refer to $M_{200}$ as ``the mass of the halo.'' This point needs
to be discussed. For the Schwarzschild lens, $M_{200}$ is indeed the mass of
the halo. The SIS density profile drops as $r^{-2}$ at large $r$, and the
mass therefore diverges unless we introduce a cutoff. The halo mass will
then be equal to $M_{200}$ only if the cutoff is chosen to be $r_{200}$.
The NFW density profile drops as $r^{-3}$, hence the total mass diverges
logarithmically, and this profile also needs a cutoff.
The TIS density profile drops asymptotically as $r^{-2}$,
but the TIS model includes a cutoff. This cutoff is located a radius
$r_t\approx1.2r_{200}$, and the mass inside the cutoff is 
$M_t=1.168M_{200}$.
In any case, a rigorous definition of the halo mass would require an
unambiguous determination of the boundary between the halo and the
background matter (such determination exists only for the TIS model),
as well as dealing with the fact that the assumption of spherical symmetry
that enters in these models most likely breaks down for real halos at large
enough radii. Treating $M_{200}$ as the actual mass of the halo
is the best compromise.

For a given halo mass $M_{200}$, redshift $z$, and cosmological background
model, the density profiles for the SIS and Schwarzschild lens are
fully determined. 
For the SIS, we integrate equation~(\ref{sisfit}) between $r=0$
and $r=r_{200}$, and get
\begin{equation}
\label{m200sis}
M_{200}={2\sigma_V^2r_{200}\over G}\,.
\end{equation}

\noindent Combining equations~(\ref{m200}) and (\ref{m200sis}), we get
\begin{equation}
\label{sigmav}
\sigma_V^{\phantom2}=\left[{100\pi G^3\rho_c(z)M_{200}^2\over3}\right]^{1/6}\,.
\end{equation}

\noindent For the Schwarzschild lens, $M_{\rm Sch}$ is simply
given by $M_{200}$. 

For the NFW profile, the concentration parameter 
$c$ must be specified in addition to the parameters $M_{200}$ and 
$z$. The value of $c$ is not
completely independent of the other parameters since there is a statistical
expectation that $c$ is correlated with $M_{200}$ and $z$. However,
for any individual halo, $c$ is not known a priori. 
To determine $c$ for a given halo, we will use
the typical value expected from
the statistical correlation of $c$ with halo mass and the redshift of
observation of the halo, $z_{\rm obs}=z_L$, according to 
the formalism of \cite{ENS01}.
Once the value of $c$ is known, we can compute the parameters of the profile.
The characteristic radius of the NFW profile is given by
\begin{equation}
\label{rnfw}
r_{\rm NFW}^{\phantom2}=r_{200}/c\,,
\end{equation}

\noindent
and the characteristic density $\rho_{\rm NFW}^{\phantom2}$ is given by
\begin{equation}
\label{rhonfw}
\rho_{\rm NFW}^{\phantom2}={200c^3\rho_c(z)\over3[\ln(1+c)-c/(1+c)]}\,.
\end{equation}

The TIS halo is uniquely specified by the central density $\rho_0$ and
core radius $r_0$. 
These parameters are functions of the mass $M_{200}$ and redshift $z$,
but in this case, $z$ is not the redshift $z_L$ where the halo is
located, but rather the collapse redshift $z_{\rm coll}$ where the
halo formed, which can be larger than $z_L$.
We will assume that a TIS
halo which formed at some $z_{\rm coll}\geq z_L$ did not evolve between
$z_{\rm coll}$ and $z_L$. 
For a given $M_{200}$ and $z_{\rm coll}$, the parameters $\rho_0$ and $r_0$
are determined as follows. The central density $\rho_0$ depends
only on $z_{\rm coll}$,
\begin{equation}
\label{rho0}
\rho_0=1.8\times10^4\rho_c(z_{\rm coll})\,,
\end{equation}

\noindent while $M_t=1.167M_{200}=772.6\rho_0r_0^3$, or
\begin{equation}
\label{r0}
r_0=1.51\times10^{-3}(M_{200}/\rho_0)^{1/3}\,.
\end{equation}

\begin{figure}[t!]
\includegraphics[height=4.0in]{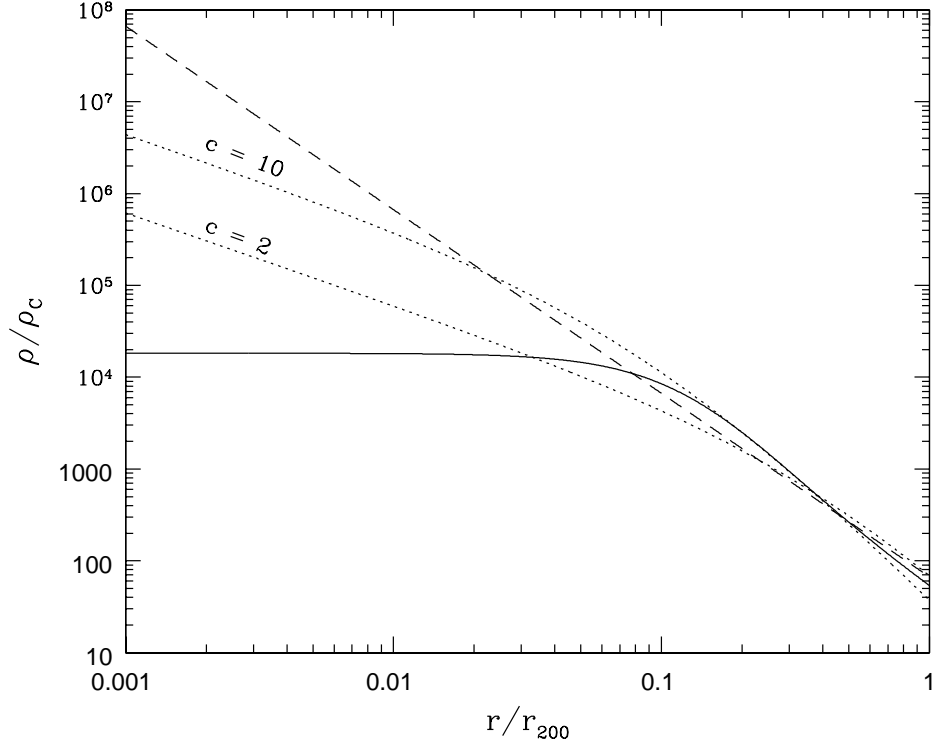}
\caption{Radial density profiles, for 4 different halos with the same
values of $r_{200}$ and $M_{200}$. Solid curve: TIS; dotted curves:
NFW profiles with concentration parameters $c=2$ and 10 (as labeled);
dashed curve: SIS.\label{radprof}}
\end{figure}

Fig.~\ref{radprof} shows a comparison
of the various profiles, for halos with the same values of
$r_{200}$ and $M_{200}$ (the TIS curve is for $z_{\rm coll}=z_L$).

The equations describing the lensing properties of the Schwarzschild lens,
the singular isothermal sphere, and the NFW profile can be
found in the literature. For brevity, we will not repeat these
equations here. We refer the reader to \cite{SEF92} for the
Schwarzschild lens (\S8.1.2) and the singular isothermal sphere (\S8.1.4),
and to \cite{CT02} and \cite{WB00} for
the NFW profile.

\subsubsection{The Interior Mass Profile}

The projected
surface density is given by
\begin{equation}
\label{proj}
\Sigma(\xi)=\int_{-\infty}^\infty\rho(r)dz\,,
\end{equation}

\noindent where $\xi$ is the projected distance from the center of the halo, 
and $z=(r^2-\xi^2)^{1/2}$. For the TIS, we substitute 
equation~(\ref{tisfit}) in equation~(\ref{proj}), and get
\begin{equation}
\label{sigma}
\Sigma_{\rm TIS}^{\phantom2}
(\xi)=\pi\rho_0r_0^2\left[{A\over(a^2r_0^2+\xi^2)^{1/2}}-
{B\over(b^2r_0^2+\xi^2)^{1/2}}\right]
\end{equation}

\noindent \cite{I00,NL97}.
For spherically symmetric lenses, one important quantity is
the interior mass $M(\xi)$ inside a cylinder of projected radius
$\xi$ centered around the center of the lens. This quantity is given by
\begin{equation}
\label{minterior}
M(\xi)=2\pi\int_0^\xi\Sigma(\xi')\xi'd\xi'\,.
\end{equation}

\noindent
we substitute equation~(\ref{sigma}) in equation~(\ref{minterior}), and get
\begin{equation}
\label{minttis}
M_{\rm TIS}^{\phantom2}(\xi)=2\pi^2\rho_0r_0^3
\left[A(a^2+\xi^2/r_0^2)^{1/2}-B(b^2+\xi^2/r_0^2)^{1/2}-Aa+Bb\right]\,.
\end{equation}

\begin{figure}[t!]
\includegraphics[height=4.0in]{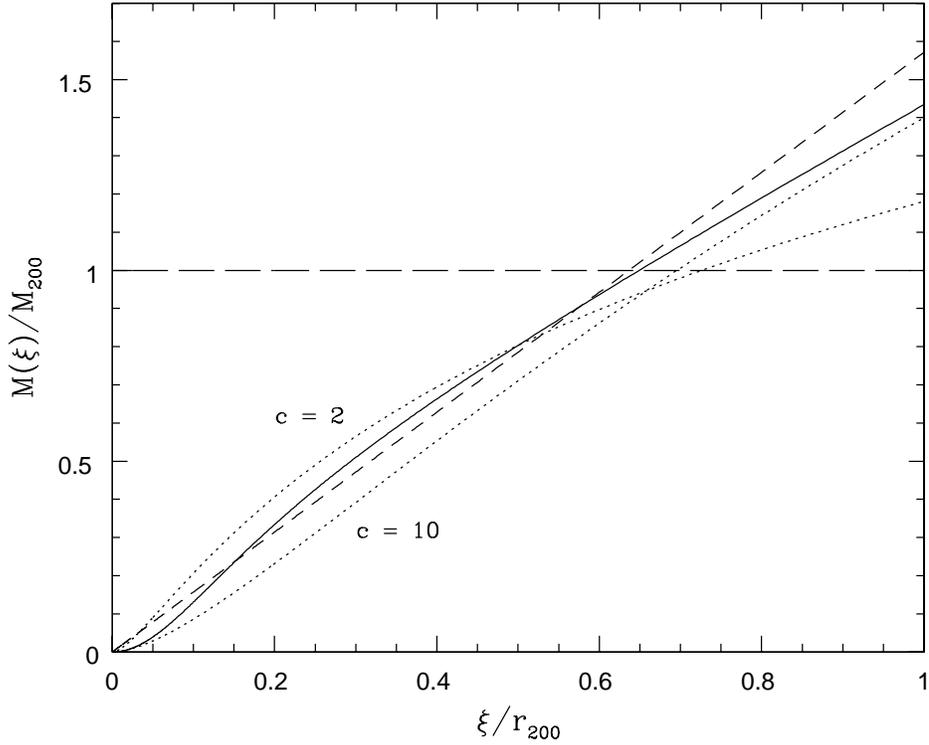}
\caption{Interior mass profiles, for 5 different halos with the same
values of $r_{200}$ and $M_{200}$. Solid curve: TIS; dotted curves:
NFW profiles with concentration parameters $c=2$ and 10 (as labeled);
short-dashed curve: SIS; long-dashed curve: Schwarzschild lens.
\label{projmass}}
\end{figure}

\noindent Fig.~\ref{projmass} 
shows a comparison of $M(\xi)$ for various halos with
the same values of $r_{200}$ and $M_{200}$.

\subsection{THE LENS EQUATION}

\begin{figure}[t!]
\includegraphics[width=5in]{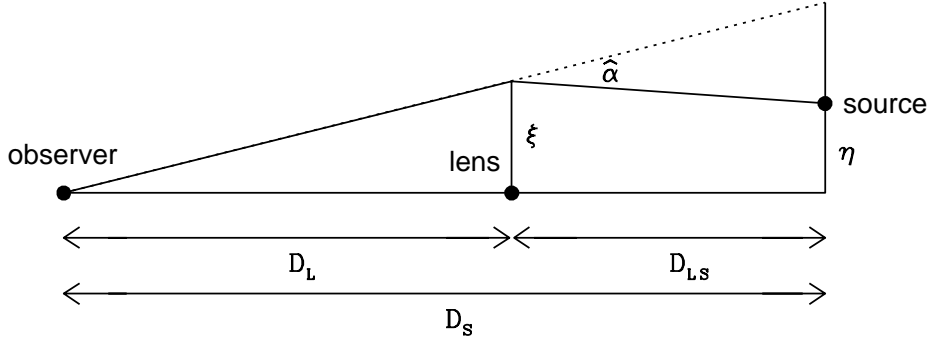}
\caption{The lensing geometry: the dots indicate the location of the 
observer, lensing galaxy, and source. $\xi$
and $\eta$ are the positions of the image and the source, respectively, and
$\hat\alpha$ is the deflection angle.
The angular diameter distances $D_L$, $D_{LS}$, and $D_S$ are also indicated.
\label{schematic}}
\end{figure}

Fig.~\ref{schematic} illustrates the lensing geometry. The quantities
$\eta$ and $\xi$ are the position of the source on the 
source plane and the image on the image plane, respectively,
$\hat\alpha$ is the deflection angle, and $D_L$, $D_S$, and $D_{LS}$ are
the angular diameter distances between observer and lens, observer and source,
and lens and source, respectively. The lens equation is
\begin{equation}
\eta={D_S\over D_L}\xi-D_{LS}\hat\alpha
\end{equation}

\noindent 
(\cite{SEF92}, eq.~[2.15b]).
Notice that since the lens is axially symmetric, we can write the
quantities $\eta$, $\xi$, and $\hat\alpha$ as scalars instead
of 2-component vectors.
We nondimensionalize
the positions and deflection angle, as follows:
\begin{eqnarray}
y&=&{D_L\eta\over D_Sr_0}\,,\\
\label{xscale}
x&=&{\xi\over r_0}\,,\\
\alpha&=&{D_LD_{LS}\hat\alpha\over D_Sr_0}\,.
\end{eqnarray}

\noindent The lens equation reduces to
\begin{equation}
\label{lensfinal}
y=x-\alpha(x)\,.
\end{equation}

\noindent
The deflection angle is given by
\begin{equation}
\label{alpha1b}
\alpha(x)={M(r_0x)\over\pi r_0^2\Sigma_{\rm crit}x}\,.
\end{equation}

\noindent (\cite{SEF92}, eq.~[8.3]), where $\Sigma_{\rm crit}$ is
the critical surface density, given by
\begin{equation}
\label{sigmacrit}
\Sigma_{\rm crit}={c^2D_S\over4\pi GD_LD_{LS}}\,,
\end{equation}

\noindent where $c$ is the speed of light. We substitute
equation~(\ref{minttis}) into equation~(\ref{alpha1b}),
and get
\begin{equation}
\label{alpha3}
\alpha_{\rm TIS}^{\phantom2}(x)={2ab\kappa_c\over(Ab-Ba)x}
\left[A(a^2+x^2)^{1/2}-B(b^2+x^2)^{1/2}-Aa+Bb\right]\,.
\end{equation}

\noindent 
where $\kappa_c$ is the central convergence of the TIS, defined by
\begin{equation}
\label{kappac}
\kappa_c\equiv{\Sigma(\xi=0)\over\Sigma_{\rm crit}}
={\pi\rho_0r_0\over\Sigma_{\rm crit}}\left({A\over a}-{B\over b}\right)
\end{equation}

\noindent (see also \cite{CT01}).

\subsection{CRITICAL CURVES AND CAUSTICS}

The lensing properties of a halo depend on its radial density
profile $\rho(r)$, the redshift $z_L$ where it is located,
the source redshift $z_S$, and the source position $\eta$. In
the remainder of this section, we will simplify the problem by fixing
the redshifts $z_L$ and $z_S$. We are primarily interested here in
lensing by cluster-scale halos. Most of
these halos are located in the redshift range $0.2\leq z_L\leq0.6$,
and none are located at redshift $z_L>0.6$.\cite{WNB99}
In what follows, we will
assume that halos are located at $z_L=0.5$. 
We will also 
assume that the source is located at redshift $z_S=3$. This is not a
a very constraining assumption, because the lensing properties vary weakly
with the source redshift for $z_S\gg1$. 

For the Schwarzschild lens, SIS, and NFW profile, the density profile
$\rho(r)$ is fully determined by the values of $M_{200}$ and $z_L$.
For the TIS, we also need to specify the collapse redshift $z_{\rm coll}$.
We will consider three particular values, $z_{\rm coll}=0.5$, 1.0, and 1.5.

\subsubsection{Solutions}

The determination of the critical curves is quite trivial for
axially symmetric lenses. 
Tangential and radial critical curves are 
defined respectively by
\begin{eqnarray}
\label{tancrit}
{m(x_t)\over x_t^2}&\equiv&{\alpha(x_t)\over x_t}=1\,,\\
\label{radcrit}
\left[{d(m/x)\over dx}\right]_{x=x_r}&\equiv
&\left[{d\alpha\over dx}\right]_{x=x_r}=1\,.
\end{eqnarray}
 
\begin{figure}[t!]
\includegraphics[width=5.5in]{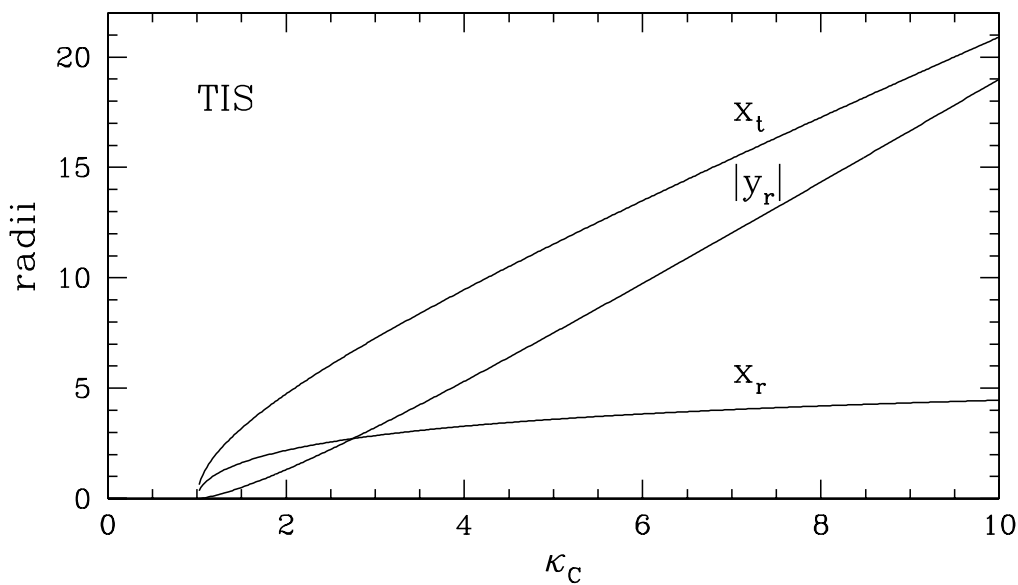}
\caption{Radii of the radial critical circle, $x_r$, tangential
critical circle, $x_t$, and radial caustic, $y_r$, versus $\kappa_c$,
for the TIS.\label{crit2}}
\end{figure}

\begin{figure}[t!]
\includegraphics[width=5.5in]{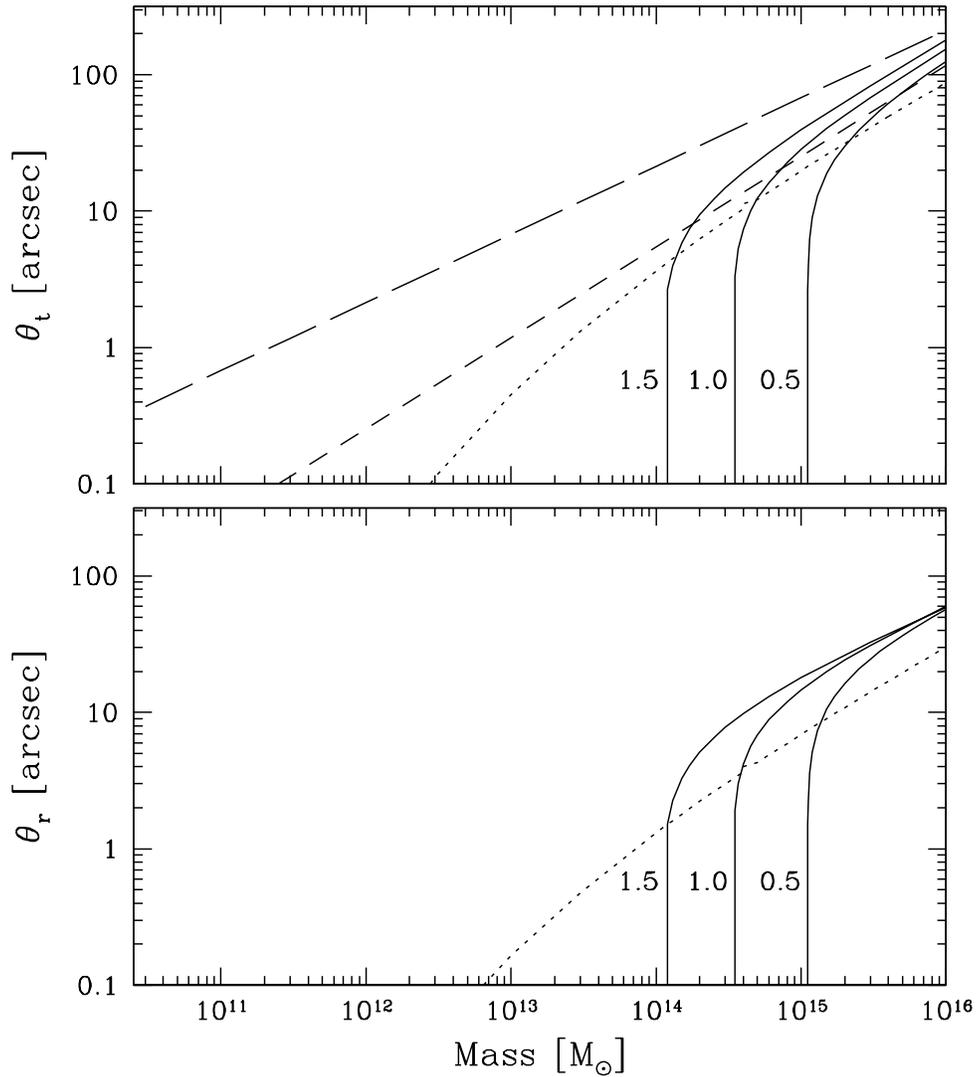}
\caption{Angular radius of the tangential critical curve (top) and
radial critical curve (bottom), versus lens mass, for the
TIS (solid curves), the NFW profile (dotted curves), the SIS 
(short dashed curves) and the Schwarzschild lens (long dashed curves).
For the TIS, the values of $z_{\rm coll}$ are indicated.
\label{critical2}}
\end{figure}

\noindent (\cite{SEF92} eq.~[8.3]), where
$m(x)=M(r_0x)/\pi r_0^2\Sigma_{\rm crit}$ is the dimensionless
interior mass.
For the Schwarzschild lens and the SIS, the solutions are $x_t=1$, and there
are no real solutions for $x_r$. For the NFW and TIS models, 
equations~(\ref{tancrit}) and~(\ref{radcrit}) have to be solved numerically 
for $x_t$ and $x_r$. The solutions for the TIS are plotted
in Fig.~\ref{crit2}, as functions of $\kappa_c$. 
Also plotted is the radial caustic
radius $y_r$, obtained by substituting the value of $x_r$ into 
equation~(\ref{lensfinal}).
(The value of $y_r$ we obtain is actually negative, because the
source and image are on opposite sides of the lens.
The actual radius of the caustic circle,
then, is the absolute value of $y_r$.)
Both $x_t$ and $|y_r|$
increase rapidly with $\kappa_c$, while the
value of $x_r$ levels off. Fig.~\ref{critical2} shows the angular radii of
the tangential and radial critical circles,
$\theta_r=\xi_r/D_L$ and $\theta_t=\xi_t/D_L$, in arc seconds as
functions of the mass of the lens.
For spherically symmetric lenses, multiple images (and thus
critical circles) are possible only if the central convergence 
$\kappa(0)\equiv\Sigma(0)/\Sigma_{\rm crit}$
exceeds unity (\cite{SEF92}, p.~236, theorem~[e]). 
For the Schwarzschild lens, SIS,
and NFW profile, the central convergence diverges, hence these profiles
can always produce multiple images. But for the TIS, the central
convergence $\kappa(0)\equiv\kappa_c$ is finite, and multiple images
can be produced only if $\kappa_c>1$. This explains the sharp low-mass
cutoff seen in Fig.~\ref{critical2} for the TIS (solid curves).

The value of $\theta_t$ for the Schwarzschild lens is called the Einstein
radius $\theta_E$. It is often used to estimate the characteristic scale of
image features caused by strong lensing (e.g. ring radius, radial location
of arc, image separations) and to estimate the size of the region within
which the mass responsible for that strong lensing must be concentrated.
Since lensing halos are not actually point masses, however, the angular
radius $\theta_{\rm ring}$ of the 
actual Einstein ring which results if the source is located along the line 
of sight through the lens center will usually
differ from the Einstein radius $\theta_E$, assuming that the lens mass
distribution is actually capable of producing a ring. As we see in 
Fig.~\ref{critical2},
$\theta_E$ significantly exceeds $\theta_t$ for all profiles considered
(TIS, NFW profile, and SIS) for all masses considered. Hence, a mass estimate
based on assuming that the scale of image features is
of order $\theta_E$ will underestimate the actual mass of the lens, unless
the lens happens to be a Schwarzschild lens.

A source located behind the lens will produce multiple images if $y<y_r$.
The angular cross section for multiple imaging is therefore
\begin{equation}
\sigma_{\rm m.i.}=\pi\left({\eta_r\over D_S}\right)^2
=\pi\left({y_r\xi_0\over D_L}\right)^2\,.
\end{equation}

\begin{figure}[t!]
\includegraphics[height=4in]{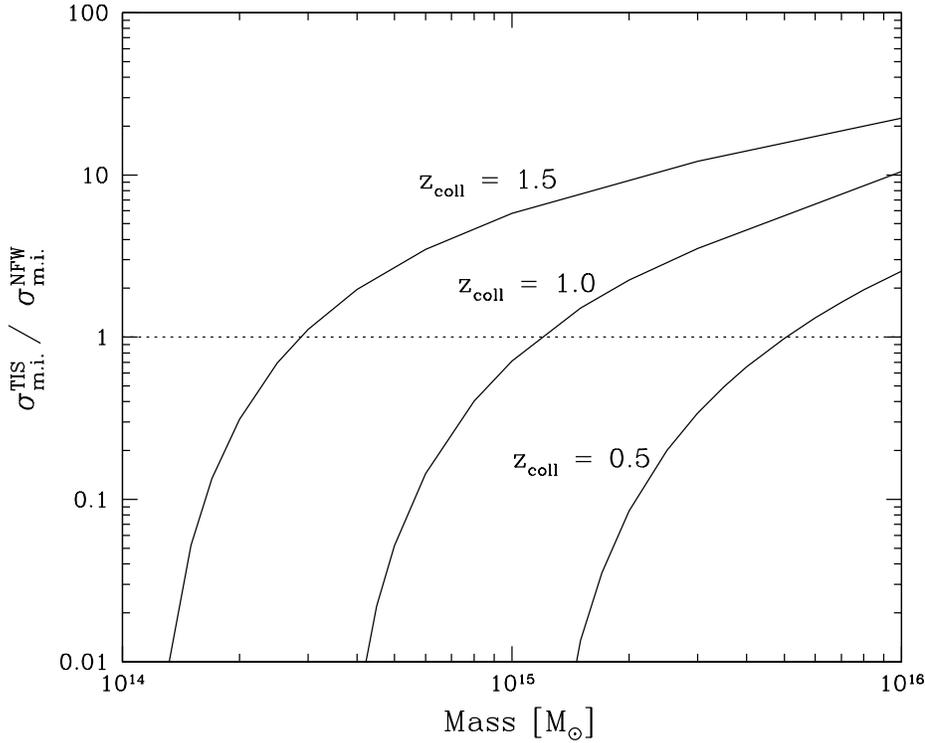}
\caption{Ratio of the cross sections for multiple imaging by the TIS
and the NFW profile, versus lens mass.
The values of $z_{\rm coll}$ for the TIS are indicated.
\label{cross}}
\end{figure}

\noindent In Fig.~\ref{cross}, 
we plot the ratio of the cross sections for the
NFW and TIS profiles. 
At low masses, $M<5\times10^{15}M_\odot$ for $z_{\rm coll}=0.5$,
$M<1.2\times10^{15}M_\odot$ for $z_{\rm coll}=1.0$, 
$M<3\times10^{14}M_\odot$ for $z_{\rm coll}=1.5$,
the ratios are less than unity, 
indicating that a distribution of lenses described by the NFW profile
would be more likely to produce cases with multiple images than if the same
distribution is described by the TIS model. This trend is reversed at
higher masses, and a TIS is 
more likely to produce multiple images than a NFW profile of
the same mass.

\subsubsection{Illustrative Example}

\begin{figure}[t!]
\includegraphics[width=5.5in]{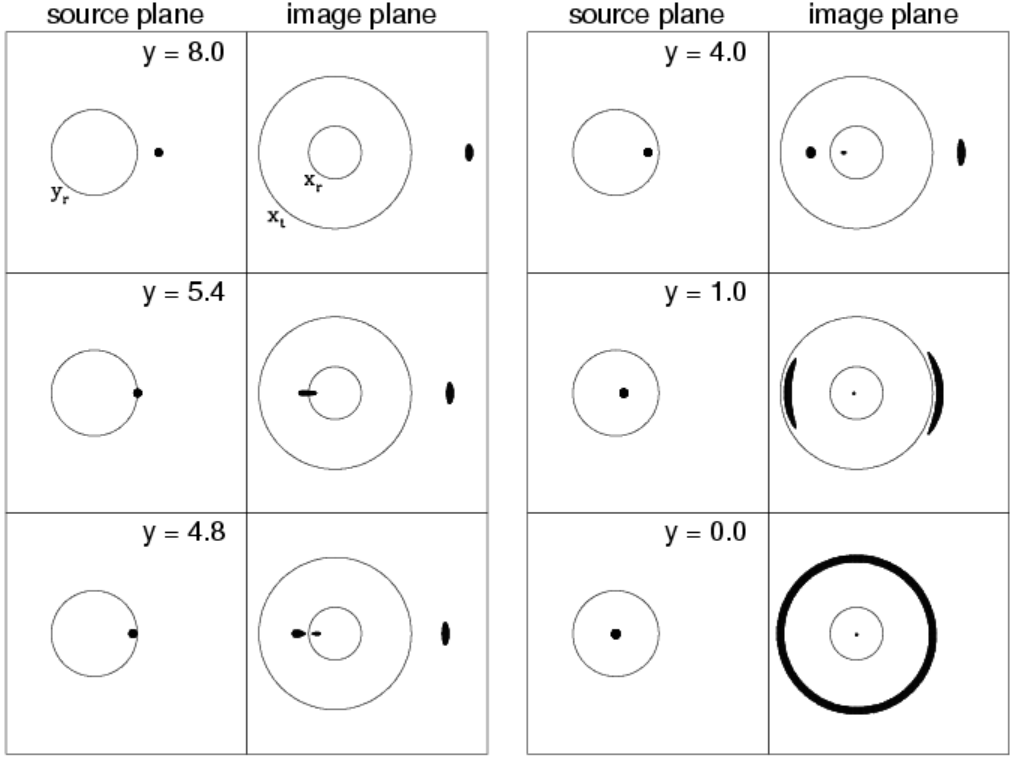}
\caption{Images of a circular source. Each pair of panels shows the
source plane in the left panel, with the caustic, and the image plane in
the right panel, with the radial (inner) and tangential (outer) critical
circles. The position $y$ of the source on the source plane is indicated.
We used $\kappa_c=4.015$, and a source of diameter $\Delta y=1$.
\label{images}}
\end{figure}

Using a simple ray-tracing algorithm, we computed the image(s) of a circular
source of diameter $\Delta y=1$, created by a TIS with central convergence
$\kappa_c=4.015$. The results are shown in Fig.~\ref{images} for 6 different
locations of the source, ranging from $y=8.0$ to $y=0.0$. For each case, the
left panel shows the source and the caustic circle ($y_r=5.640$) on
the source plane, and the right panel shows the images(s), the radial
critical circle ($x_r=3.334$), and the tangential critical circle
($x_t=9.783$) on the image plane. At $y=8.0$,
only one image appears. At $y=5.4$, the source overlaps the caustic, and a
second, radially-oriented image appears on the radial critical circle.
At $y=4.8$, the source is entirely inside the caustic, and the second
image splits in two images, located on opposite sides of the radial
critical circle, forming with the original image a system of 3 aligned images.
As the source moves toward $y=0$, the central image moves toward $x=0$
and becomes significantly fainter, while the other images move toward the
tangential critical circle and become bright, elongated arcs. At $y=0$,
the two arcs have merged to form an Einstein ring located on top of
the tangential critical circle, while the central image, very faint,
is still visible in the center.

\subsection{IMAGE SEPARATION AND MAGNIFICATION}

\begin{figure}[t!]
\includegraphics[width=5.5in]{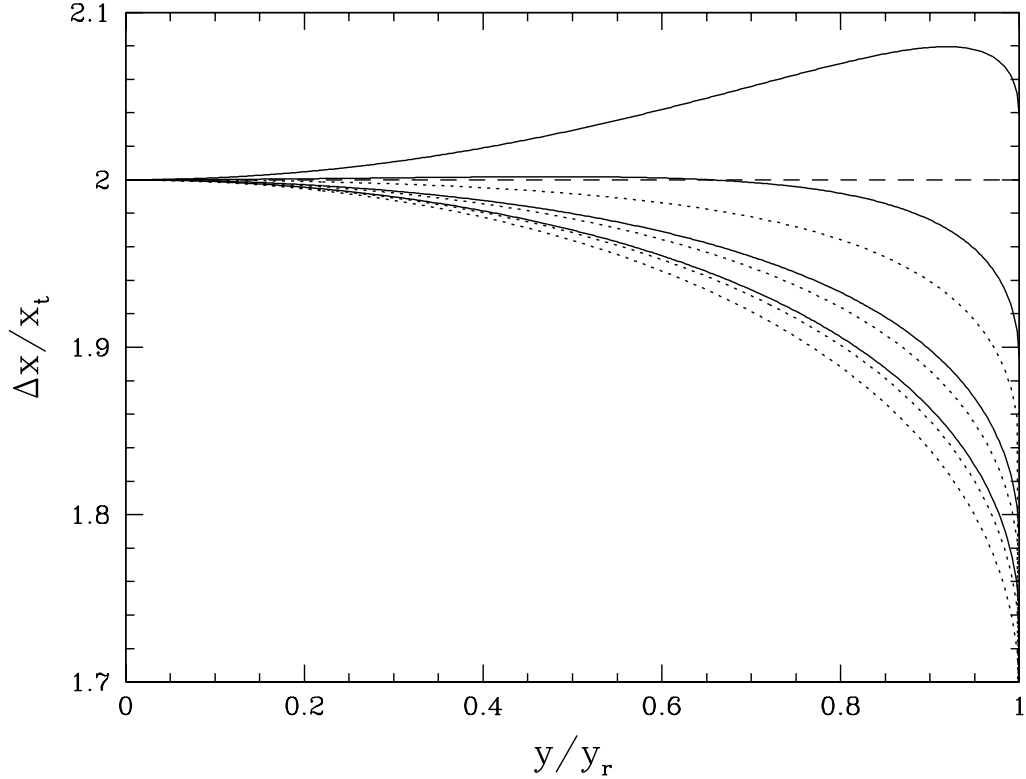}
\caption{Separation $\Delta x$ between the two outer images,
in units of the tangential critical radius $x_t$, versus source
location $y$ in units of the caustic radius $y_r$. The solid curves, from top
to bottom, corresponds to TIS with $\kappa_c=10$, 5, 2.5, and 1.2, 
respectively.
The dotted curves, from top
to bottom, corresponds to NFW profiles 
with $\kappa_s=1.0$, 1.0, 0.5, and 0.2, respectively. 
The dashed line corresponds to the SIS. Results for the Schwarzschild lens are
not plotted.
\label{separation}}
\end{figure}

The locations of the images are computed by solving the lens 
equation~(\ref{lensfinal}). For the TIS, this
equation must be solved numerically. In Fig.~\ref{separation}, 
we plot the separation between the two outer images as
a function of the source location. The plot only extends to $y/y_r=1$, since
larger values of $y$ only produce one image. 
The solid and dotted curves show the separations for the TIS and NFW profile,
respectively, with various values of $\kappa_c$ and 
$\kappa_s\equiv\rho^{\phantom2}_{\rm NFW}
r^{\phantom2}_{\rm NFW}/\Sigma_{\rm crit}$.
The separation is fairly
insensitive to the source location, and stays within $\sim15\%$ of the
Einstein ring diameter $\Delta x=2x_t$ for all values of 
$\kappa_c$ and $\kappa_s$ considered.
The dashed line in Fig.~\ref{separation} shows the separation for the SIS,
which is independent of $y$.
For all profiles considered, the image separation is always of order 
the Einstein ring diameter, independently of the source location.
This is particularly convenient for theoretical studies, when
the actual source location can be ignored \cite{MPM02}.

The magnification of an image located at position $x$ on the lens
plane is given by
\begin{equation}
\label{magni}
\mu=
\left(1-{m\over x^2}\right)^{-1}
\left[1-{d\over dx}\left({m\over x}\right)\right]^{-1}
=\left(1-{\alpha\over x}\right)^{-1}
\left(1-{d\alpha\over dx}\right)^{-1}
\end{equation}

\begin{figure}[t!]
\includegraphics[width=5.5in]{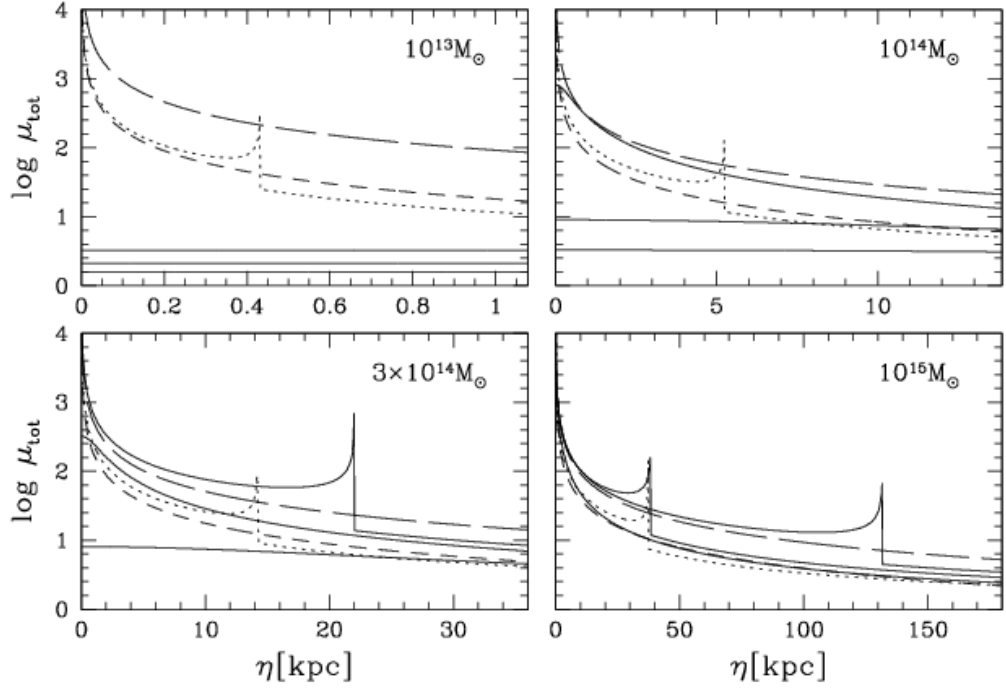}
\caption{Total magnification $\mu_{\rm tot}$ versus source location $\eta$
for lenses of masses $10^{13}-10^{15}M_\odot$, for the
TIS (solid curves), the NFW profile (dotted curves), the SIS 
(short dashed curves) and the Schwarzschild lens (long dashed curves).
For the TIS, the curves, from top to bottom, correspond to
$z_{\rm coll}=1.5$, 1.0, and 0.5, respectively.
\label{magnification2}}
\end{figure}

\noindent (\cite{SEF92}, eq.~[8.17]).
We computed the magnification of the images produced by halos of masses
$10^{13}$, $10^{14}$, $3\times10^{14}$, 
and $10^{15}M_\odot$. Fig.~\ref{magnification2} shows
the total magnification.
The dotted curves show the results for a NFW profile. As $y$ (or $\eta$)
decreases, the magnification
slowly increases, until the source reaches the radial caustic $y=y_r$.
At that moment, a second image, with infinite magnification 
appears on the radial critical curve (for clarity, we truncated those infinite
``spikes'' in Fig.~\ref{magnification2}). As $y$
keeps decreasing, that second image splits into two images, and the 
total magnification becomes finite again, until the source reaches
$y=0$, and an Einstein ring with infinite magnification
appears on the tangential critical curve.
Of course, these infinite magnifications are not physical, since they can only
occur for point sources.
The total magnification is always larger than unity, and always larger
when 3 images are present.

The solid curves in Fig.~\ref{magnification2} shows the results for the TIS.
At low masses, there is no radial caustic (See Fig.~\ref{critical2}), 
and only one image
appears. Because of the presence of a flat density core, the magnification
is nearly constant if the path of the rays goes near the center of the core.
For instance, for the case $M_{200}=10^{13}M_\odot$ (top left
panel of Fig.~\ref{magnification2}), $r_0>10\,{\rm kpc}$, hence,
over the range of $\eta$ being plotted, we are way inside the core. As
the mass increases, the magnification increases, until a radial
caustic forms. This happens at $M_{200}=1.11\times10^{15}M_\odot$
for $z_{\rm coll}=0.5$, $M_{200}=3.5\times10^{14}M_\odot$ for 
$z_{\rm coll}=1.0$, and $M_{200}=1.2\times10^{14}M_\odot$ for 
$z_{\rm coll}=1.5$
according to Fig.~\ref{critical2}. 
At this point, the TIS has the ability to form
three images, and the results become qualitatively similar to the ones
for the NFW profile.

The short-dashed and long-dashed curves in Fig.~\ref{magnification2} 
shows the
results for the SIS and the Schwarzschild lens, respectively. Because of
the absence of radial caustic, the magnification always varies smoothly with
source position, and the only divergence occurs at $\eta=0$, where
an Einstein ring forms.

For a given mass, the Schwarzschild lens always produces a stronger 
magnification than a SIS or NFW profile,
unless the source is very close to the radial caustic of
the NFW profile, where a ``spike'' of infinite magnification
forms. The magnifications produced by the SIS and NFW profile tend to 
be similar.
If the NFW profile produces only one image (that is, we are on the right
hand side of the dotted spike in Fig.~\ref{magnification2}), the SIS
produces a larger magnification than the NFW profile. But if the NFW profile
produces 3 images, then the total magnification exceeds the one produced by
the SIS. As for the TIS, at low masses, where only one image forms, the
magnification is much smaller than for the other profiles. But at large
masses, where multiple images can form, the magnification becomes comparable
to the one for the other profiles, and can even exceed the
magnification produced by the Schwarzschild lens.

\subsection{WEAK LENSING}

Weak lensing usually refers to the magnification and distortion of the
image of a background source by a foreground lens. Unlike strong lensing,
weak lensing normally does not produce multiple images of single sources.
The detection of coherent distortion patterns in the sky has been used
to constrain the mass of clusters. The first detections were reported by
\cite{BMF94,DML94,Fetal94,Setal95,TWV90},
followed by many others (see \cite{BS01}
and references therein). More recently, the distortion pattern produced
by individual galaxies has also been detected
\cite{BBS96,DT96,E98,Fetal00,Getal96,Hetal98,Netal96}.

Following the approach of \cite{WB00}, we use as measure
of the distortion produced by a lens the average shear $\bar\gamma$ inside
a distance $\xi=r_{200}$ from the lens center. In practice, this quantity
would be evaluated by averaging the shear of all images observed inside
$r_{200}$, after having eliminated foreground sources. We estimate this
quantity by integrating the shear over the projected area of the cluster.
The average shear inside radius $x$ is given by
\begin{equation}
\label{shearavg1}
\bar\gamma(x)={2\over x^2}\int_0^xx'\gamma(x')dx'
={2\over x^2}\int_0^xx'\left[{m(x')\over{x'}^2}-\kappa(x')\right]dx'\,,
\end{equation}

\begin{figure}[t!]
\includegraphics[width=5.5in]{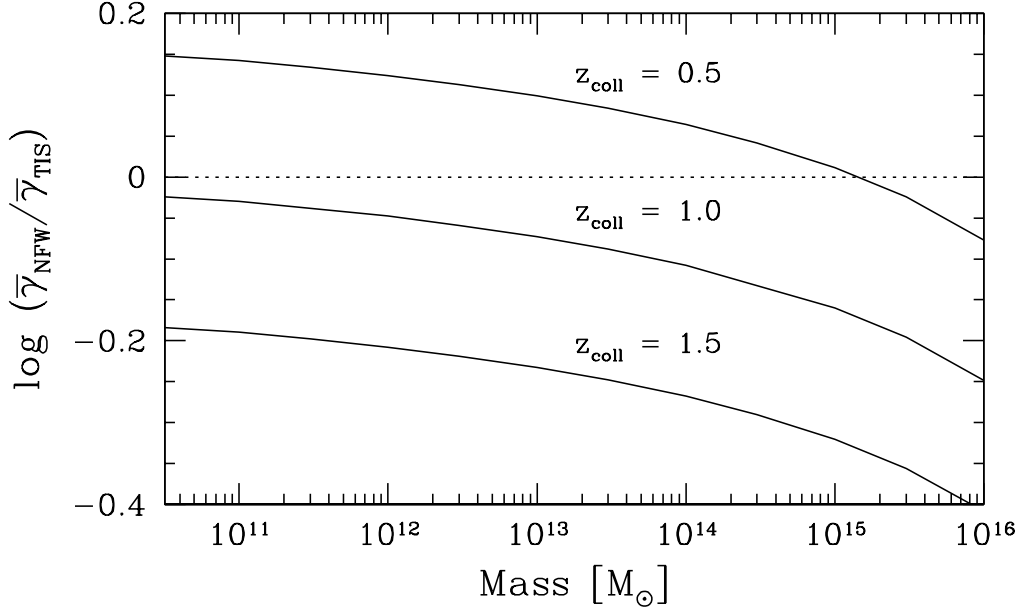}
\caption{Ratio of the average shear inside radius $r_{200}$
for the NFW profile and the TIS profile, versus mass of the halo.
The values of $z_{\rm coll}$ for the TIS are indicated.
\label{shearavg}}
\end{figure}

\noindent where $\kappa(x')=\Sigma(x')/\Sigma_{\rm crit}$ is the
convergence. After some algebra, we get
\begin{eqnarray}
\label{shearavgtis}
\bar\gamma_{\rm TIS}^{\phantom2}(x)
={2ab\kappa_c\over(Ab-Ba)x^2}&&\kern-20pt\Bigg\{
Aa\left[\left(1+{x^2\over a^2}\right)^{1/2}-1-2\ln{1+(1+x^2/a^2)^{1/2}\over2}
\right]\nonumber\\
&-&
Bb\left[\left(1+{x^2\over b^2}\right)^{1/2}-1-2\ln{1+(1+x^2/b^2)^{1/2}\over2}
\right]\Bigg\}\,.
\end{eqnarray}

\noindent
We evaluate this expression at $r=r_{200}$, or equivalently 
$x=r_{200}/r_0\equiv\eta_{200}=24.2$. It reduces to
\begin{equation}
\label{shearavgtis2}
\bar\gamma_{\rm TIS}^{\phantom2}(r_{200})
=408.67{\rho_cr_{200}\over\Sigma_{\rm crit}}\,,
\end{equation}

\noindent where we used equation~(\ref{rho0}) to eliminate $\rho_0$.

In Fig.~\ref{shearavg} we plot the ratio 
$\bar\gamma_{\rm NFW}^{\phantom2}/\bar\gamma_{\rm TIS}^{\phantom2}$
versus mass. The ratios are larger at lower collapse
redshifts $z_{\rm coll}$, and
also decrease with increasing mass. This reflects the
fact that as the mass increases, the concentration parameter $c$ of the NFW
profile decreases, while the ratio $r_{200}/r_0$ for the TIS remains
fixed at 24.2. This figure can be {\it qualitatively} compared with the
top left panel of Fig.~3 in \cite{WB00}.
We reach essentially the same conclusion as these authors,
namely that using the average shear to estimate the mass of lensing halos
can lead to considerable errors if the wrong density profile is
assumed. For $z_{\rm coll}=0.5$, we find
$\bar\gamma_{\rm NFW}^{\phantom2}/\bar\gamma_{\rm TIS}^{\phantom2}>1$
at small mass,
and therefore the mass of a TIS would be underestimated if the lens
is incorrectly assumed to follow a NFW profile. At high mass, the
true mass of a TIS would be overestimated. For $z_{\rm coll}=1.0$ and 1.5,
$\bar\gamma_{\rm NFW}^{\phantom2}/\bar\gamma_{\rm TIS}^{\phantom2}<1$
at all masses considered, down to $3\times10^{10}M_\odot$, 
and therefore the true mass of a TIS would always be overestimated. 

\subsection{DISCUSSION}

In the previous two sections, we derived the effects of strong and 
weak lensing, respectively. Here, ``strong lensing'' refers to case
with multiple images, arcs, or rings, while ``weak lensing'' refers
to the magnification and shear of single images. We can divide the
observed cases of strong lensing in two groups. The first group contains 
the ``arc second'' cases: multiple-image systems with image separations
of order arc seconds, or rings with radii of that order
\cite{Ketal98}.\footnote{See http://cfa-www.harvard.edu/castles.} 
In most cases, the lens is a
single, massive galaxy, with possibly some additional
contribution from the environment in which this galaxy is located
(\cite{TOG84}; see, however, \cite{PM04}).
A classic example is Q0957+561, the first gravitational lens
to be discovered. 
The second group contains the ``arc minute'' cases, in which the lens
is an entire cluster of galaxies. These lenses produce mostly giant arcs,
with radii in the range $15''-60''$ (see Table~1 of \cite{WNB99}). 
The most famous case is the cluster
CL~0024+1654, which produces multiple arcs.

We showed that for all profiles considered, the image separation is weakly
dependent on the source location (Fig.~\ref{separation}), 
when multiple images actually
form. If we neglect this dependence, the image separation
$\Delta\theta\approx2\theta_t$ can be read off the top panels of 
Fig.~\ref{critical2}.
We see immediately that galaxy-size objects cannot produce
arc-second separations if they are described by the TIS or the NFW profile
(under the assumption described in \S2 that
$c$ for NFW halos is the typical value for halos at $z_{\rm obs}=z_L$).
The TIS does not produce multiple images unless the mass is of order
$10^{14}M_\odot$ or above. The NFW profile can produce multiple images for
any mass, but a separation $\Delta\theta>1''$ requires a mass
of order $10^{13}M_\odot$. Even the SIS needs a mass in excess of
$10^{12}M_\odot$ to produce arc-second separations.  
At the cluster scale, all profiles are capable of producing arc-minute
separations. In this limit, for a given mass, the separation is larger for
the Schwarzschild lens and smaller for the NFW profile (because of the
small concentration parameter at large mass). The separations are comparable
for the SIS and TIS for $z_{\rm coll}=z_L$, but for $z_{\rm coll}>z_L$
the separation tends to be larger for the TIS than the SIS. 
As we indicated in
the introduction, the TIS and NFW profile are applicable to dwarf
galaxies and clusters of galaxies, but might not be applicable to 
ordinary
galaxy-scale objects because baryonic processes are neglected. Hence, the
inability of the TIS and NFW profiles to produce arc seconds separations
with galaxy-size lenses is not a concern.

Let us now focus on the TIS, and perform a simple calculation to
estimate the probability that a TIS can produce multiple images.
To compute $\kappa_c$, we substitute
equations~(\ref{m200}), (\ref{rho0}), and (\ref{r0}) in
equation~(\ref{kappac}). For the particular cosmological model ($\Lambda$CDM)
and redshifts ($z_S=3$, $z_L=0.5$) we have considered in this paper,
we get
\begin{equation}
\label{kappacoll2}
\kappa_c=480.75{\rho_c^{2/3}(z_L)\over\Sigma_{\rm crit}(z_L)}
\left[{\rho_c(z_{\rm coll})\over\rho_c(z_L)}\right]^{2/3}M_{200}^{1/3}
=0.968\left[{\rho_c(z_{\rm coll})\over\rho_c(z_L)}\right]^{2/3}M_{15}^{1/3}\,,
\end{equation}

\noindent where $M_{15}\equiv M_{200}/10^{15}M_\odot$. 
Strong lensing requires $\kappa_c\geq1$. 
Setting $\kappa_c=1$, $z_{\rm coll}=z_L$, and solving for $M_{200}$,
we get $M_{200}=1.102\times10^{15}M_\odot$. The requirement for strong
lensing then becomes
\begin{equation}
M_{15}\geq M_{15,\min}\,,
\end{equation}

\noindent where
\begin{equation}
M_{15,\min}\leq1.102\,.
\end{equation}

\noindent Hence, in the worst case scenario, 
a halo that formed at the latest possible
collapse redshift $z_{\rm coll}=z_L$, the required mass is about
$10^{15}M_\odot$. That value drops quite rapidly with increasing 
$z_{\rm coll}$,
down to about $1.2\times10^{14}M_\odot$ for $z_{\rm coll}=1.5$ 
(see Fig.~\ref{critical2}).
For the cosmological model we consider, 1-$\sigma$ density
fluctuations collapsing at redshift $z_{\rm coll}=(0.5,1.0,1.5)$ have masses of
about $M_{15}=(2\times10^{-3},4\times10^{-4},9\times10^{-5})$.
Such ``typical'' objects will not be capable of
producing multiple images of a source at redshift $z_S=3$, since the
resulting values $\kappa_c=(0.122,0.127,0.120)$ are smaller than unity. 
This simply indicates that
multiple images are not produced by typical objects,
which is certainly consistent with the fact that fewer than 30 
arc-minutes, multiple-image
systems have been observed.  It is remarkable that the value of 
$\kappa_c$ for typical objects is nearly independent of $z_{\rm coll}$.
It just happens the the dependences of $\rho_c(z_{\rm coll})$ and
$M_{15}$ in equation~(\ref{kappacoll2}) nearly cancel out when $M_{15}$ is
set to the typical value at that redshift.

Increasing $\kappa_c$ above
unity would require an object about 500 times
more massive than a typical object at the same redshift. Objects of this
mass are rare but do exist. We can make a simple estimate of how atypical
such a massive object is. Over most of the mass range of cosmological
interest (from small
galaxies to clusters of galaxies) the CDM power spectrum can
be roughly approximated
by a power law $P(k)\propto k^n$, where $k$ is the wavenumber
and $n\approx-2$. The rms density 
fluctuation $\delta_{\rm rms}$ is then given by 
$\delta_{\rm rms}\approx k^{3/2}P^{1/2}(k)\propto k^{1/2}$.
At a given redshift, different values of
the wavenumber $k$ correspond to different mass scales 
$M$ according to $M\propto k^{-3}$. The relation between rms
density fluctuation and mass scale at fixed epoch is therefore approximated by
\begin{equation}
\delta_{\rm rms}\propto M^{-1/6}\,.
\end{equation}

\noindent Increasing the mass by a factor of 500 therefore reduces
$\delta_{\rm rms}$ 
by a factor of $500^{1/6}\approx2.8$. Because of the reduction
in $\delta_{\rm rms}$, 
a 1-$\sigma$ fluctuation ($\delta=\delta_{\rm rms}$) at this higher mass
will no longer collapse by the same redshift (it will collapse later),
but a 2.8-$\sigma$ fluctuation ($\delta=2.8\delta_{\rm rms}$)
will. Such fluctuations are rare, but not
vanishingly rare. In Gaussian statistics, the probability that a randomly
located point in space is inside a 2.8-$\sigma$ density fluctuation
(i.e. $\delta\geq2.8\delta_{\rm rms}$) is about 1/200.
Hence, one of every 200 halos would be capable of producing multiple images
(of course, whether any halo actually produces multiple images
depends on the location of the sources).

To give a specific example of a cluster-mass lens with $z_{\rm coll}>z_L$,
consider the case of the cluster CL~0025+1654.
\cite{SI00} showed that the
mass-model derived by \cite{TKD98} to explain their lensing data for
CL~0024+1654 at $z=0.39$ is very well-fitted by a TIS halo with
$\rho_0\approx0.064h^2M_\odot\,{\rm pc}^{-3}$ and
$r_0\approx20h^{-1}{\rm kpc}$. This central density implies that
the halo collapse redshift is $z_{\rm coll}\approx2.5$ (i.e.
$z_{\rm coll}\gg z_L$). The halo mass in then 
$M_{200}=661.6\rho_0r_0^3=4.84\times10^{14}M_\odot$ (for $h=0.7$),
and equation~(\ref{kappacoll2}) gives $\kappa_c=4.8\gg1$. It is therefore
not surprising that this cluster produces strong lensing.

\subsection{SUMMARY}

We have derived the lensing properties of cosmological halos
described by the TIS model. The solutions depend
on the background cosmological model through the critical surface
density $\Sigma_{\rm crit}$, which is a function of the cosmological
parameters and the source and lens redshifts, and the TIS parameters
$\rho_0$ and $r_0$, which are functions of the mass and collapse redshift
of the halo, and the cosmological parameters. By expressing the surface
density of the halo in units of $\Sigma_{\rm crit}$ and the distances
in units of $r_0$, all explicit dependences on the cosmological
model disappear, and the solutions are entirely expressible in terms of two
dimensionless parameters, the central convergence $\kappa_c$ and
the scaled position $y$ of the source. We have computed
solutions for the critical curves and caustics, the image separations, 
and the total magnification.
The ability of the TIS to produce strong lensing (multiple
images and rings) depends entirely on $\kappa_c$. If $\kappa_c<1$,
only one image can form. If $\kappa_c>1$,
either one or three images can form, depending on
whether the source is located outside or inside the radial caustic.
When three images are produced, the central one is usually very faint, being
highly demagnified. 
The angular separation between the
two outermost images depends strongly on $\kappa_c$, but only
weakly on the source location.

For comparison, we derived (or extracted from the literature) the
lensing properties of three comparison models: the NFW profile,
the singular isothermal sphere, and the Schwarzschild lens.
Unlike the TIS, all of these profiles have a central singularity,
which allows them to produce multiple images at any mass, provided
that the source is sufficiently aligned with the lens. In practice,
image separations large enough to be resolved can be achieved by
galactic-mass objects only for the Schwarzschild lens, and by
supergalactic-mass objects for all profiles.

We applied the TIS model to the currently-favored
$\Lambda$CDM universe, to calculate the central convergence $\kappa_c$
expected for TIS halos of different masses and collapse epochs.
We found that high-redshift sources (e.g. $z_S\approx3$) will be
strongly lensed by TIS halos (i.e. $\kappa_c>1$) only for cluster-mass halos,
assuming that these halos formed at the redshift they are observed. 
As equation~(\ref{kappacoll2}) shows, the halo mass required for
strong lensing can be decreased by increasing the formation redshift
of the halo. However, as the formation redshift increases, the typical
halo mass decreases, leading to a near-cancellation of the effect: the
value of $\kappa_c$ for a typical halo is about 0.12, independent of
the collapse redshift. From this, we showed that a halo described by a TIS
must form out of a $\sim2.8-\sigma$ fluctuation to be capable of producing
strong lensing ($\kappa_c>1$), no matter when the halo formed. Notice
that weak lensing (magnification and shear of single images)
does not require $\kappa_c>1$, and can be caused by halos of any mass.


\vspace{1cm}

This work was supported by the Research and Training Network
"The Physics of the Intergalactic Medium" set up by the 
European Community RTN under the contract HPRN-CT2000-00126 RG29185,
 NASA grants NAG5-10825, NAG5-10826, NNG04G177G and Texas Advanced Research 
Program grant 3658-0624-1999.  MAA is 
grateful for the support of a DOE Computational Science Graduate Fellowship.


\begin{thebibliography}{}

\bibitem{abn}
Abadi, M. G., Bower, R. G. \& Navarro, J. F. 2000, MNRAS, 314, 759

\bibitem{AS}
Ahn, K. \& Shapiro, P. R. 2003, JKAS, 36, 89

\bibitem{as04}
Ahn, K. \& Shapiro, P. R. 2004, in preparation

\bibitem{AAS} 
Alvarez, M. A., Ahn, K., \& Shapiro, P. R. 2003, RevMexAA SC, 18, 4

\bibitem{ASM00} 
Alvarez, M. A., Shapiro, P. R., \& Martel, H. 2000, RevMexAA SC, 10, 214

\bibitem{ASM01} 
Alvarez, M. A., Shapiro, P. R., \& Martel, H. 2001, AIP Conf.Proc., 586, 938

\bibitem{ASM03} 
Alvarez, M. A., Shapiro, P. R., \& Martel, H. 2003, RevMexAA SC, 17, 39

\bibitem{AVI} 
Avila-Reese, V., Colin, P., Valenzuela, O., D'Onghia, E., Firmani, C. 
2001, ApJ, 559, 516

\bibitem{afh}
Avila-Rees, V., Firmani, C. \& Hern\'{a}ndez, X. 1998, ApJ, 505, 37

\bibitem{BSI}
Balberg, S., Shapiro, S. L. \& Inagaki, S. 2002, ApJ, 568, 475

\bibitem{bbks}
Bardeen, J. M., Bond, J. R., Kaiser, N. \& Szalay, A. S. 1986, ApJ, 304, 15 

\bibitem{BE}
Barnes, J., \& Efstathiou, G. 1987, ApJ, 319, 575

\bibitem{BS01}
Bartelmann, M., \& Schneider, P. 2001, Phys.Rep., 340, 291

\bibitem{BERT}
Bertschinger, E. 1985, ApJ, 58, 39

\bibitem{bert-cool}
Bertschinger, E. 1989, ApJ, 340, 666

\bibitem{BT}
Binney, J., \& Tremaine, S.: \emph{Galactic Dynamics} (Princeton University Press,
Princeton 1987)

\bibitem{Betal91}
Blandford, R. D., Saust, A. B., Brainerd, T. G., \& Villumsen, J. V.
1991, MNRAS, 251, 600

\bibitem{BOT} Bode, P., Ostriker, J.P., Turok, N. 2001, ApJ, 556, 93

\bibitem{BMF94}
Bonnet, H., Mellier, Y., \& Fort, B. 1994, ApJ, 427, L83

\bibitem{BBS96}
Brainerd, T. G., Blandford, R. D., \& Smail, I. 1996, ApJ, 466, 623

\bibitem{BN}
Bryan, G.L. \& Norman, M.L. 1998, ApJ, {  495}, 80

\bibitem{BUL} Bullock, J.S., Kolatt, T.S., Sigad, Y., Somerville, R.S., 
Kravtsov, A.V., Klypin, A.A., Primack, J.R., Dekel, A., 2001, MNRAS, 321, 559

\bibitem{BUL2} Bullock, J. S., Dekel, A., Kolatt, T. S., Kravtsov, A. V.,
Klypin, A. A., Porciani, C., \& Primack, J. R. 2001, ApJ, 555, 240

\bibitem{B}
Burkert, A. 1995, ApJ, 447, L25

\bibitem{B00}
Burkert, A. 2000, ApJ, 534, L143

\bibitem{BS99}
Burkert, A. \& Silk, J. 1999, in
Dark Matter in Astrophysics and Particle Physics, Proceedings of the 
second International Conference on Dark Matter in Astrophysics and Particle 
Physics, eds. H. V. Klapdor-Kleingrothaus and L. Baudis
(Philadelphia: Institute of Physics Publishers), p. 375

\bibitem{carlberg} Carlberg, R. G., Yee, H. K. C., Ellingson, E.,
Morris, S. L., Abraham, R., Gravel, P., Pritchet, C. J., Smecker-Hane,
T., Hartwick, F. D. A., Hesser, J. E., Hutchings, J. B. \& Oke,
J. B. 1997, ApJ, 485, L13

\bibitem{CT01}
Chiba, T. \& Takahashi, R. 2001, unpublished (astro-ph/0106273 v1)

\bibitem{CT02}
Chiba, T. \& Takahashi, R. 2002, Prog.Theor.Phys., 107, L625

\bibitem{CL96}
Cole, S. \& Lacey, C. 1996, MNRAS, 281, 716

\bibitem{CKK}
Colin, P., Klypin, A. A., \& Kravtsov, A. V. 2000, ApJ, 539, 561

\bibitem{CAV} Colin, P., Avila-Reese, V., Valenzuela, O. 2000, ApJ, 542, 
622

\bibitem{CKVG} Colin, P., Klypin, A., Valenzuela, O., \& Gottl\"{o}ber, S. 2003, astro-ph/0308348

\bibitem{CMKS02}
Czoske, O., Moore, B., Kneib, J.-P., Soucail, G. 2002, A\&A, 386, 31  

\bibitem{DML94}
Dahle, H., Maddox, S. J., \& Lilje, P. B. 1994, ApJ, 435, L79

\bibitem{Dave}
Dav\'{e}, R., Spergel, D. N., Steinhardt, P. J. \& Wandelt, B. D. 2001, ApJ, 547, 574

\bibitem{DEB}
de Blok, W.J.G., McGaugh, S.S., Bosma, A., \& Rubini, V.C. 2001, ApJ, 552, L23

\bibitem{DB} de Blok, W.J.G., \& Bosma, A. 2002, A\&A, 385, 816

\bibitem{DT96}
Dell'Antonio, I., \& Tyson, J. A. 1996, ApJ, 473, L17

\bibitem{popolo}
Del Popolo, A., Gembera, M., Recami, E. \& Spedicato, E. 2000, A\&A, 353, 427

\bibitem{DSPBCEO}
Dressler, A., \& Smail, I., Poggianti, B.M., Butcher, H., Couch, W.J.,
Ellis, R.S., Oemler, A., Jr. 1999, APJS,  122, 51 

\bibitem{E98}
Ebbels, T. 1998, Ph.D. thesis, Cambridge Univ.

\bibitem{ENF} Eke, V.R., Navarro, J.F., Frenk, C.S. 1998, ApJ, 503, 569

\bibitem{ENS01}
Eke, V. R., Navarro, J. F., \& Steinmetz, M. 2001, ApJ, 554, 114

\bibitem{Eetal02}
Evrard, A.E. et al. 2002, ApJ, 573, 7 

\bibitem{EMN}
Evrard, A.E., Metzler, C.A., \& Navarro, J.F. 1996, ApJ, 469, 494

\bibitem{Fetal94}
Fahlman, G., Kaiser, N., Squires, G., \& Woods, D. 1994, ApJ, 437, 56

\bibitem{FG} 
Fillmore, J.A., Goldreich, P. 1984, ApJ, 281, 1

\bibitem{FDCHA}
Firmani, C., D'Onghia, E., Chincarini, G., Hernandes, X., \& Avila-Reese, V. 
2000, MNRAS, 321, 713

\bibitem{Fetal00}
Fischer, P. et al. 2000, AJ, 120, 1198

\bibitem{FP} 
Flores, R.A., \& Primack, J.R. 1994, ApJ, 427, L1

\bibitem{SantaBarbara}
Frenk, C. S. et al. 1999, ApJ, 525, 554

\bibitem{FM97}
Fukushige, T., \& Makino, J. 1997, ApJ, 477, L9

\bibitem{FM01}
Fukushige, T., \& Makino, J. 2001, ApJ, 557, 533

\bibitem{FM03}
Fukushige, T., \& Makino, J. 2003, ApJ, 588, 674

\bibitem{FKM}
Fukushige, T., Kawai, A., \& Makino, J. 2004, ApJ, 606, 625

\bibitem{GAV}
Gavazzi, R., Fort, B., Mellier, Y., Pello\', R., \& Dantel-Fort, M. 2003,
A\&A, 403, 11

\bibitem{Getal00}
Ghigna, S., Moore, B., Governato, F., Lake, G., Quinn, T., \& Stadel, J. 2000,
ApJ, 544, 616

\bibitem{G}
Girardi, M., Giuricin, G., Mardirossian, F., 
Mezzetti, M., Boschin, W. 1998, ApJ,  505, 74

\bibitem{Getal96}
Griffiths, R. E., Casertano, S., Im, M., \& Ratnatunga, K. U. 1996,
MNRAS, 282, 1159

\bibitem{gg}
Gunn, J. E. \& Gott, J. R. 1972, ApJ, 176, 1

\bibitem{HK87}
Hinshaw, G., \& Krauss, L. M. 1987, ApJ, 320, 468

\bibitem{hiotelis}
Hiotelis N. 2002, A\&A, 382, 84

\bibitem{HS}
Hoffman, Y. \& Shaham, J. 1985, ApJ, 297, 16

\bibitem{hbf}
Hozumi, S., Burkert, A. \& Fujiwara, T. 2000, MNRAS, 311, 377

\bibitem{Hetal98}
Hudson, M. J., Gwyn, S. D. J., Dahle, H., \& Kaiser, N. 1998, ApJ, 503, 531

\bibitem{HJS99}
Huss, A., Jain, B., \& Steinmetz, M. 1999, MNRAS, 308, 1011

\bibitem{I00}
Iliev, I. T. 2000, PhD Thesis (University of Texas at Austin)

\bibitem{ISMS03}
Iliev, I. T., Scannapieco, E., Martel, H., \& Shapiro, P. R. 2003,
MNRAS, 341, 81

\bibitem{ISa}
Iliev, I.T. \& Shapiro, P.R. 2001a, ApJ, 546, L5

\bibitem{ISb}
Iliev, I.T., \& Shapiro, P.R. 2001b, MNRAS, 325, 468 (Paper II)

\bibitem{J91}
Jaroszy\'nski, M. 1991, MNRAS, 249, 430

\bibitem{J92}
Jaroszy\'nski, M. 1992, MNRAS, 255, 655

\bibitem{JS00}
Jing, Y., \& Suto, Y. 2000, ApJ, 526, L69

\bibitem{JS02}
Jing, Y., \& Suto, Y. 2002, ApJ, 574, 538

\bibitem{JF}
Jones, C. \& Forman, W. 1999, ApJ, 511, 65

\bibitem{JKH01}
Jang-Condell, H., \& Hernquist 2001, ApJ, 548, 68

\bibitem{KKT}
Kaplinghat, M., Knox, L., \& Turner, M.S. 2000, Phys.Rev.Lett., 85, 3335

\bibitem{KM01}
Keeton, C. R., \& Madau, P. 2001, ApJ, 549, L29

\bibitem{KKVP}
Klypin, A. A., Kravtsov, A. V., Valenzuela, O., \& Prada, F. 1999, ApJ, 522, 82

\bibitem{Ketal01}
Klypin, A., Kravtsov, A. V., Bullock, J. S., \& Primack, J. R. 2001,
ApJ, 554, 903

\bibitem{KNE} Knebe, A., Devriendt, J.E.G., Mahmood, A., Silk J. 2002, 
MNRAS, 329, 813

\bibitem{K95}
Kochanek, C. S. 1995, ApJ, 453, 545

\bibitem{Ketal98}
Kochanek, C. S., Falco, E. E., Impey, C., Lah\'ar, J., McLeod, B.,
\& Rix, H.-W. 1998, CASTLE Survey Gravitational Lens Data Base
(Cambridge:CfA)

\bibitem{KW}
Kochanek, C. S. \& White, M. 2001, ApJ, 559, 531

\bibitem{KF96}
Kormendy, J. \& Freeman, K. C.: 'Scaling laws for
dark matter halos in late-type and dwarf spheroidal galaxies'.
In \emph{Ringberg Proceedings 1996 Workshop},
eds. R. Bender, T. Buchert, P. Schneider, \& F. von Feilitzsch (MPI, Munich
1996), pp. 13 - 15

\bibitem{KF01}
Kormendy, J. \& Freeman K.C. 2004, in preparation

\bibitem{KKBP}
Kravtsov, A.V., Klypin, A.A., Bullock, J.S., \& Primack, J.R. 1998 
ApJ, 502, 48

\bibitem{kull}
Kull, A. 1999, ApJ, 516, L5

\bibitem{LO02}
Li, L.-X., \& Ostriker, J. P. 2002, ApJ, 566, 652

\bibitem{LH} 
Lokas, E.L., Hoffman, Y. 2000, ApJ, 542, L139 

\bibitem{LM} 
Lokas, E.L., Mamon, G.A. 2001, MNRAS, 321, 155

\bibitem{MPM02}
Martel, H., Premadi, P., \& Matzner, R. 2002, ApJ, 570, 17

\bibitem{MS} Martel, H., Shapiro, P.R. 2001, RevMexSC, 10, 101

\bibitem{MOVK}
Mateo, M., Olszewski, E.W., Vogt, S.S., \& Keane, M.J. 1998, ApJ, 116, 2315 

\bibitem{ME}
Mathiesen, B.F., \& Evrard A.E. 2001, ApJ, 546, 100

\bibitem{MME}
Mohr, J.J., Mathiesen, B.F., \& Evrard A.E. 1999, ApJ, 517, 627

\bibitem{MOO}
Moore, B. 1994, Nature, 370, 629

\bibitem{M01}
Moore, B., 2001, in AIP Conf. Series 586, 
Relativistic Astrophysics, Proceedings of the 20th
Texas Symposium, eds. J. C. Wheeler \& H. Martel, p. 73

\bibitem{Metal98}
Moore, B., Governato, F., Quinn, T., Stadel, J., \& Lake, G. 1998, ApJ, 499, L5

\bibitem{MGGLQST}
Moore, B., Ghigna, S., Governato, F., Lake, G., Quinn, T., Stadel, J., \& Tozzi, P. 
1999, ApJ, 524, L19

\bibitem{Metal99}
Moore, B., Quinn, T., Governato, F., Stadel, J., \& Lake, G. 1999, MNRAS, 310, 1147

\bibitem{MB00}
Mori M. \& Burkert A. 2000, 538, 559

\bibitem{NW88}
Narayan, R., \& White, S. D. M. 1988, MNRAS, 231, 97

\bibitem{Netal96}
Natarajan, P., Kneib, J.-P., Smail, I., \& Ellis, R. S. 1998, ApJ, 499, 600

\bibitem{NL97}
Natarajan, P., \& Lynden-Bell, D. 1997, MNRAS, 286, 268

\bibitem{NFW96}
Navarro, J.F, Frenk, C. S., White, S. D. M. 1996, ApJ, 462, 563

\bibitem{NFW97}
Navarro, J. F., Frenk, C. S., \& White, S. D. M. 1997, ApJ, 490, 493

\bibitem{Navarro03}
Navarro, J. F., Hayashi, E., Power, C., Jenkins, A. R., Frenk, C. S.,
White, S. D. M., Springel, V., Stadel, J., \& Quinn, T. R. 2004, MNRAS, 349,
1039

\bibitem{owv}
Owen, J. J., Weinberg, D. H. \& Villumsen, J. V. 1998, astro-ph/9805097

\bibitem{OWE} Owen, J.M., Villumsen, J.V., Shapiro, P.R., Martel, H., 
1998, ApJS, 116, 155

\bibitem{Petal03}
Power, C., Navarro, J. F., Jenkins, A., Frenk, C. S., 
White, S. D. M., Springel, V., Stadel, J., \& Quinn, T. 2003,
MNRAS, 338, 14

\bibitem{PM04}
Premadi, P., \& Martel, H. 2004, ApJ, in press

\bibitem{PMM98}
Premadi, P., Martel, H., \& Matzner, R. 1998, ApJ, 493, 10

\bibitem{Petal01a}
Premadi, P., Martel, H., Matzner, R., \& Futamase, T. 2001a, 
ApJ Suppl., 135, 7

\bibitem{Petal01b}
Premadi, P., Martel, H., Matzner, R., \& Futamase, T. 2001b, 
P.A.S.Aus., 18, 201

\bibitem{PS}
Press, W.H., \& Schechter, P. 1974, ApJ, 187, 425

\bibitem{Petal99}
Primack, J. R., Bullock, J. S., Klypin, A. A., \& Kravtsov, A. V. 1999,
in ASP Conf. Ser. 182,
Galaxy Dynamics, ed. D. R. Merritt, M. Valluri, and J. A. 
Sellwood (San Francisco: ASP)

\bibitem{Ricotti} Ricotti, M. 2003, MNRAS, 344, 1237

\bibitem{RT} Richstone, D.O. \& Tremaine, S. 1984, ApJ, 286, 27

\bibitem{RM01}
Rusin, D., \& Ma, C.-P. 2001, ApJ, 549, L33

\bibitem{rg}
Ryden, B. S. \& Gunn, J. E. 1987, ApJ, 318, 15

\bibitem{SAN}
Sand, D.J., Treu, T., Smith, G.P., \& Ellis, R.S. 2004, ApJ, 604, 88

\bibitem{SEF92}
Schneider, P., Ehlers, J., \& Falco, E. E. 1992, Gravitational Lenses
(New York: Springer)

\bibitem{S01}
Shapiro P.R.: `Cosmological Reionization'. In: Proceedings 
of the 20th Texas Symposium on
Relativistic Astrophysics and Cosmology, eds. H. Martel and 
J. C. Wheeler, (AIP Conference Series, v.586, 2001), pp. 219-232

\bibitem{SI00} 
Shapiro, P.R., \& Iliev, I.T. 2000, ApJ, 542, L1

\bibitem{SI02} 
Shapiro, P.R., \& Iliev, I.T. 2002,  ApJ, 565, L1

\bibitem{SIR}
Shapiro, P.R., Iliev, I.T., \& Raga, A.C. 1999,  MNRAS, 307, 203 (Paper I)

\bibitem{SHA} Shapiro, P.R., Martel, H., Villumsen, J.V., Owen, J.M., 
1996, ApJS, 103, 269

\bibitem{SSM} Shapiro, P.R. \& Struck-Marcell 1985, ApJS, 57, 205

\bibitem{SIM}
Simon, J.D., Bolatto, A.D., Leroy, A., \& Blitz, L. 2003, ApJ, 596, 957

\bibitem{Setal95}
Smail, I., Ellis, R. S., Fitchett, M. J., \& Edge, D. 1995, MNRAS, 273, 277

\bibitem{So01}
Soucail, G. 2001, in AIP Conf. Series 586, 
Relativistic Astrophysics, Proceedings of the 20th
Texas Symposium, eds. J. C. Wheeler \& H. Martel, p. 233

\bibitem{Setal03}
Spergel, D. N. et al.\ 2003, ApJS, 148, 175

\bibitem{SS}
Spergel, D. N. \& Steinhardt, P. J. 2000, PRL, 84, 3760

\bibitem{SPR}
Springel, V., Yoshida, N., White, S.D.M. 2001, NewA, 6, 79

\bibitem{sco}
Subramanian, K., Cen, R. \& Ostriker, J. P. 2000, ApJ, 538, 528

\bibitem{SWA1}
Swaters, R.A., Madore, B.F., van den Bosch, F.C., \& Balcelss, M. 2003, ApJ, 583, 732

\bibitem{SWA2}
Swaters, R.A., Verheijen, M.A.W., Bershady, M.A., \& Anderson, D.R. 2003, ApJ, 587, L19

\bibitem{TC01}
Takahashi, R., \& Chiba, T. 2001, ApJ, 563, 489

\bibitem{TKGK}
Tasitsiomi, A., Kravtsov, A. V., Gottl\"{o}ber, S., \& Klypin, A. A. 2004, ApJ, 607, 125

\bibitem{tca}
Teyssier, R., Chieze, J.-P. \& Alimi, J.-M. 1997, ApJ, 480, 36 

\bibitem{TW} Thoul, A. A. \& Weinberg, D.H. 1995, ApJ, 442, 480

\bibitem{TBW97}
Tormen, G. , Bouchet, F. R., \& White, S. D. M. 1997, MNRAS, 286, 865

\bibitem{TOG84}
Turner, E. L., Ostriker, J. P., \& Gott, J. R. 1984, ApJ, 284, 1

\bibitem{TKD98}
Tyson J.A., Kochanski, G.P., \& Dell'Antonio I.P. 1998, ApJ, 498, L107

\bibitem{TWV90}
Tyson, J. A., Wenk, R. A., \& Valdes, F. 1990, ApJ, 349, L1

\bibitem{VAL} Valinia, A., Shapiro, P.R., Martel, H., Vishniac, E.T. 
1997, ApJ, 479, 46

\bibitem{VDB} van den Bosch, F. C. 2002, MNRAS, 331, 98

\bibitem{WEC} Wechsler R.H., Bullock, J.S., Primack, J.R., Kravtsov, A.V., \& 
Dekel, A. 2002, ApJ, 568, 52

\bibitem{WNB99}
Williams, L. L. R., Navarro, J. F., \& Bartelmann, M. 1999, ApJ, 527, 535

\bibitem{WB00}
Wright, C. O., \& Brainerd, T. G. 2000, ApJ, 534, 34

\bibitem{WTS01}
Wyithe, J. S. B., Turner, E. L., \& Spergel, D. N. 2001, ApJ, 555, 504

\bibitem{Yoshida}
Yoshida, N., Springel, V., White, S. D. M. \&
Tormen, G. 2000, ApJ, 544, L87 

\bibitem{Yetal80}
Young, P., Gunn, J. E., Kristian, J., Oke, J. B., \& Westphal, J. A. 1980,
ApJ, 241, 507

\bibitem{ZHA}
Zhao, D.H., Mo, H.J., Jing, Y.P., \& Borner, G., 2003, 339, 12

\end{thebibliography}
\end{document}